\documentclass[twocolumn]{aastex63}

\graphicspath{{./}{figures/}}

\usepackage{color}
\usepackage{amsmath}
\usepackage{xcolor}
\usepackage{mathrsfs}
\usepackage{amssymb}
\usepackage{natbib}
\usepackage{gensymb}
\usepackage{xspace}
\usepackage{subfigure}
\usepackage{placeins}
\usepackage{graphicx}

\newcommand{\Rj}{\ensuremath{R_{\rm{Jup}}}\xspace}
\newcommand{\Mj}{\ensuremath{M_{\rm{Jup}}}\xspace}

\newcommand{\Msun}{\ensuremath{M_{\odot}}\xspace}
\newcommand{\Lsun}{L_\odot}
\newcommand{\Teff}{\ensuremath{T_\mathrm{eff}}\xspace}
\newcommand{\logg}{\ensuremath{\log{g}}\xspace}
\newcommand{\lbollsun}{\ensuremath{\log(L_\mathrm{bol}/\Lsun)}}
\newcommand{\lbol}{\ensuremath{L_\mathrm{bol}}}
\newcommand{\vsini}{\ensuremath{v \sin i}\xspace} 

\newcommand{\cratio}{\ensuremath{\mathrm{^{12}C / ^{13}C}\xspace}}

\newcommand{\logco}{\ensuremath{\mathrm{log(^{12}CO / ^{13}CO)}\xspace}}

\newcommand{\co}{\ensuremath{\mathrm{^{12}CO / ^{13}CO}\xspace}}

\newcommand{\kms}{km~s$^{-1}$\xspace}
\newcommand{\kapand}{$\kappa$\,And}
\newcommand{\kapandb}{$\kappa$\,And\,b}

\newcommand{\caltech}{Department of Astronomy, California Institute of Technology, Pasadena, CA 91125, USA}
\newcommand{\gps}{Division of Geological \& Planetary Sciences, California Institute of Technology, Pasadena, CA 91125, USA}
\newcommand{\ucsc}{Department of Astronomy \& Astrophysics, University of California, Santa Cruz, CA95064, USA}
\newcommand{\keck}{W. M. Keck Observatory, 65-1120 Mamalahoa Hwy, Kamuela, HI, USA}
\newcommand{\ucla}{Department of Physics \& Astronomy, 430 Portola Plaza, University of California, Los Angeles, CA 90095, USA}
\newcommand{\jpl}{Jet Propulsion Laboratory, California Institute of Technology, 4800 Oak Grove Dr.,Pasadena, CA 91109, USA}
\newcommand{\ucsd}{Department of Astronomy \& Astrophysics,  University of California, San Diego, La Jolla, CA 92093, USA}

\newcommand{\osu}{Department of Astronomy, The Ohio State University, 100 W 18th Ave, Columbus, OH 43210 USA}

\newcommand{\arizona}{James C. Wyant College of Optical Sciences, University of Arizona, Meinel Building 1630 E. University Blvd., Tucson, AZ. 85721}
\newcommand{\mpia}{Max-Planck-Institut für Astronomie, Königstuhl 17, 69117 Heidelberg, Germany}
\newcommand{\carnegiew}{Earth and Planets Laboratory, Carnegie Institution for Science, Washington, DC, 20015}
\newcommand{\toronto}{David A. Dunlap Institute Department of Astronomy \& Astrophysics, University of Toronto, 50 St. George Street, Toronto, ON M5S 3H4, Canada}

\newcommand{\upenn}{Department of Physics and Astronomy, University of Pennsylvania, Philadelphia PA 19104}
\newcommand{\kansas}{Department of Physics and Astronomy, University of Kansas, Lawrence, KS 66045, USA}

\begin{document}

\title{Are these planets or brown dwarfs? Broadly solar compositions from high-resolution atmospheric retrievals of $\sim$10--30~\Mj companions}

\correspondingauthor{J. Xuan}
\email{wxuan@caltech.edu}

\author[0000-0002-6618-1137]{Jerry W. Xuan}
\affiliation{\caltech}

\author[0000-0002-5370-7494]{Chih-Chun Hsu}
\affil{Center for Interdisciplinary Exploration and Research in Astrophysics (CIERA), Northwestern University,
1800 Sherman, Evanston, IL, 60201, USA}

\author[0000-0002-1392-0768]{Luke Finnerty}
\affiliation{\ucla}

\author[0000-0003-0774-6502]{Jason J. Wang}
\affiliation{Center for Interdisciplinary Exploration and Research in Astrophysics (CIERA) and Department of Physics and Astronomy, Northwestern University, Evanston, IL 60208, USA}

\author[0000-0003-2233-4821]{Jean-Baptiste Ruffio}
\affiliation{\ucsd}

\author[0000-0003-0097-4414]{Yapeng Zhang}
\affiliation{\caltech}

\author[0000-0002-5375-4725]{Heather A. Knutson}
\affiliation{\gps}

\author{Dimitri Mawet}
\affiliation{\caltech}
\affiliation{\jpl}

\author[0000-0003-2008-1488]{Eric E. Mamajek}
\affiliation{\jpl}

\author[0000-0001-9164-7966]{Julie Inglis}
\affiliation{\gps}

\author[0000-0003-0354-0187]{Nicole L. Wallack}
\affiliation{\carnegiew}

\author[0000-0002-6076-5967]{Marta L. Bryan}
\affiliation{\toronto}

\author{Geoffrey A. Blake}
\affiliation{\gps}

\author{Paul Mollière}
\affiliation{\mpia}

\author[0000-0001-5541-6087]{Neda Hejazi}
\affiliation{\kansas}

\author{Ashley Baker}
\affiliation{\caltech}

\author{Randall Bartos}
\affiliation{\jpl}

\author[0000-0003-4737-5486]{Benjamin Calvin}
\affiliation{\ucla}

\author{Sylvain Cetre}
\affiliation{\keck}

\author[0000-0001-8953-1008]{Jacques-Robert Delorme}
\affiliation{\keck}

\author{Greg Doppmann}
\affiliation{\keck}

\author[0000-0002-1583-2040]{Daniel Echeverri}
\affiliation{\caltech}

\author[0000-0002-0176-8973]{Michael P. Fitzgerald}
\affiliation{\ucla}

\author[0000-0001-5213-6207]{Nemanja Jovanovic}
\affiliation{\caltech}

\author[0000-0002-4934-3042]{Joshua Liberman}
\affiliation{\caltech}
\affiliation{\arizona}

\author[0000-0002-2019-4995]{Ronald A. L\'opez}
\affiliation{\ucla}

\author{Evan Morris}
\affiliation{\ucsc}

\author{Jacklyn Pezzato}
\affiliation{\caltech}

\author[0000-0003-1399-3593]{Ben Sappey}
\affiliation{\ucsd}

\author{Tobias Schofield}
\affiliation{\caltech}

\author{Andrew Skemer}
\affiliation{\ucsc}

\author[0000-0001-5299-6899]{J. Kent Wallace}
\affiliation{\jpl}

\author[0000-0002-4361-8885]{Ji Wang}
\affiliation{\osu}

\author[0000-0003-2429-5811]{Shubh Agrawal}
\affiliation{\upenn}

\author{Katelyn Horstman}
\affiliation{\caltech}

\begin{abstract}
Using Keck Planet Imager and Characterizer (KPIC) high-resolution ($R$$\sim$35000) spectroscopy from $2.29-2.49~\mu$m, we present uniform atmospheric retrievals for eight young substellar companions with masses of $\sim$10--30~\Mj, orbital separations spanning $\sim$50--360~au, and $\Teff$ between $\sim$1500--2600~K. We find that all companions have solar C/O ratios, and metallicities, to within the 1-2$\sigma$ level, with the measurements clustered around solar composition. Stars in the same stellar associations as our systems have near-solar abundances, so these results indicate that this population of companions is consistent with formation via direct gravitational collapse. Alternatively, core accretion outside the CO snowline would be compatible with our measurements, though the high mass ratios of most systems would require rapid core assembly and gas accretion in massive disks. On a population level, our findings can be contrasted with abundance measurements for directly imaged planets with $m<10~\Mj$, which show tentative atmospheric metal enrichment compared to their host stars. In addition, the atmospheric compositions of our sample of companions are distinct from those of hot Jupiters, which most likely form via core accretion. For two companions with $\Teff$$\sim$1700--2000~K (\kapandb~and GSC~6214-210~b), our best-fit models prefer a non-gray cloud model with $>3\sigma$ significance. The cloudy models yield $2-3\sigma$ lower $\Teff$ for these companions, though the C/O and [C/H] still agree between cloudy and clear models at the $1\sigma$ level. Finally, we constrain \co~for three companions with the highest S/N data (GQ~Lup~b, HIP~79098~b, and DH~Tau~b), and report $\vsini$ and radial velocities for all companions.

\end{abstract}


\section{Introduction} \label{sec:intro}
High-contrast imaging surveys have revealed a population of substellar companions, generally classified as giant planets ($\sim2-13~\Mj$) or brown dwarfs ($\sim13-75~\Mj$), orbiting at large separations ($\sim3-1000$ au) from their host stars \citep[see reviews by][]{bowler_imaging_2016, Currie_review_2023}. Between giant planets and brown dwarfs, there are also dozens of low-mass substellar companions ($m\sim$10--30~\Mj) at wide orbital separations (dozens to hundreds of au). These objects have often been termed `planetary-mass companions' \citep[e.g.][]{ireland_two_2010, kraus_three_2013}, though there is no conclusive evidence as to whether they form like planets. Insights into their formation processes would help provide more physically-based definitions for giant planets and brown dwarfs \citep{schlaufman_evidence_2018}, with giant planets being the product of bottom-up core accretion \citep{pollack_formation_1996}, and brown dwarfs the product of top-down gravitational collapse either in a disk or molecular cloud \citep[e.g.][]{offner_Formation_2010, bate_stellar_2012, Kratter2016}.  

These widely-separated 10--30~\Mj companions have occurrence rates of only a few percent \citep[e.g.][]{nielsen_gemini_2019}. The rarity of these companions aligns with the difficulties that they pose to both planet-like and star-like formation processes. The current orbital locations of many of these companions are too far for either core accretion or disk instability to operate efficiently given low surface densities at large distances ($>100$ au) in the disk \citep[e.g.][]{Dodson-Robinson2009}. On the other hand, cloud fragmentation has issues explaining the extreme mass ratios (a few percent) of these systems \citep[e.g.][]{bate_stellar_2012}. If these companions form via core accretion at closer distances followed by outward scattering, there should be close-in companions that served as the scatters, which have not yet been detected \citep{bryan_searching_2016, pearce_gsc_2018}. 

To understand the nature of directly imaged companions, the field has focused on two complementary approaches. The first examines their orbital architectures as a function of companion mass. Such studies have found evidence for distinct distributions of semi-major axis, orbital eccentricity, and stellar obliquity around a dividing mass of $\sim10-20~\Mj$ \citep{nielsen_gemini_2019, bowler_Populationlevel_2020, Bowler2023, Nagpal2023}, though the exact results can be sensitive to the specific dividing mass \citep{DoO2023}. This suggests a fuzzy boundary between giant planets and brown dwarfs and highlights the importance of further understanding the intermediate-mass companions with $\sim$10--30~\Mj. The second approach relies on the analysis of spectro-photometry, which contains information about the physical processes and chemical inventory of their atmospheres. Indeed, the atmospheric abundances of substellar companions encode fossil information about their accretion histories, and could potentially inform different formation scenarios \citep[e.g.][]{nowak_Peering_2020, molliere_Interpreting_2022a}. 

Early studies highlighted the carbon and oxygen abundances of the atmosphere as informative observables \citep[e.g.][]{oberg_effects_2011, madhusudhan_c/o_2012}. To first order, the C/O ratios of solids in the disk are predicted to vary as a function of disk radius. Planets that form via core accretion in a protoplanetary disk, which is a relatively slow process occurring on Myr timescales, can incorporate varying quantities of gas and solids into their atmospheres, potentially resulting in a wide range of atmospheric metallicities and C/O ratios. On the other hand, companions that form rapidly on dynamical timescales via direct gravitational collapse are expected to inherit C/O and metallicities similar to those of their host stars, analogous to the case of binary star systems \citep{hawkins_Identical_2020}. 

However, these predictions can be complicated by a range of effects. In particular, if widely-separated ($\gtrsim100$ au), $\sim$10--30~\Mj companions can form via core accretion outside the CO snowline, they are expected to have stellar C/O and metallicities as the solids at these locations are of stellar composition and the bulk of the metals is in the solid phase \citep{Chachan2023}. We may also see systematically lower atmospheric metallicities for objects that form via core accretion in the outer disk, as small grains could rapidly drain inward to the star in the absence of pressure gaps. On the other hand, if these companions form via disk instability, pressure bumps and spiral structures can lead to local enhancements or reductions in the disk metallicity that can be inherited by the companions \citep[e.g.][]{Boley2011}. Therefore, mapping the measured composition of a single planet/companion to a specific formation pathway is far from a simple one-to-one process and requires sophisticated disk models to fully disentangle the intricacies \citep{molliere_Interpreting_2022a}.

While there could be significant uncertainty in interpreting the composition for a single object, a dominant process would be more apparent as a trend in the population. In this regard, \citet{Wang2023b_arxiv} noted that several imaged planets with $m\approx3-13~\Mj$ have metallicities higher than their star's by $\approx$0.1--0.7 dex, with typical errors of 0.2 dex \citep{nowak_Peering_2020, Petrus2021, BrownSevilla2023, Wang2023, molliere_Retrieving_2020, Wang2023b_arxiv}, suggesting that they may have formed via core accretion. Recently, \citet{Zhang2023} also reported a potential $\sim$30--170$\times$ metal enrichment for AF~Lep~b ($\approx3\Mj$) relative to its star. These measurements are not without caveats. For example, \citet{Zhang2023} could not reliably constrain the C/O of AF Lep b from their low-resolution data and \citet{Landman2023} showed that high-resolution retrievals of $\beta$ Pic b lead to sub-stellar metallicity that disagree with the super-stellar metallicity found by \citet{nowak_Peering_2020} using low-resolution data. Despite these caveats, there is a possible trend of super-stellar metallicities for at least some directly imaged planets. In contrast, higher mass ($m\sim50-70~\Mj$) brown dwarf companions generally exhibit both C/O and bulk metallicities consistent with their stellar values \citep[e.g.][]{line_Uniform_2015, Xuan2022, wang_Retrieving_2022, Phillips2024}, which is expected given their presumed binary-star like formation pathways. A few exceptions to this trend of chemical homogeneity between high mass BDs and their stars have been attributed to missing physics in the modeling, rather than real differences \citep[e.g.][]{calamari_Atmospheric_2022, Balmer2023}. 

Systematically measuring the atmospheric compositions of $\sim$10--30~\Mj companions could help determine their nature. To date, only a few of these companions have reported abundance measurements \citep{Hoch2020, zhang_13COrich_2021, Hoch2022, Palma2023, Demars2023, Inglis2024}, with a trend of approximately solar C/O values \citep{Hoch2023}. With the exception of \citet{Inglis2024} however, all these studies employed medium-resolution spectroscopy ($R$$\sim$4000) and used self-consistent grid models to estimate the companion's abundances. Grid models with low dimensionality can provide poor fits to data but yield unrealistically tight constraints on the model parameters \citep[e.g.][]{ruffio_Deep_2021}, though this could be accounted for by inflating the uncertainties \citep{Hoch2020}. In addition, older grid models may contain outdated line lists, while the retrieval approach enables the incorporation of new line lists more easily. Retrievals also allow more flexibility in defining the cloud models and fitting for isotopic abundances. However, retrievals are not without caveats either. For example, retrieval studies often produce unphysically small radii \citep[e.g.][]{Gonzales2020, burningham_Cloud_2021, Lueber2022, Hood2023}, and overly isothermal profiles which might suggest inadequacies in the cloud models \citep[e.g.][]{burningham_retrieval_2017, BrownSevilla2023}. Ultimately, it is important to compare both approaches, for example, by using the information from self-consistent thermal profiles as informed priors in retrievals \citep{Zhang2023, Xuan2024}.

In this paper, we present systematic atmospheric retrievals for a sample of eight young ($\sim1-100$ Myr) companions with $m$$\sim$10--30~\Mj using Keck Planet Imager and Characterizer (KPIC) high-resolution spectroscopy ($R$$\sim$35,000, $K$ band). To ensure physical solutions, our retrievals are informed by mass and radius priors from evolutionary models, and self-consistent thermal profiles following \citet{Xuan2024}. We measure the C and O abundances of all companions and constrain the \cratio~for three companions with the highest signal-to-noise (S/N) data. With the statistical leverage of our sample and uniform analysis framework, we aim to understand whether this population of objects is more akin to high-mass giant planets or low-mass brown dwarfs. 

This paper is organized as follows: \S~\ref{sec:system_prop} overviews the known properties of our eight systems and uniformly estimates the stellar ages and companion bulk properties from evolutionary models. In \S~\ref{sec:obs_reduce}, we describe the KPIC observations and data reduction. \S~\ref{sec:model_framework} lays out our spectral analysis framework, including the atmospheric retrieval setup. The results of our retrievals are summarized in \S~\ref{sec:results}, and we discuss the implications of our measurements and analysis in \S~\ref{sec:discuss}. Finally, we present our conclusions in \S~\ref{sec:conclude}.

\section{System Properties} \label{sec:system_prop}

\begin{figure}[t]
    \centering
    \includegraphics[width=\linewidth]{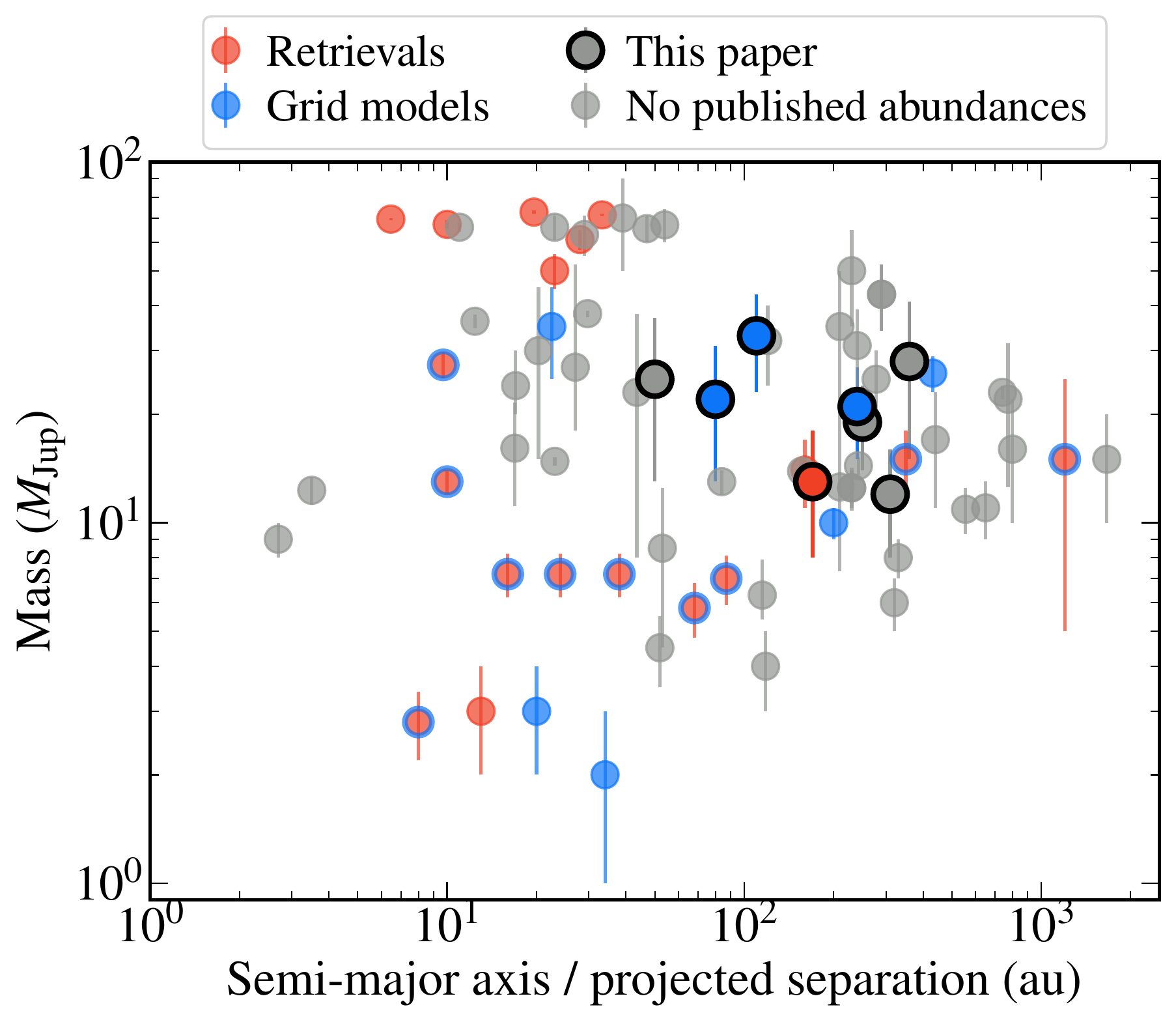}
    \caption{Confirmed directly imaged substellar companions that have published C/O and metallicity values from retrievals (red) and/or grid model fits (blue). Objects that have abundance measurements from both grid models and retrievals are shown as red points with a blue outline. The eight companions studied in this paper are denoted with a black outline. As shown, some of our objects have previous abundance measurements, which we summarize in \S~\ref{sec:overview_sys}.}
    \label{fig:DI_companions}
\end{figure}

In Fig.~\ref{fig:DI_companions}, we place our sample in the context of directly imaged companions with both C/O and metallicity measurements.\footnote{Before this work, directly imaged companions with measured abundances from both retrievals and grid model fits are: AF Lep b \citep{Zhang2023, Palma2024}, HR 8799 b, c, d, e \citep{lavie_HELIOS_2017, wang_Chemical_2020a, molliere_Retrieving_2020, Wang2023, konopacky_detection_2013, barman_simultaneous_2015, ruffio_Deep_2021}, $\beta$ Pic b \citep{nowak_Peering_2020, Landman2023}, HIP 62426 b \citep{Wang2023_hip65426, Petrus2021}, HD 206893 B \citep{Kammerer2021}, VHS~J1256–1257~b \citep{Gandhi2023, Petrus2023, Petrus2024}. 
The objects with abundances from retrievals are 51 Eri b \citep{Whiteford2023, BrownSevilla2023}, YSES 1 b \citep{zhang_13COrich_2021}, HD 4747 B \citep{Xuan2022}, HR 7276 B \citep{wang_Retrieving_2022}, Gl 229 B \citep{calamari_Atmospheric_2022, howe_GJ_2022}, HD 72426 B \citep{Balmer2023}, HD 33632 B \citep{Hsu2024_submitted}, HD 984 B \citep{Costes2024}. Finally, objects with abundances from grid models are PDS 70 b and c \citep{wang_Constraining_2021},  \kapandb~\citep{Hoch2020}, HD 284149 b \citep{Hoch2022}, AB Pic b \citep{Palma2023}, GQ Lup b and GSC 6214-210 b \citep{Demars2023}.} Our sample consists of six late K to early M type stars (GQ Lup, DH Tau, ROXs 12, ROXs 42B, 2M0122, GSC 6214-610) and two B stars (HIP~79098, \kapand), and are either confirmed or likely members of various nearby star-forming regions and/or young moving groups. Indeed, five of the systems are located in the Scorpius–Centaurus association. Below, we summarize the properties of the systems, with a focus on parameters relevant to our retrieval study.

\subsection{Stellar Ages}
The stellar, and by extension, system ages can inform the evolutionary states of substellar companions, including their radius and mass. To estimate ages of our stars, we either perform isochrone fitting or adopt a literature age when isochrone fitting is complicated by factors such as unresolved binarity. For isochrone fitting, we use the \citet{Baraffe2015} models (BHAC15) and fit using literature bolometric luminosity and \Teff measurements for the stars. We inflate the \Teff error bars to $150~K$ to be conservative, which is larger or equal to the reported \Teff uncertainties for our stars. For \lbol, we apply a correction based on the stars' Gaia DR3 parallaxes, as several measurements were reported using a pre-Gaia distance. We implement a rejection-sampling method to interpolate the models in mass and age space following \citet{Dupuy2017}. In short, for each mass-age pair, we compare the interpolated \Teff and \lbol~to the measured values and accept values based on their probability. The median and 68\% credible interval of the resulting age posteriors are listed in Table~\ref{tab:sys_prop}, and as we discuss in \S~\ref{sec:overview_sys}, are all consistent with previous age estimates in the literature.  

\begin{figure*}[t!]
    \centering
    \includegraphics[width=1.0\linewidth]{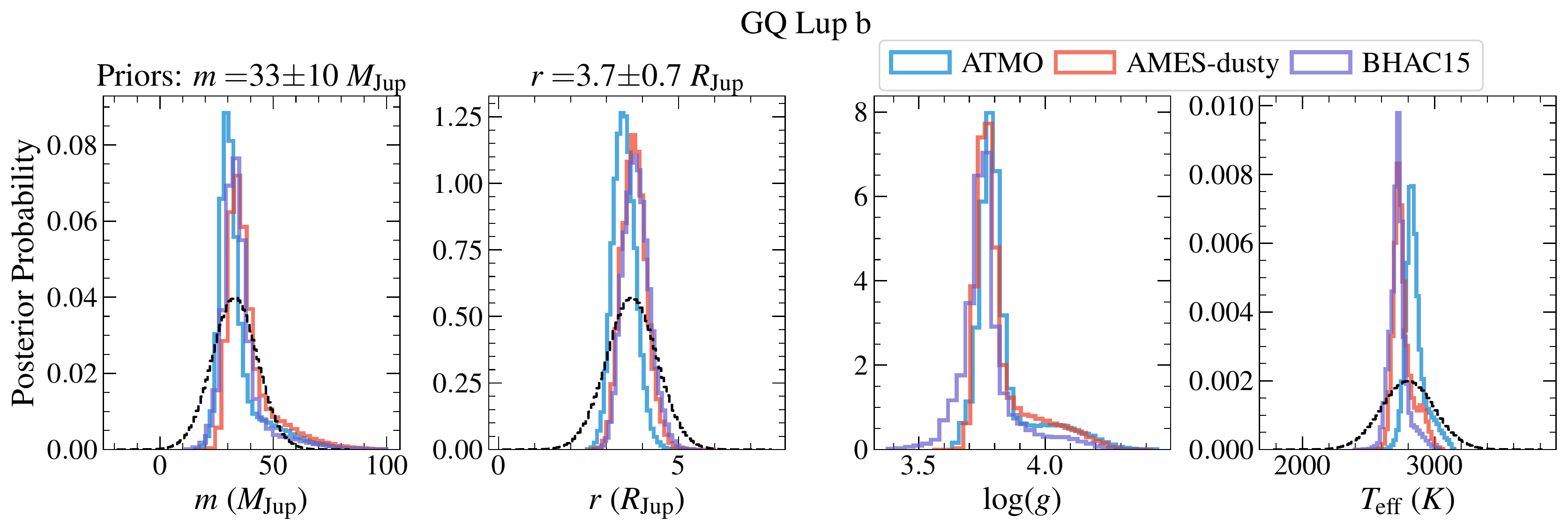}
    \caption{The interpolated mass, radius, \logg, and $\Teff$ from three different evolutionary models (ATMO 2020 in blue, \citealt{Phillips2020}; AMES-Dusty in red, \citealt{Allard2001}; BHAC15 in purple, \citealt{Baraffe2015}) for GQ~Lup~b. The dashed black lines indicate the adopted priors for mass and radius in the retrievals. Plots for the other companions are shown in Appendix~\ref{app:evol_model}.}
    \label{fig:mr_prior}
\end{figure*}

\subsection{Companion Mass and Radius}\label{sec:m_r_priors}
Given the stellar ages and literature measurements of the companions' \lbol, we can derive the expected companion mass and radius from substellar evolutionary models that have been shown to reasonably reproduce the bulk properties of benchmark substellar companions with dynamical masses \citep[e.g.][]{Dupuy2017} -- although the models can differ from each other. To derive these priors while accounting for model uncertainty, we consider four different evolutionary models: ATMO 2020 \citep{Phillips2020, Chabrier2023}, SM08 \citep{Saumon_2008}, AMES-Dusty \citep{Allard2001}, and BHAC15 \citep{Baraffe2015}.\footnote{Specifically, we use the chemical equilibrium models in ATMO 2020 and the hybrid cloud models in SM08. We note that the BHAC15 and SM08 grids do not cover the necessary parameter space for some of our companions, so we only use them when possible.}

With the same rejection-sampling technique as described above, we use the stellar ages and companion \lbol~to derive posteriors for mass, radius, $\Teff$, and $\logg$. As described in \S~\ref{sec:retr_setup}, we will use the evolutionary-model derived mass and radius as priors in our atmospheric retrievals. Since different models predict slightly different mass and radius, we visually determine mass and radius priors that encompass most of the posterior range for different models. These priors are listed in Table~\ref{tab:sys_prop}, and we show an example in Fig.~\ref{fig:mr_prior}. Plots for other companions are included in Appendix~\ref{app:evol_model}. We note that our derived masses and radii are consistent with previous estimates in the literature for each companion, and we rederive them for the sake of uniformity.

\subsection{Overview of the Systems}\label{sec:overview_sys}
Below, we summarize the properties of each system, including our derived stellar ages and companion properties. Note that most of our companions show little orbital motion since their discovery, so we quote their projected orbital separations at the epoch of the KPIC observations. Statistically, the most likely orbital semi-major axis is similar to the observed projected separation \citep{Yelverton2019}.

\subsubsection{ROXs 42B}
ROXs 42B is a resolved binary with $K$-band flux ratio $\sim$3 \citep{Ratzka2005} and a member of the $\rho$ Ophiuchus cloud ($\rho$ Oph). Our isochrone fits for the primary component yield an age of $2.2^{+1.8}_{-1.0}$ Myr, consistent with the $\rho$ Oph age from \citet{miretroig2022}, who confirmed the star's $\rho$ Oph membership from Gaia DR3 kinematics. A candidate companion was identified around the binary by \citet{Ratzka2005} and later confirmed by \cite{kraus_three_2013}. The companion is located at a projected separation $\rho\approx1.2\arcsec$ from the central binary, or about 170 au.\footnote{We report the projected separation at time of the KPIC observation for all companions, which is between 2020 to 2023.} The companion has a spectral type of $L1\pm1$ \citep{bowler_spectroscopic_2014}. Using the companion's \lbol~from \citet{Currie2014}, we estimate $m=13\pm5~\Mj$ and $r=2.10\pm0.35~\Rj$ for ROXs 42B b, consistent with previous studies \citep[e.g.][]{kraus_three_2013, Currie2014}. An early retrieval study was performed by \citet{Daemgen2017}, who used only 1--5~$\mu$m photometry and did not provide constraints on the companion's chemical abundances. Recently, \citet{Inglis2024} performed retrievals on pre-upgrade Keck/NIRSPEC spectra ($R$$\sim$25,000) of the companion, finding $\rm C/O=0.50\pm0.05$, $\rm [Fe/H]=-0.67\pm0.35$, and $\vsini=10.5\pm0.9~$\kms. ROXs 42 b has a mid-IR excess from Spitzer indicative of a circumsubstellar disk \citep{Martinez2021}, although the companion does not show any accretion features. 

\subsubsection{ROXs 12}
ROXs 12 is likely a member of $\rho$ Oph \citep{miretroig2022}. However, \citet{Luhman2022} assign the star memberships of either $\rho$ Oph or Upper Scorpius (Upper Sco). For consistency, we perform isochrone fitting to obtain $6.5^{+3.8}_{-2.6}$ Myr, consistent with the age estimate from \citet{kraus_three_2013}. A candidate companion to ROXs 12 was first noted by \citet{Ratzka2005} and later confirmed by \citet{kraus_three_2013}. The companion (ROXs~12~b) is located at a projected separation of $\approx$1.8$\arcsec$ or about 250 au. Based on the companion's \lbol~estimated by \citet{bowler_young_2017}, we find $m$=19$\pm5~\Mj$ and $r$=2.2$\pm0.35~\Rj$. \citet{bowler_young_2017} perform a detailed characterization of the system, and we summarize the results. They determine a spectral type of $L0\pm2$ for the companion and find that the companion is likely on a misaligned orbit relative to the host star's spin axis. These authors also find evidence of an outer tertiary component in the system at $5000$ au, which shares common proper motion and radial velocity as ROXs 12. A lack of Pa$\beta$ emission indicates there is no evidence of a disk around ROXs~12~b. \citet{bryan_Worlds_2020} use pre-upgrade Keck/NIRSPEC to measure $\vsini=8.4^{+2.1}_{-1.4}~$\kms for ROXs~12~b.

\subsubsection{DH Tau}
DH Tau is a member of Taurus. Given the large age scatter in Taurus \citep[e.g.][]{Luhman2023}, we perform isochrone fits to derive $0.7^{+0.3}_{-0.1}$ Myr. The companion to DH Tau was discovered by \citet{itoh_young_2005} and \citet{Luhman2006}, and orbits at a projected separation of $\approx$2.3$\arcsec$ or $\approx$310 au from DH Tau, which is part of an ultra-wide binary system (2210 au) with DI Tau \citep{Kraus_Hillenbrand_2009}. \citet{bonnefoy_library_2014} determine a spectral type of M$9-9.5$ for DH~Tau~b. Based on the companion's \lbol~from \citet{Luhman2006}, we estimate $m$=12$\pm4~\Mj$ and $r$=2.6$\pm0.6~\Rj$ for the companion. \citet{xuan_Rotation_2020} measured $\vsini=9.6\pm0.7$ \kms for DH~Tau~b using pre-upgrade Keck/NIRSPEC data and detected CO and H$_2$O in its spectrum. DH~Tau~b is likely accreting via a circumsubstellar disk, as evidenced by the presence of the $H\alpha$ line, excess optical continuum emission \citep{zhou_accretion_2014}, the Pa$\beta$ line \citep{bonnefoy_library_2014}, detection of linear polarization \citep{vanHolstein2021}, and mid-IR excess emission seen in Spitzer \citep{Martinez2021}.

\subsubsection{GQ Lup}
GQ Lup is an on-cloud member of Lupus 1. \citet{Galli2020} estimate an age of $1.2-1.8$ Myr for Lupus 1. For consistency, we carry out isochrone fits and obtain $2.5^{+1.5}_{-0.9}$ Myr. The companion GQ~Lup~b was discovered by \citet{Neuhauser2005} and has a projected separation of $\approx0.7\arcsec$ or $\approx110$ au. \citet{Alcala2020} found a wide $\sim0.15\Msun$ component at $2400$ au, which they conclude to be most likely bound to GQ Lup A, making this a likely triple system. Like DH~Tau~b, GQ~Lup~b likely hosts a circumsubstellar disk, as indicated by $H\alpha$ and Pa$\beta$ emission lines and an elevated optical continuum \citep{Seifahrt2007, zhou_accretion_2014, Demars2023}. \citet{Stolker2021} fitted 0.6--5 $\mu$m spectro-photometry of the companion and determined a spectral type of M9. They also found excess emission at 4--5~$\mu$m that can be explained from a blackbody with $T\approx460~K$, which they attribute to a disk around GQ~Lup~b. \citet{Demars2023} used VLT/SINFONI medium-resolution data and grid models based on ATMO to estimate C/O and metallicity for GQ~Lup~b. Their values are broadly consistent with a solar composition, but discrepant between different observing epochs at the $\sim60\%$ level in C/O and $>0.4$ dex in metallicity. From VLT/CRIRES spectroscopy of GQ~Lup~b, \citet{Schwarz2016} measured $\vsini=5.3^{+0.9}_{-1.0}$ \kms and made detections of CO and H$_2$O in the companion's atmosphere. Based on the companion's \lbol~from \citet{Stolker2021}, we estimate $m=33\pm10~\Mj$ and $r=3.7\pm0.7~\Rj$ for GQ~Lup~b. 

\subsubsection{GSC 06214-00210}
GSC 06214-00210 (hereafter GSC 6214-210) is an Upper Sco member according to \citet{miretroig2022}, who determine the star to be in the slightly older ``$\pi$ Sco'' sub-group. Our isochrone fits yield $22.2^{+10.7}_{-8.0}$ Myr, consistent with the results from \citet{pearce_gsc_2018} who found $16.9^{+1.9}_{-2.9}$ Myr. The companion was discovered by \citet{ireland_two_2010} and is separated by $2.2\arcsec$ on the sky or $\approx240$ au. \citet{bowler_PLANETS_2014} determine a spectral type of M$9.5\pm1$, and \citet{Bowler2011} detect Pa$\beta$ line emission, indicating GSC~6214-210~b possesses a circumsubstellar disk. \citet{Demars2023} used VLT/SINFONI data and ATMO grid models to estimate C/O and [M/H] for GSC~6214-210~b. Their values are broadly consistent with solar, but discrepant between different epochs at the $\sim70\%$ level in C/O and $\sim0.3$ dex in metallicity. \citet{bryan_constraints_2018} use pre-upgrade Keck/NIRSPEC to measure $\vsini=6.1^{+4.9}_{-3.8}~$\kms for this companion. Using the companion's \lbol~from \citet{pearce_gsc_2018}, we estimate $m=21\pm6~\Mj$ and $r=1.55\pm0.25~\Rj$ for GSC~6214-210~b. 

\subsubsection{2MASS J01225093-2439505}
2MASS J01225093-2439505 (hereafter 2M0122) is a member of AB Dor \citep{Malo2013}. Our isochrone fitting yields $144^{+105}_{-82}$ Myr, where the large error bars are due to the grid spacing. This age is consistent with the AB Dor age of $149^{+51}_{-19}$ Myr from \citet{bell_self-consistent_2015}. The companion 2M0122~b was detected by \citet{bowler_planets_2013} at a projected separation of $\approx1.4\arcsec$ or $\approx50$ au, who determine a spectral type of $L5\pm1$. Using the companion's \lbol~from \citet{Hinkley2015}, we estimate $m=25\pm12~\Mj$ and $r=1.2\pm0.2~\Rj$ for 2M0122~b.\footnote{The large mass error incorporates a bi-modal distribution in the inferred masses for this companion, which is located at an age-luminosity space where degeneracies in mass exist due to deuterium burning \citep{bowler_planets_2013}.}  \citet{Bryan_obliquity_2020} use pre-upgrade Keck/NIRSPEC to measure $\vsini=13.4^{+1.4}_{-1.2}~$\kms for 2M0122~b, which enabled a measurement of the companion's obliquity when combined with its photometric rotation period of $6.0^{+2.6}_{-1.0}$ hr from Hubble Space Telescope \citep{zhou_cloud_2019}.  

\subsubsection{\kapand}
\kapand~is a probable member of Columba \citep{Zuckerman2011}, and has a range of previous age measurements as summarized by \citet{Hoch2020}. Most recently, isochrone fitting from \citet{Jones2016} aided by an interferometric radius measurement of the star yields an age of $47^{+27}_{-40}$ Myr, broadly consistent with a Columba age. The BHAC15 grid does not go to high enough \Teff~for \kapand~($\Teff\approx11000~$K), so we adopt a uniform age prior between 5-100 Myr for this star based on the \citet{Jones2016} result to estimate the companion mass and radius. The companion \kapandb~was discovered by \citet{Carson2013}, has spectral type of $L0-1$ \citet{currie_SCExAO_2018}, and was at a projected separation of $\approx0.8\arcsec$ at time of KPIC observations. \citet{Hoch2020} present a spectral analysis with $R\sim4000$ $K$ band Keck/OSIRIS spectra and report C/O$=0.70^{+0.09}_{-0.24}$ and a metallicity of $-0.2-0.0$. They also carry out orbit fits to estimate $a=80^{+50}_{-20}$ au. With the companion's \lbol~from \citet{currie_SCExAO_2018}, we estimate $m=22\pm9~\Mj$ and $r=1.35\pm0.25~\Rj$. 

\subsubsection{HIP 79098}
HIP 79098 is likely an unresolved binary in Upper Sco \citep{Janson2019, Luhman_Esplin2020}. \citet{miretroig2022} find HIP 79098 to be in the ``$\sigma$ Sco'' sub-group of Upper Sco. Because the binary properties are unknown and the star exceeds the BHAC15 grid limits, we adopt the Upper Sco age of $10\pm3$ Myr for this system from \citet{Pecaut2016}. The companion HIP~79098~b was discovered by \citet{Janson2019} at a projected separation of $\approx2.4\arcsec$ or $\approx360$ au. This is the least studied companion in our sample, with only $J$, $H$, and $K$ band photometry. Unlike for previous companions, we use HIP~79098~b's absolute $K$ magnitude to estimate its \lbol~with the empirical $K$-$\lbol$~relation for young brown dwarfs from \citet{Sanghi2023}. From this, we estimate $m=28\pm13~\Mj$ and $r=2.6\pm0.6~\Rj$ for the companion.

\begin{deluxetable*}{cccccccccc}
\tablecolumns{11}
\renewcommand{\arraystretch}{1.2}
\tabletypesize{\footnotesize}
\tablewidth{\textwidth}
\tablecaption{\normalsize{System Properties for Young Companion Sample} \label{tab:sys_prop}}
\tablehead{\colhead{Target Name} & \colhead{Host SpT} & \colhead{Host $\lbol$\tablenotemark{b}} & \colhead{Host \Teff} & \colhead{Age} & \colhead{Comp. $\lbol$\tablenotemark{b}} & \colhead{Comp. Mass} & \colhead{Comp. Radius} & \colhead{References}\\
& \colhead{} & \colhead{($\lbollsun$)} & \colhead{($K$)} & \colhead{(Myr)} & \colhead{($\lbollsun$)} & \colhead{($\Mj$)} & \colhead{($\Rj$)} & \colhead{}
}
\startdata
\tableline
ROXs 42B\tablenotemark{a} b & M0 & $-0.23\pm0.10$ & $3850\pm150$ & $2.2^{+1.8}_{-1.0}$ & $-3.00\pm0.10$ & $13\pm5$ & $2.1\pm0.35$ & 3,8,11 \\
ROXs~12~b  & M0 & $-0.51\pm0.06$ & $3900\pm150$ & $6.5^{+3.8}_{-2.6}$ & $-2.81\pm0.10$ & $19\pm5$ & $2.2\pm0.35$ & 3,13,18 \\
DH~Tau~b & M2.3 & $-0.11\pm0.02$ & 3600$\pm$150 & $0.7^{+0.3}_{-0.2}$ & $-2.76\pm0.12$ & $12\pm4$ & $2.6\pm0.6$ & 1,15,17 \\
GSC~6214-210~b & K5 & $0.66\pm0.05$ & $4200\pm150$ & $22.2^{+10.7}_{-8.0}$ & $-3.35\pm0.10$ & $21\pm6$ & $1.55\pm0.25$ & 4,10,21,22 \\
2M0122-2439 b & M3.5 & $-1.78\pm0.11$ & $3400\pm150$ & $144^{+105}_{-82}$ & $-4.22\pm0.10$ & $25\pm12$ & $1.2\pm0.2$ & 2,12,19 \\
GQ~Lup~b & K7 & $0.02\pm0.10$ & $4300\pm150$ & $2.8^{+1.8}_{-1.1}$ & $-2.15\pm0.10$ & $33\pm10$ & $3.7\pm0.7$ & 6,9,14 \\
\kapandb & B9 & $1.88\pm0.03$ & $11100\pm150$ & $5-100$ & $-3.78\pm0.10$ & $22\pm9$ & $1.35\pm0.25$ & 7,8,16,23 \\
HIP~79098~b & B9 & $2.33\pm0.03$ & $11650\pm150$ & $10\pm3$ & $-2.60\pm0.20$ & $28\pm13$ & $2.6\pm0.6$ & 5,20,23 \\
\enddata
\tablenotetext{a}{ROXs 42B is a resolved binary \citep{Ratzka2005}. The ``B'' symbol here indicates it is the second brightest optical counterpart in the circle of the X-ray source ROXs 42; i.e. ROXs 42B is not physically associated with ROXs 42A. The \lbol~and \Teff~refer to those of the primary star, as calculated by \citet{kraus_three_2013} after accounting for the binary flux ratio.}
\tablenotetext{b}{Bolometric luminosities have been updated with Gaia DR3 parallaxes. Some of the literature companion \lbol~measurements have very small error bars ($<0.05$ dex) despite the limited wavelength coverage from which they are derived from. To be conservative, we adopt 0.1 dex uncertainties on the \lbol~when the quoted uncertainty is smaller than this.}
\tablerefs{(1) \citet{itoh_young_2005}, (2) \citet{bowler_planets_2013}, (3) \citet{kraus_three_2013}, (4) \citet{ireland_two_2010}, (5) \citet{Janson2019}, (6) \citet{Neuhauser2005}, (7) \citet{Carson2013}, (8) \citet{currie_SCExAO_2018}, (9) \citet{Stolker2021}, (10) \citet{pearce_gsc_2018}, (11) \citet{bowler_spectroscopic_2014}, (12) \citet{Hinkley2015}, (13) \citet{bowler_young_2017}, (14) \citet{Donati2012}, (15) \citet{Luhman2006}, (16) \citet{Jones2016}, (17) \citet{Yu2023}, (18) \citet{Ratzka2005}, (19) \citet{Sebastian2021}, (20) \citet{Pecaut2016}, (21) \citet{bowler_PLANETS_2014}, (22) \citet{Bowler2011}, (23) \citet{Gaia2022_DR3}
}
\end{deluxetable*}

\section{Observations and Data Reduction} \label{sec:obs_reduce}

\subsection{KPIC Observations}
We observed the companions in this study using the upgraded Keck/NIRSPEC \citep{martin_overview_2018, lopez_Characterization_2020}. The data were collected using both the first version of the KPIC fiber injection unit (FIU) \citep[2019-2021;][]{delorme_Keck_2021a}, and the upgraded phase 2 system \citep[2022-2023;][]{Echeverri2022}. The FIU is located downstream of the Keck II adaptive optics system and is used to inject light from a selected target into one of the single-mode fibers connected to NIRSPEC. For all targets, we obtained $R\sim35,000$ spectra in $K$ band, which is broken up into nine echelle orders from 1.94-2.49~$\mu$m. The observing strategy is similar to that of \citet{wang_Detection_2021}, although in some datasets we `nod' between two fibers to enable background subtraction between adjacent frames. The relative astrometry of each companion was computed using \href{http://whereistheplanet.com/}{whereistheplanet.com} \citep{wang_where_2021}, based on literature orbital solutions and unpublished data for \kapandb~(J. Wang, private communication). For calibration purposes, we also acquire spectra of the host stars before observing the companions, and spectra of a nearby telluric standard star (A0 or B9 spectral type) at similar airmass as the science target. The standard star is observed right before or after the associated science observations. Table~\ref{tab:obs} summarizes the observations reported in this paper. 

\begin{deluxetable*}{ccccccccc}[t]
    \caption{KPIC observations presented in this work. The throughput is end-to-end throughput measured from top of the atmosphere, and varies with wavelength due to differential atmospheric refraction and the instrumental blaze function. We report the $95\%$ percentile throughput over the $K$ band, averaged over all frames. We also report the median spectral SNR per pixel from 2.29-2.49~$\mu$m.}
    \tablehead{\colhead{Target} & \colhead{UT Date} & \colhead{Exposure Time [min]} & \colhead{Airmass} & \colhead{Throughput} & \colhead{Median SNR/pixel} & Proj. Sep. [arcsec]\tablenotemark{b} & $K$mag}
    \startdata
        GQ~Lup~b & 2023 June 23 & 99 & 1.8-2.4 & $\sim2.8\%$ & $\sim12$\tablenotemark{a} & 0.71 & 13.5 \\
        GSC~6214-210~b & 2023 June 20 & 105 & 1.3-1.7 & $\sim3.4\%$ & $\sim2$ & 2.19 & 15.0 \\
        HIP~79098~b & 2022 July 18 & 70 & 1.4-1.5 & $\sim3.7\%$ & $\sim6$ & 2.36 & 14.2 \\
        DH~Tau~b & 2022 Oct 12 & 50 & 1.2-1.5 & $\sim1.9\%$ & $\sim4$ & 2.34 & 14.2 \\
        ROXs~12~b & 2020 July 3 & 110 & 1.4-1.6 & $\sim2.3\%$ & $\sim4$ & 1.79 & 14.1 \\
        ROXs~42~Bb & 2020 July 2 & 160 & 1.4-1.6 & $\sim2.1\%$ & $\sim2$ & 1.17 & 15.0 \\
        2M0122-2439 b & 2021 Nov 19 & 80 & 1.4-1.5 & $\sim1.1\%$ & $\sim1$ & 1.45 & 14.5 \\
        \kapandb & 2022 Nov 12 & 180 & 1.1-1.4 & $\sim1.8\%$ & $\sim5$\tablenotemark{a} & 0.77 & 14.6 \\
    \enddata
    \tablenotetext{a}{For these two companions we quote the SNR from the companion light only; the SNR values are determined after fitting for the speckle contribution in the data.}
    \label{tab:obs}
    \tablenotetext{b}{For each companion, we quote the separation at the time of KPIC observations presented in this paper.}
\end{deluxetable*}

\subsection{Data Reduction}
We only briefly summarize the data reduction procedure in this paper and refer to \citet{Xuan2024} for additional details. For datasets using a single science fiber, we remove the thermal background from the raw images using combined instrument background frames taken before or after the night of observation. For datasets where we perform fiber-nodding, we apply nod-subtraction between adjacent frames, as the spectral traces of each fiber land on a different location in the detector. We also remove persistent bad pixels identified from the background frames. Then, we use data from a telluric standard star to fit the trace of each column in the four fibers and nine spectral orders, which gives us the position and standard deviation of the point spread function (PSF) in the spatial direction at each column. The trace positions and widths are smoothed using a cubic spline to mitigate random noise. We adopt the trace locations and widths as the line spread function (LSF) positions and widths in the spectral dispersion dimension.

For every frame, we then extracted the 1D spectra in each column of each order. To remove residual background light, we subtracted the median of pixels that are at least 5 pixels away from every pixel in each column. Finally, we used optimal extraction to sum the fluxes using weights defined by the 1D Gaussian LSF profiles calculated from spectra of the telluric star.

For our analysis, we use three spectral orders from 2.29-2.49~$\mu$m, which contain strong CO and H$_2$O absorption lines from the companions. The three spectral orders have gaps in between, and cover wavelengths of $2.29-2.34~\mu$m (order 33), $2.36-2.41~\mu$m (order 32), and $2.44-2.49~\mu$m (order 31), respectively.

\section{Spectral analysis} \label{sec:model_framework}
\subsection{Forward Model of KPIC High-resolution Spectra}\label{sec:forward_m}
Our forward model for KPIC data follows the framework of previous KPIC papers \citep[e.g.][]{wang_Detection_2021, Xuan2024}. Here, we present a brief summary. First, we generate atmospheric templates with \texttt{petitRADTRANS} \citep{molliere_petitRADTRANS_2019, molliere_Retrieving_2020}, which are shifted in RV and rotationally broadened using the function from \citet{Carvalho_2023}. Then, we convolve the RV-shifted and rotationally-broadened templates with the instrumental LSF determined from spectral trace widths in the spatial direction.\footnote{Following \citet{Xuan2024}, we allow the LSF width to vary between 1.0 and 1.2 times the width measured in the spatial direction when generating the instrument-convolved companion templates.}

Next, the atmospheric template is multiplied by the telluric and instrumental response ($T$ in Eq.~\ref{eqn:comp}), which is determined by dividing the standard star spectra by a PHOENIX-ACES model \citep{husser_new_2013} matching the standard star's \Teff and \logg. Since our standard stars have A0 or B9 spectral types, they have nearly no spectral lines in the wavelength region we use in our analysis (2.29-2.49~$\mu$m), mitigating errors due to an imperfect stellar spectrum. For six of our companions, which have projected separations $\sim$1.2--2.5$\arcsec$ and generally low companion-star contrasts, we find that the speckle intensity is negligible from preliminary analysis; when allowing for a speckle contribution the fit quality does not improve. The two exceptions are GQ~Lup~b and \kapandb, which are separated by $<$0.8$\arcsec$ from their host stars (see Table~\ref{tab:obs}). For these two datasets, we account for the significant speckle flux in the companion spectra using the on-axis observations of their host stars, taken immediately before the companion exposures.

The last step is to flux-normalize the companion and/or stellar models and multiply them by flux scale factors, which are in units of NIRSPEC counts. After scaling, the companion and speckle models are added in the case of GQ~Lup~b and \kapandb, while for the other companions, we only consider the companion flux. To summarize, the forward model is:
\begin{equation}
    \text{FM}_b = \alpha_b T M_b + \alpha_s D_s
    \label{eqn:comp}
\end{equation}
where $\text{FM}_b$ denotes the forward model fitted to the data, $\alpha_b$ and $\alpha_s$ are the flux scaling factors of the companion and speckle, $T$ is the combined telluric and instrumental response, $M_b$ is the companion template from \texttt{petitRADTRANS}, and $D_s$ is the observed KPIC spectra of the host star, which already has $T$ factored in. Note that for the six other companions besides GQ~Lup~b and \kapandb, $\alpha_s$ is taken to be zero.

Lastly, to remove the smoothly varying continuum in the KPIC spectra, we apply high-pass filtering with a median filter of 100 pixels ($\sim0.002~\mu$m) on the data and forward model (FM$_b$) before computing the residuals. The choice of 100 pixels was determined from a series of injection-recovery tests by \citet{Xuan2022} as the optimal size for accurately retrieving molecular abundances in KPIC data. In Appendix~\ref{app:spline}, we also show results from an alternative continuum treatment with a spline model \citep{Ruffio2023, Agrawal2023}. 

\subsection{Preliminary Analysis for Molecular Detection}\label{sec:ccf}
\begin{figure*}[t!]
\centering
\begin{subfigure}
  \centering
  \includegraphics[width=.32\linewidth]{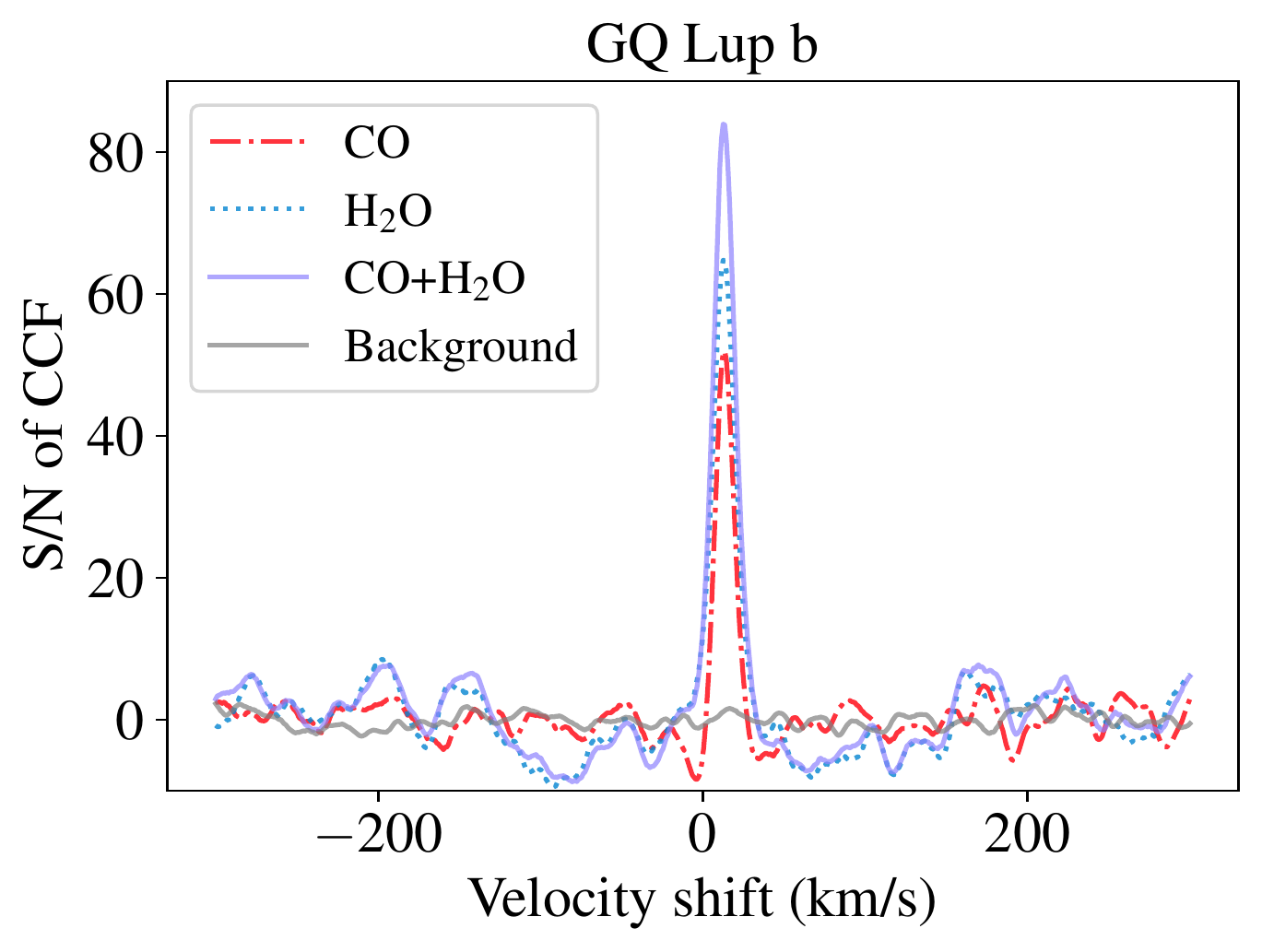}
\end{subfigure}
\begin{subfigure}
  \centering
  \includegraphics[width=.32\linewidth]{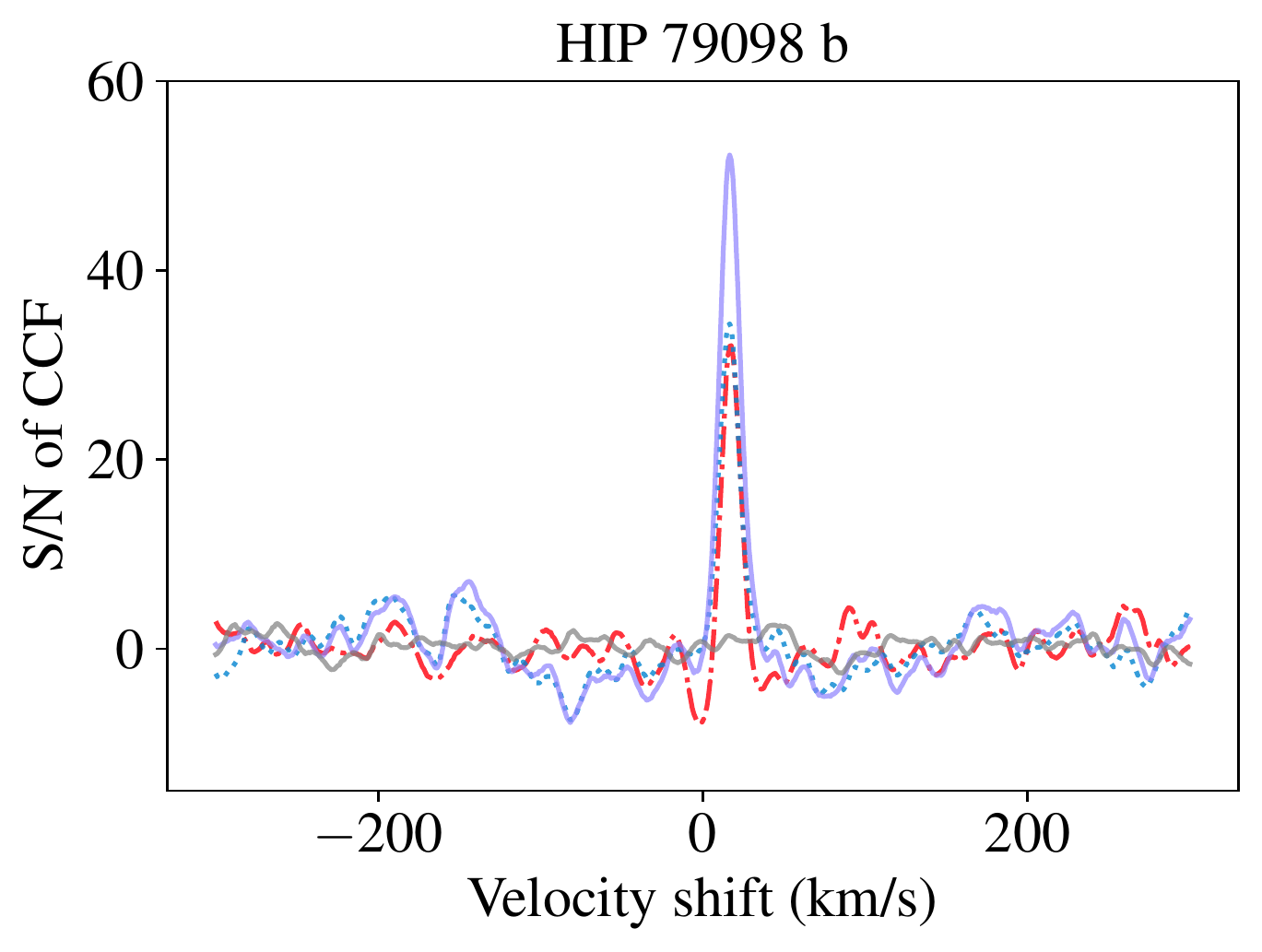}
\end{subfigure}
\begin{subfigure}
  \centering
  \includegraphics[width=.32\linewidth]{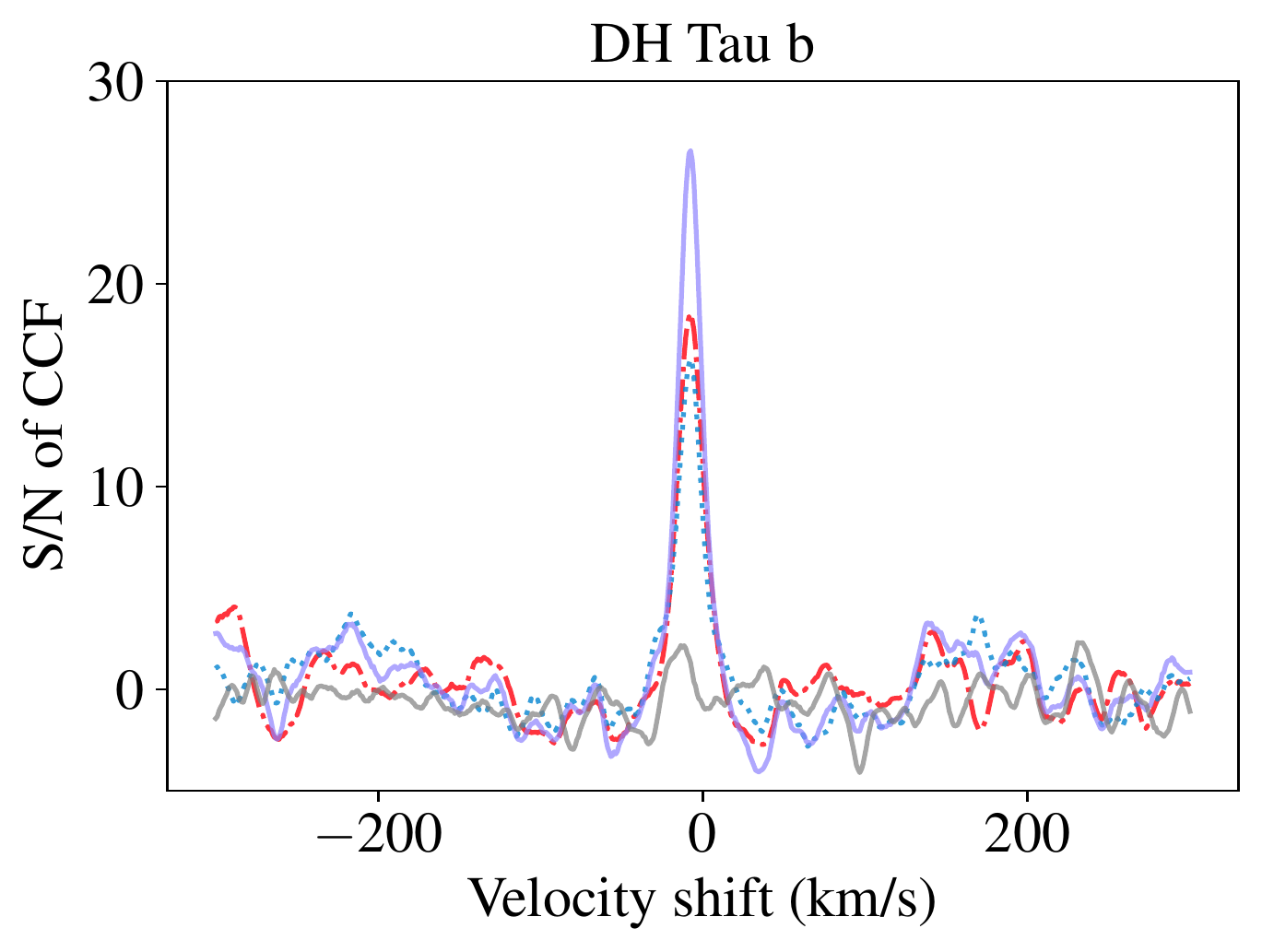}
\end{subfigure}
\begin{subfigure}
  \centering
  \includegraphics[width=.32\linewidth]{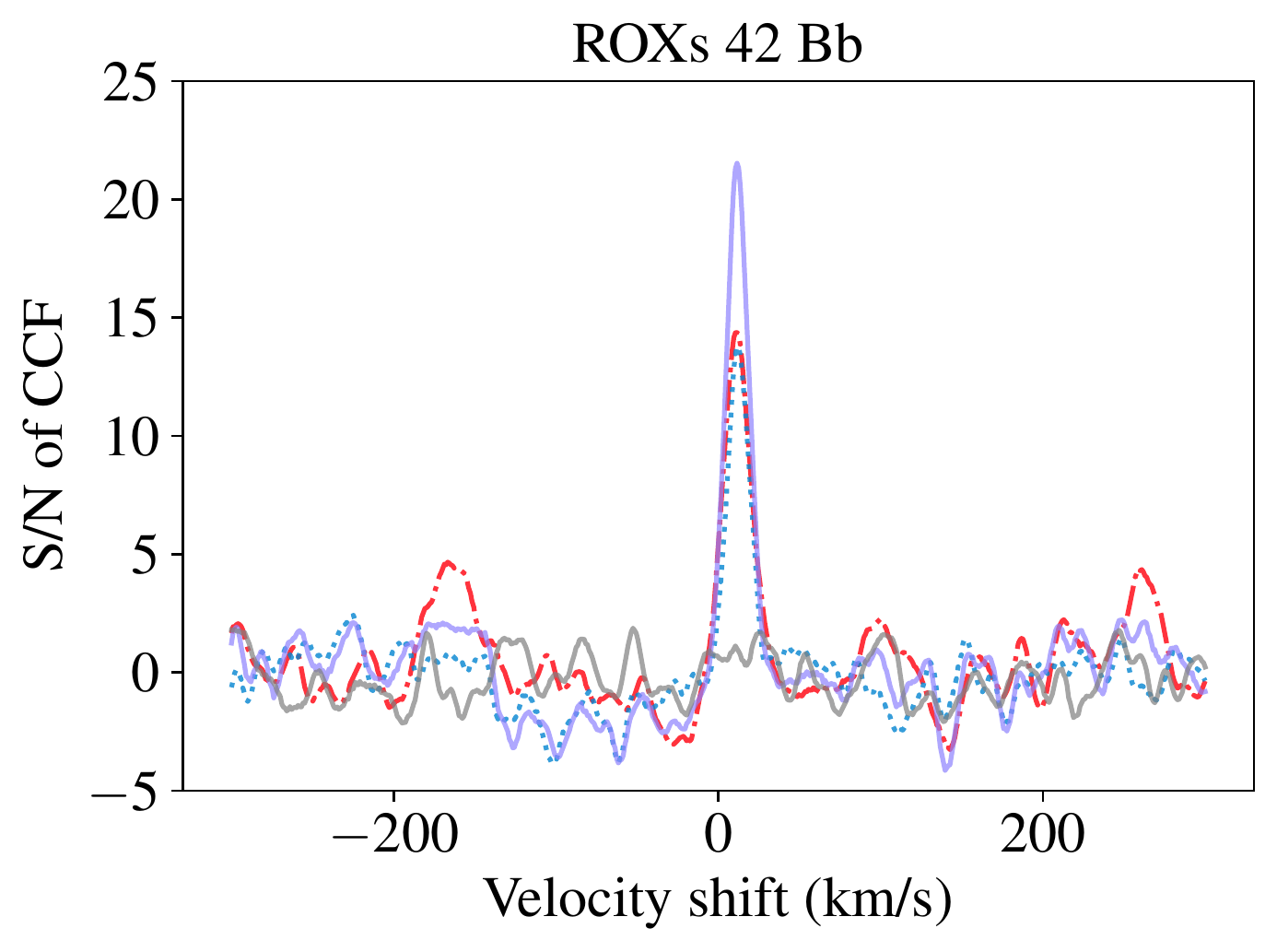}
\end{subfigure}
\begin{subfigure}
  \centering
  \includegraphics[width=.32\linewidth]{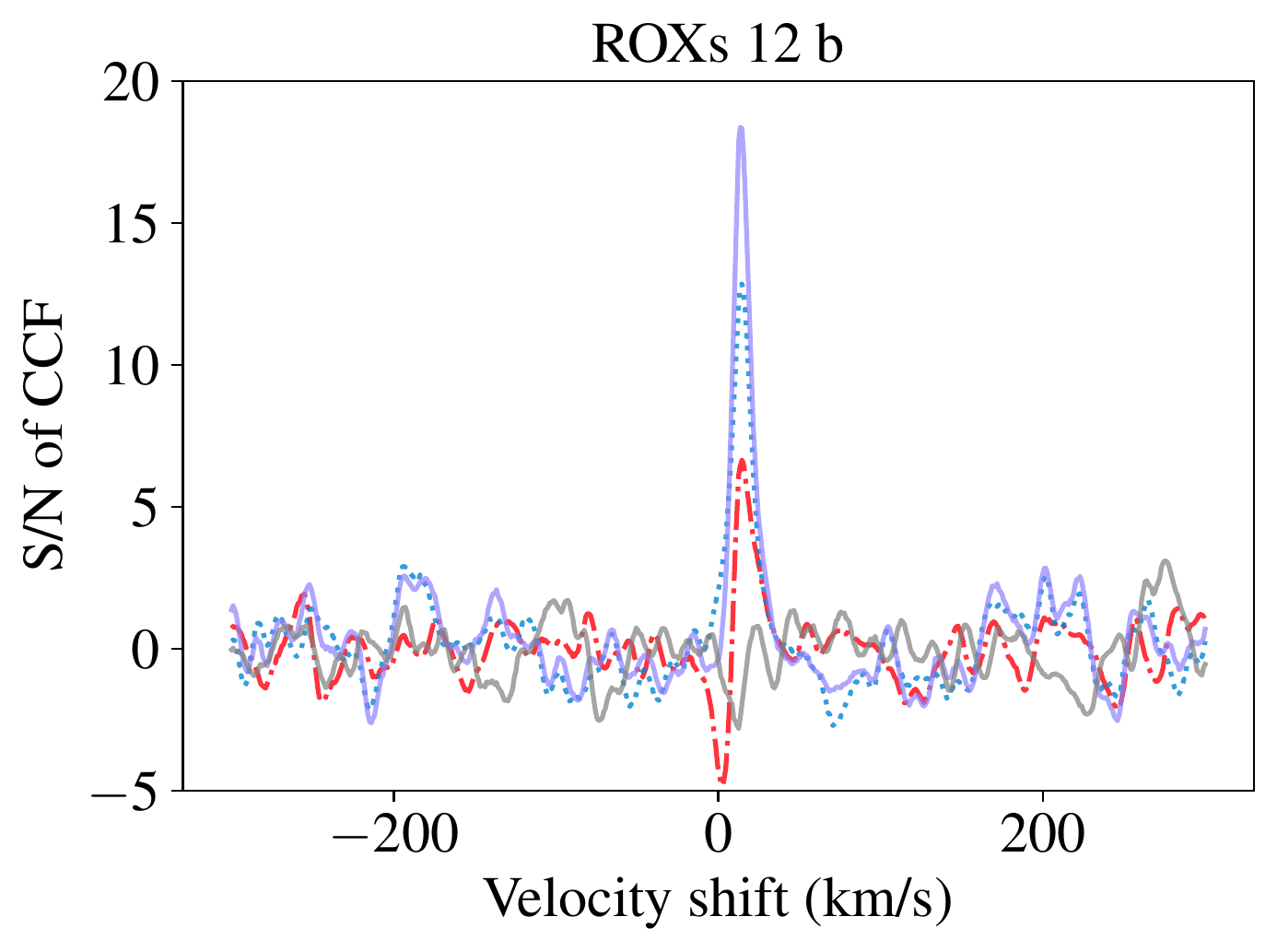}
\end{subfigure}
\begin{subfigure}
  \centering
  \includegraphics[width=.32\linewidth]{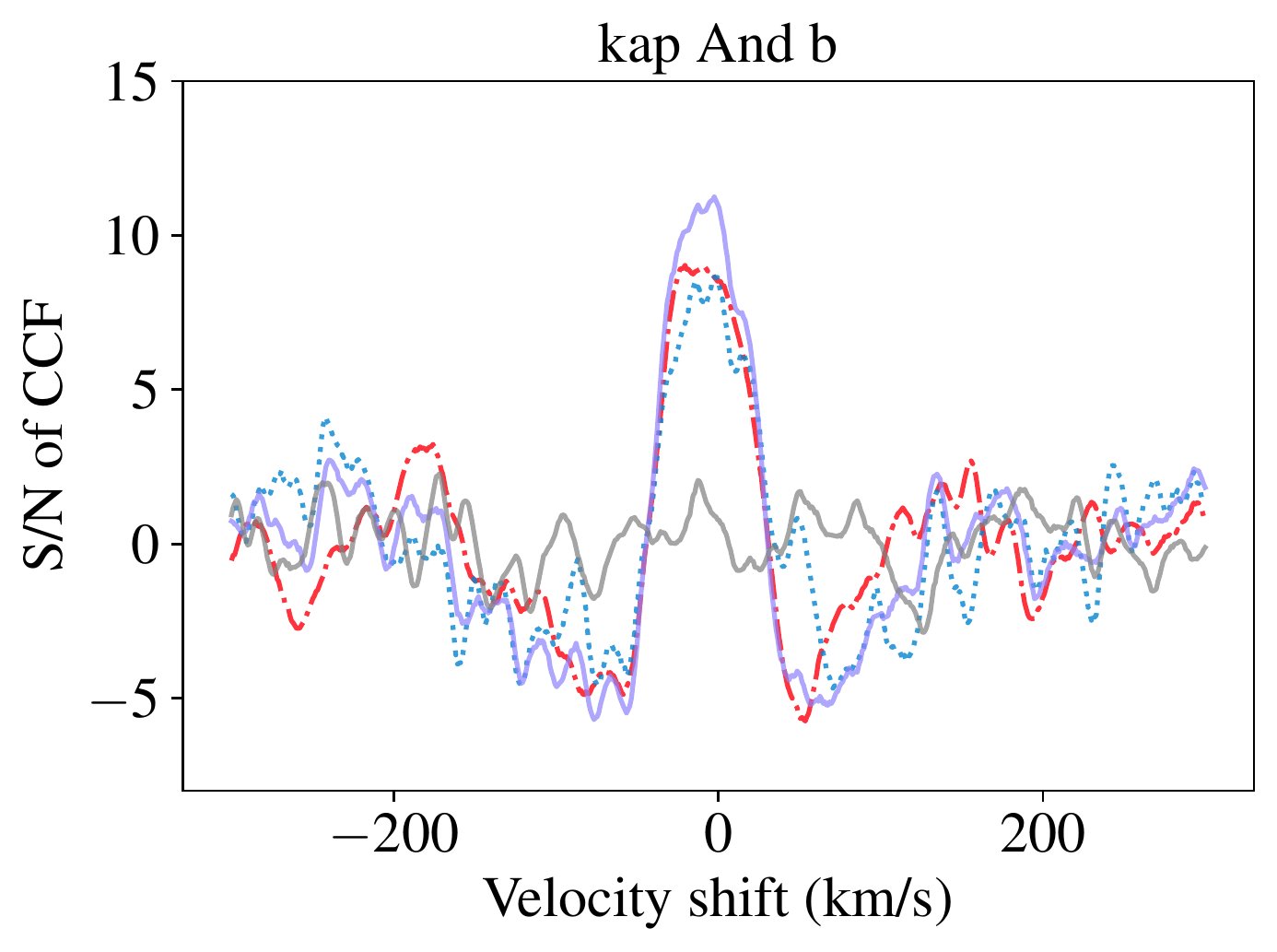}
\end{subfigure}
\begin{subfigure}
  \centering
  \includegraphics[width=.32\linewidth]{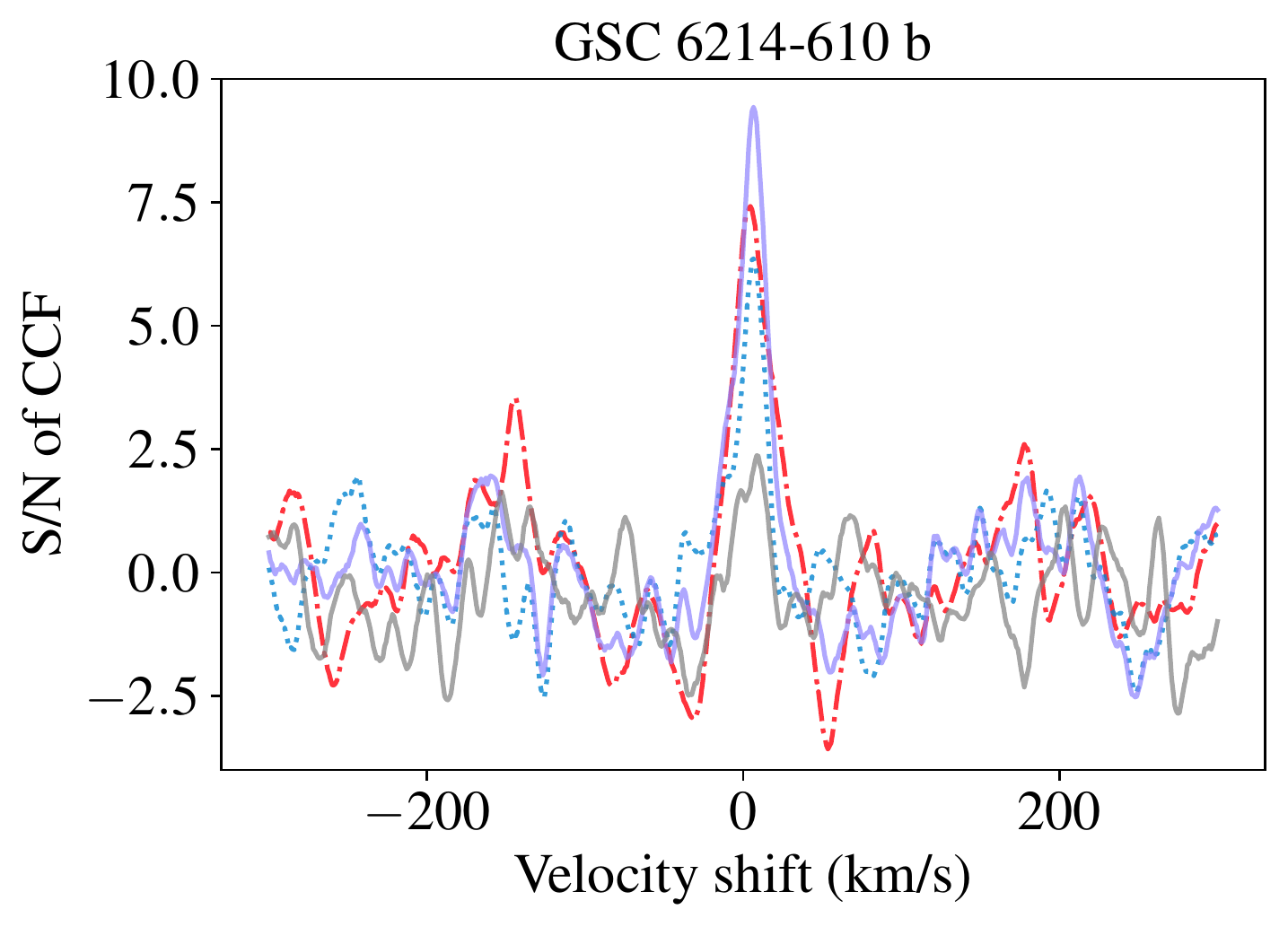}
\end{subfigure}
\begin{subfigure}
  \centering
  \includegraphics[width=.32\linewidth]{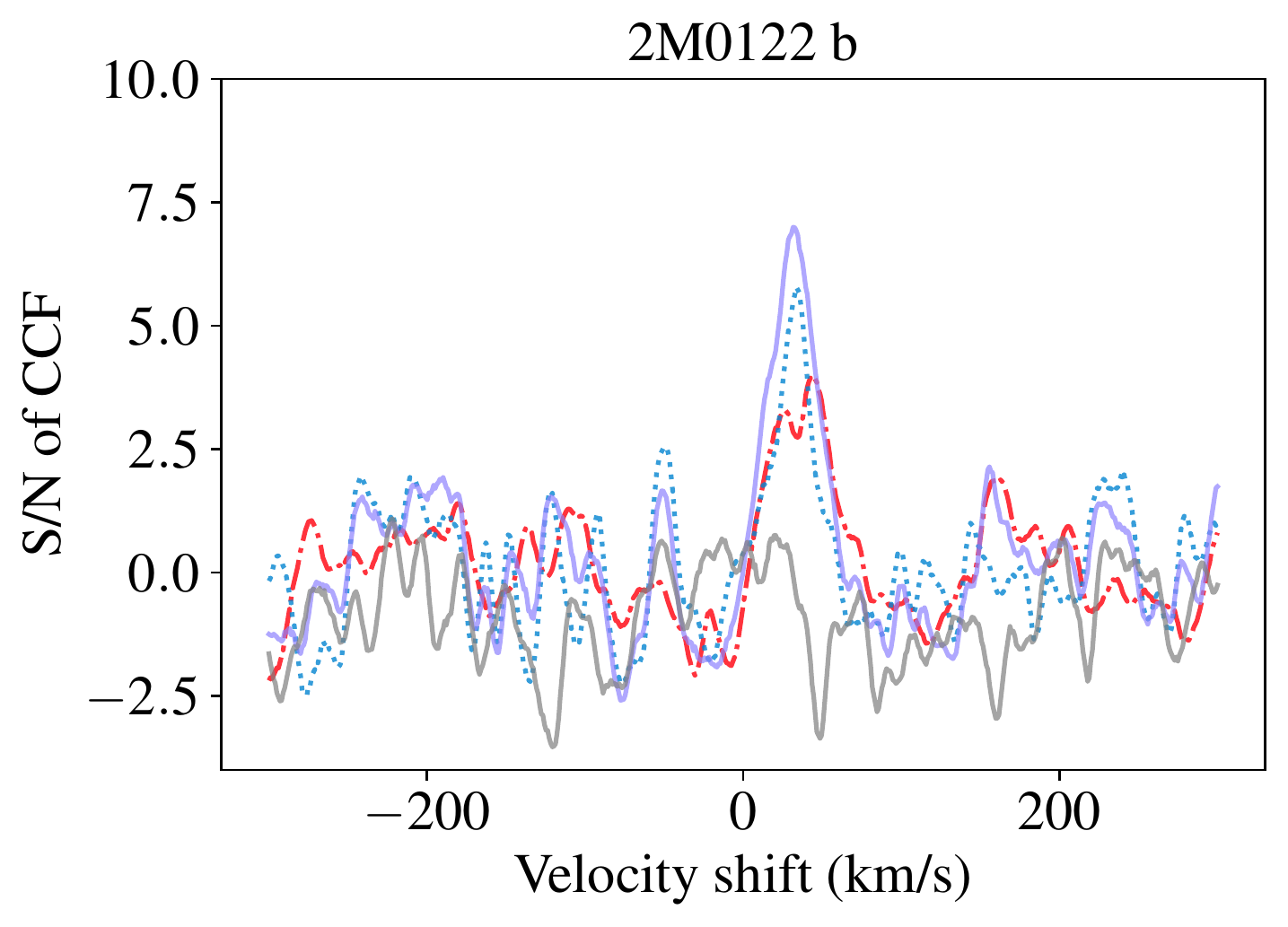}
\end{subfigure}
\caption{Cross-correlation functions (CCFs) with CO (red, dashdot lines), H$_2$O (blue, dotted lines), and CO+H$_2$O (purple, solid lines) molecular templates from \citet{marley_Sonora_2021}. Each panel is for a different companion, but shares the same legend. These CCFs are computed using 3 spectral orders from $2.29-2.49~\mu$m. The gray lines are CCFs of the CO+H$_2$O templates with background flux in the slit, and we use the standard deviations of the gray lines to estimate the CCF noise. The effect of rotational broadening is clearly visible in the \kapandb~CCF.}
\label{fig:ccf_detect}
\end{figure*}

To confirm detection of the companion signal in our data, we fit the KPIC spectra of each companion using atmospheric models from the cloudless \texttt{Sonora} grid \citep{marley_Sonora_2021}. We select \texttt{Sonora} models with $\Teff$ and \logg~that best match each companion's bulk properties, as estimated from the evolutionary models in \S~\ref{sec:m_r_priors}. Using the forward model framework described above, we estimate the maximum likelihood value for both the companion flux and speckle flux in the data as a function of RV shift, following \citet{ruffio_thesis_2019} and \citet{wang_Detection_2021}. This forward model framework allows us to estimate the companion flux (in data counts) as a function of RV, which can be interpreted as a cross-correlation function (CCF). In this paper, we refer to this as the CCF for simplicity, but note that it is not the same as the traditional CCF (e.g. as in eq.1 of \citealt{xuan_Rotation_2020}). To estimate the CCF S/N, we perform the same fitting procedure using a spectral trace that contains only background flux, and take the standard deviation of this background CCF as noise. We calculate CCFs for templates with CO, H$_2$O, and CO + H$_2$O, as shown in Fig.~\ref{fig:ccf_detect}. The companions are detected with CCF S/N between 7 to 80 when using the combined CO+H$_2$O template, and both CO and H$_2$O are individually detected with $\rm S/N>3$ for all companions. By detecting the major carbon and oxygen-bearing species in their atmospheres, we can constrain their atmospheric compositions with retrievals.

\subsection{Atmospheric Retrieval Setup and Inputs} \label{sec:retr_setup}
We use the radiative-transfer code \texttt{petitRADTRANS} to generate synthetic companion templates for use in atmospheric retrievals. These synthetic templates represent $M_b$ in Eq.~\ref{eqn:comp} of the forward model. We use the line-by-line opacity sampling mode, and down-sample the native $R=10^6$ opacities by a factor of 3 to speed up the retrievals. In our retrievals, we fit for the chemical abundances (\S~\ref{sec:chem}), cloud structure (\S~\ref{sec:cloud_choice}), and temperature profile (\S~\ref{sec:pt_setup}). We impose mass and radius priors motivated by evolutionary models in the retrievals (\S~\ref{sec:m_r_priors_retr}). Other parameters such as RV, \vsini, and flux scales ($\alpha_b$, $\alpha_s$) are also fitted for in our forward model. As an example, we summarize the fitted parameters in Table~\ref{tab:param_prior} for \kapandb.

\begin{deluxetable*}{llll}[t!]
\tablecaption{Fitted Parameters and Priors for \kapandb~Retrievals}\label{tab:param_prior}
\tabletypesize{\small}
\tablehead{Parameter & Prior & Parameter & Prior}
\startdata
Mass ($\Mj$) & $\mathcal{N}(22.0, 9.0)$ & Radius ($\Rj$) & $\mathcal{N}(1.35, 0.25)$  \\
$T_{\rm anchor}$ [log($P$)=-0.1)]$^{\rm (a)}$ (K) & $\mathcal{U}(1600, 2800)$ & RV (\kms) & $\mathcal{U}(-50 , 50)$ \\
$\Delta T_1$ [1.0 to 0.5] (K) & $\mathcal{U}(200, 700)$ & $v\sin{i}$ (\kms) & $\mathcal{U}(0, 80)$ \\
$\Delta T_2$ [0.5 to 0.3] (K) & $\mathcal{U}(50, 400)$ & $\rm C/O$ & $\mathcal{U}(0.1,1.0)$  \\
$\Delta T_3$ [0.3 to -0.1] (K) (K) & $\mathcal{U}(50, 600)$ & $\rm [C/H]$ & 
$\mathcal{U}(-1.5,1.5)$ \\
$\Delta T_4$ [-0.1 to -0.4] (K) & $\mathcal{U}(0, 500)$ & \logco & $\mathcal{U}(0, 6)$ \\
$\Delta T_5$ [-0.4 to -1.0] (K) & $\mathcal{U}(100, 750)$ & $f_{\rm sed}^{\rm (d)}$ (one for each cloud) & $\mathcal{U}(0, 10)$ \\
$\Delta T_6$ [-1.0 to -2.0] (K) & $\mathcal{U}(150, 650)$ & ${\rm log}(K_{\rm zz}/\rm{cm^2 s^{-1}})^{\rm (d)}$ & $\mathcal{U}(5, 13)$ \\ 
$\Delta T_7$ [-2.0 to -4.0] (K) & $\mathcal{U}(50, 700)$  & $\sigma_{\rm g}^{\rm (d)}$ & $\mathcal{U}(1.05, 3)$ \\ 
log(gray opacity/$\rm{cm}^2 g^{-1}$)$^{\rm (b)}$ & $\mathcal{U}(-6, 6)$  &  ${\rm log}(\tilde{X}_{\rm MgSiO_3})^{\rm (d)}$ & $\mathcal{U}(-2.3, 1)$ \\
Error multiple$^{\rm (c)}$ & $\mathcal{U}(1, 5)$ & ${\rm log}(\tilde{X}_{\rm Fe})^{\rm (d)}$ & $\mathcal{U}(-2.3, 1)$ \\
Comp. flux, $\alpha_c$ (counts) & $\mathcal{U}(0, 300)$ & Speckle flux, $\alpha_s$ (counts) & $\mathcal{U}(0, 300)$  \\ 
\enddata
\tablecomments{$\mathcal{U}$ stands for a uniform distribution, with two numbers representing the lower and upper boundaries. $\mathcal{N}$ stands for a Gaussian distribution, with numbers representing the mean and standard deviation. The $P$--$T$ parameters are described in \S~\ref{sec:pt_setup} and the cloud parameters are described in \S~\ref{sec:cloud_choice}.}
\tablenotetext{a}{The pressure at $T_{\rm anchor}$ and pressure points between which we fit $\Delta T$ values in our $P$--$T$ profile are given in square brackets. They are in log(bar) units.}
\tablenotetext{b}{Parameter for the gray opacity cloud model.}
\tablenotetext{c}{An error multiple term is fitted for KPIC data to account for any underestimation of the uncertainties.}
\tablenotetext{d}{Parameters for the EddySed cloud model.}
\end{deluxetable*}

\subsubsection{Opacities}
For the hottest companions in our sample, we find that there is contribution to the emission spectrum from regimes with $T>3000~$K (see Fig.~\ref{fig:ptprofile}), which exceeds $T_{\rm max}=3000~$K of default \texttt{petitRADTRANS} opacity tables. Therefore, we adopt the opacities generated by \citet{Xuan2024}, which go up to $4500~K$. We include the line opacities of H$_2^{16}$O \citep{Polyansky2018}, C$^{16}$O, $^{13}$CO \citep{Rothman2010}, OH \citep{Brooke2016}, FeH \citep{Dulick2003, Bernath2020}, TiO \citep{McKemmish2019}, AlH \citep{Yurchenko2018}, VO \citep{McKemmish2016}, H$_2$S \citep{Azzam2016}, and CH$_4$ \citep{hargreaves_Accurate_2020}. In addition, we include atomic opacities from Na, K, Mg, Ca, Ti, Fe, Al, and Si \citep{Kurucz2011}. For continuum opacities, we include the collision induced absorption (CIA) from H$_2$-H$_2$ and H$_2$-He, as well as the H- bound-free and free-free opacity. 

\subsubsection{Chemistry}\label{sec:chem}
We parameterize the chemical abundances with C/O and [C/H], where [C/H] is equivalent to the bulk metallicity. In other words, we assume [C/H]=[Fe/H]=[N/H] and so on,\footnote{We note that the assumption that $\rm [C/H]=[Fe/H]$ is only valid for companions that form outside the CO snowline \citep{Chachan2023}. Our companions are found at projected separations between $\sim50-360$ au from their stars. Therefore, assuming no significant orbital migration took place, these companions are generally outside the inferred CO snowline locations of $\sim30-80$ au for T Tauri stars with K spectral types and $\gtrsim 80$ au for Herbig Ae stars \citep[e.g.][]{Qi2013, Qi2015, Qi2019, Zhang2019}.} while the oxygen abundance is determined by [C/H] and C/O. We denote the bulk metallicity as [C/H] since we are only sensitive to C-, O-bearing species (i.e. CO, H$_2$O) in the companion atmospheres. 

As in \citet{Xuan2022}, we use an equilibrium chemistry grid in our retrievals computed with \texttt{easyCHEM}, a Gibbs free energy minimizer described in \citet{molliere_Observing_2017}. In this paper, we update the chemical grid to use the updated solar elemental abundances from \citet{Asplund2021}. Our chemical grid stores the mass-mixing ratios of both gas-phase species and condensates (which we use for the cloud model in the next section). We tested the option of including a quench pressure ($P_{\rm quench}$) to allow for carbon disequilibrium chemistry, which fixes the abundances of H$_2$O, CO, and CH$_4$ where $P < P_{\rm quench}$ using the equilibrium values found at $P_{\rm quench}$ \citep{Zahnle_methane_2014}. However, most of our companions are too hot ($\Teff\gtrsim1800~K$) for CH$_4$ to be detectable, and simultaneous detection of CO and CH$_4$ is necessary to constrain carbon quenching \citep{Xuan2022}. Since we find that $P_{\rm quench}$ is unconstrained for all companions from preliminary tests, we do not include it in the reported retrievals.

\subsubsection{Clouds} \label{sec:cloud_choice}
Condensate cloud opacity is expected to gradually decrease with increasing temperature from L to M spectral types, as important cloud particles such as MgSiO$_3$, Fe, and Al$_2$O$_3$ start to evaporate between $\sim$1600--1900~K at a pressure of 0.1 bars \citep{molliere_Interpreting_2022a}. Since our companions have spectral types ranging from L to early M spectral types, we consider both clear and cloudy models in order to explore the sensitivity of our retrieved abundances to the assumed cloud models. 

For the cloudy models, we use 1) a simple gray opacity model which adds a constant opacity to the atmosphere, and the 2) EddySed model \citep{ackerman_Precipitating_2001} as implemented in \texttt{petitRADTRANS}, which includes the effect of scattering from clouds \citep{molliere_Retrieving_2020}. We now describe fitted parameters of the EddySed model. First, $\rm{log} (\tilde{X}_{\rm Mg_SiO_3})$ is the scaling factor for the cloud mass fraction, so that $\rm{log} (\tilde{X}_{\rm Mg_SiO_3})=0$ is equal to the equilibrium mass fraction. The equilibrium mass fraction is determined by the chemical grid (see \S~\ref{sec:chem}). For each cloud condensate, this scaling factor along with $f_{\rm sed}$, $K_{\rm zz}$, and $\sigma_{\rm g}$ set the cloud mass fraction as a function of pressure and the cloud particle sizes \citep[for details, see][]{ackerman_Precipitating_2001, molliere_Retrieving_2020}. Here, $f_{\rm sed}$ is the sedimentation efficiency, $K_{\rm zz}$ is the eddy diffusion coefﬁcient, and $\sigma_{\rm g}$ is the width of the lognormal cloud particle size distribution. Following \citet{Zhang2023}, we fit a different $f_{\rm sed}$ for each cloud species but a global $K_{\rm zz}$ and $\sigma_{\rm g}$. Therefore, when including two different cloud species, there are a total of six cloud parameters.

Given the range in $\Teff$ of our objects, we consider models with MgSiO$_3$ + Fe for the colder objects ($\Teff$$\lesssim$2000~K) and Fe + Al$_2$O$_3$ for the hotter objects. Since the cross sections of these cloud species have similar slopes over the small wavelength range ($2.29-2.49~\mu$m) we are modeling \citep{wakeford_Transmission_2015}, our choice of the cloud species primarily serves to set the cloud base locations in the EddySed model. For each companion, we choose two cloud species which intersect their $P$--$T$ profile at the deepest pressures (i.e. closer to the photosphere), as these clouds would more meaningfully impact the emission spectra.

\subsubsection{Temperature Structure} \label{sec:pt_setup}
We adopt the pressure-temperature ($P$--$T$) parameterization from \citet{Xuan2024}, which is motivated by \citet{Piette2020}. Our profile is parameterized by seven $\Delta T/ \Delta P$ values between eight pressure points and the temperature at one of these pressures, $T_{\rm anchor}$. Because the photosphere for each companion is located at slightly different locations, we manually customize the pressure extent for each companion's retrieval to optimally encompass the companion's emission contribution. Specifically, we set at least four pressure points in the region where 90\% of the flux originates, which we determine by computing the wavelength-weighted emission contribution function. We choose the other points to be approximately equally spaced in log pressure. The selected pressure points are labeled in Fig.~\ref{fig:ptprofile} for a few companions, and listed in Table~\ref{tab:param_prior} and Appendix~\ref{app:priors} for all companions. For the radiative transfer, the eight $P$--$T$ points from our profile are interpolated onto a finer grid of 100 $P$--$T$ points using a monotonic cubic interpolation as recommended by \citet{Piette2020}. We do not apply smoothing to our profiles as \citet{Rowland2023} showed that smoothing can bias retrieval results.

\begin{deluxetable*}{ll|cccccc|l}
\tablecaption{Results of Spectral Retrievals for Eight Substellar Companions \label{table:bayes_factors}}
\tabletypesize{\small}
\tablehead{
Target & Model & C/O & C/H ($\times$ solar) & \co & \vsini (\kms) & retr. $T_\textrm{eff}$ (K) & evol. $T_\textrm{eff}$ (K) & ln($B$)\tablenotemark{a}}
\startdata
GQ~Lup~b & \textbf{Clear} & $0.70^{+0.01}_{-0.02}$ & $2.5^{+1.5}_{-1.0}$ & $153^{+43}_{-31}$ & $6.4^{+0.3}_{-0.4}$ & $2350\pm50$ & $2800\pm200$ & 0 \\
 & Clear (no $^{13}$CO) & $0.70^{+0.01}_{-0.02}$ & $2.2^{+1.1}_{-0.7}$ & ... & $6.4\pm0.4$ & $2330\pm50$ & - & -14.2 \\
 & Gray & $0.70^{+0.01}_{-0.02}$ & $2.4^{+1.1}_{-0.8}$ & $147^{+40}_{-29}$  & $6.4^{+0.4}_{-0.3}$ & $2340\pm50$ & - & +2.4 \\ 
 & EddySed & $0.70^{+0.01}_{-0.02}$ & $2.3^{+1.2}_{-0.8}$ & $165^{+50}_{-36}$ & $6.5\pm0.4$ & $2330\pm50$ & - & +2.2 \\
\hline
HIP~79098~b  & \textbf{Clear} & ${0.54\pm0.03}$ & $0.7^{+0.7}_{-0.3}$ & ${52^{+22}_{-17}}$ & ${4.0 ^{+0.8}_{-1.0}}$ & $2360^{+70}_{-80}$ & $2650\pm250$ & 0 \\ 
 & Clear (no $^{13}$CO) & ${0.54\pm0.03}$ & $0.60^{+0.6}_{-0.3}$ & ... & ${4.2 ^{+0.9}_{-1.1}}$ & $2360^{+70}_{-80}$ & - & -7.6 \\ 
 & Gray & ${0.54\pm0.03}$ & $0.8^{+1.0}_{-0.4}$ & $53^{+22}_{-17}$ & $4.0^{+0.9}_{-1.2}$ & $2350^{+70}_{-80}$ & - &  +0.6 \\
 & EddySed & $0.53\pm0.03$ & $0.7^{+0.7}_{-0.3}$ & $50^{+20}_{-16}$ & $3.9^{+0.9}_{-1.3}$ & $2370\pm70$ & - & +1.4 \\
 \hline 
DH~Tau~b  & \textbf{Clear} & $0.54^{+0.06}_{-0.05}$ & $0.5^{+0.6}_{-0.2}$ & $53^{+50}_{-24}$ & ${5.7 ^{+0.8}_{-1.0}}$ & $2050^{+120}_{-100}$ & $2350\pm200$ & 0 \\ 
 & Clear (no $^{13}$CO) & $0.55^{+0.06}_{-0.05}$ & $0.5^{+0.5}_{-0.2}$ & ... & ${5.6\pm1.0}$ & $2080^{+120}_{-100}$ & - & -4.4 \\ 
 & Gray & $0.54^{+0.06}_{-0.05}$ & $0.5^{+0.6}_{-0.3}$ & $54^{+50}_{-23}$ & $5.9^{+0.8}_{-0.9}$ & $2040^{+120}_{-100}$ & - & +0.2 \\
 & EddySed & $0.55\pm0.05$ & $0.4^{+0.5}_{-0.2}$ & $55^{+65}_{-25}$ & $5.6\pm0.8$ & $2040^{+120}_{-100}$ & - & +1.3  \\
 \hline
 \kapandb & Clear & $0.64\pm0.04$ & $0.5^{+0.5}_{-0.2}$ & ... & $39.4\pm1.3$ & $2050^{+130}_{-140}$ & $1810\pm200$ & 0 \\  
 & Gray & $0.63\pm0.05$ & $0.9^{+1.0}_{-0.5}$ & ... & $40.2\pm1.2$  & $1870^{+170}_{-130}$ & - & +0.2 \\
 & \textbf{EddySed} & $0.58^{+0.05}_{-0.04}$  & $0.8^{+0.8}_{-0.3}$ & ... & $39.4\pm0.9$ & $1680^{+60}_{-50}$ & - & +5.3 \\
\hline
GSC~6214-210~b  & Clear & $0.67^{+0.08}_{-0.09}$ & $0.7^{+1.2}_{-0.3}$ & ... & $13.2\pm1.7$ & $2420\pm150$ & $2160\pm200$ & 0 \\ 
 & Gray & $0.68^{+0.07}_{-0.08}$ & $1.1^{+2.6}_{-0.7}$ & ... & $13.1^{+2.1}_{-1.8}$ & $2340^{+180}_{-220}$ & - & +1.0 \\ 
 & \textbf{EddySed} & $0.70^{+0.07}_{-0.06}$ & $1.4^{+1.6}_{-0.7}$ & ... & $11.6^{+1.9}_{-2.1}$ & $1860^{+170}_{-110}$ & - & +4.3 \\
\hline
ROXs~12~b  & \textbf{Clear} & $0.54\pm0.05$ & $0.5^{+0.4}_{-0.2}$ & ... & $3.6^{+1.2}_{-1.6}$ & $2500\pm140$ & $2470\pm200$ & 0 \\ 
 & Gray & $0.56\pm0.04$ & $0.7^{+0.8}_{-0.3}$ & ... & $3.6^{+1.2}_{-1.6}$  & $2440\pm140$ & - & +0.1 \\ 
 & EddySed & $0.54\pm0.04$  & $0.4^{+0.4}_{-0.2}$ & ... & $3.8^{+1.0}_{-1.3}$ & $2480^{+130}_{-160}$ & - & +0.6 \\
\hline
ROXs 42B b  & \textbf{Clear} & $0.48\pm0.08$ & $1.0^{+2.8}_{-0.7}$ & ... & $4.4^{+1.6}_{-2.1}$ & $2270^{+170}_{-140}$ & $2240\pm150$ &  0 \\
 & Gray & $0.48^{+0.08}_{-0.07}$ & $1.1^{+2.7}_{-0.7}$ & ... & $4.3^{+1.6}_{-2.0}$  & $2240^{+160}_{-150}$ & - & +0.1 \\
 & EddySed & $0.48^{+0.08}_{-0.07}$  & $1.1^{+2.3}_{-0.7}$ & ... & $4.5^{+1.5}_{-2.1}$ & $2220\pm190$ & - & +0.2 \\
  \hline
 2M0122~b  & \textbf{Clear} & $0.37\pm0.08$ & $0.5^{+0.6}_{-0.2}$ & ... & $19.6^{+3.0}_{-2.5}$ & $1710^{+170}_{-160}$ & $1500\pm150$ & 0 \\ 
 & Gray & $0.37\pm0.07$ & $0.6^{+0.8}_{-0.3}$ & ... & $19.8^{+3.5}_{-2.7}$ & $1660^{+200}_{-170}$ & - & +0.1 \\ 
 & EddySed & $0.37\pm0.08$  & $0.5^{+0.7}_{-0.3}$ & ... & $20.0^{+3.2}_{-2.8}$ & $1640^{220}_{-170}$ & - & +1.0 \\ 
\enddata
\tablecomments{Selected atmospheric parameters and their central 68\% credible interval with equal probability above and below the median are listed. These values only account for statistical error. For clarity, we list the metallicity as C/H ($\times$ solar) instead of the fitted [C/H]. The retr. \Teff is the retrieved \Teff, while the evol. \Teff refers to the expected \Teff derived from a set of evolutionary models (\S~\ref{sec:m_r_priors}). The rightmost column lists the log Bayes factor ln($B$) for each retrieval. We compute ln($B$) with respect to the clear model for each companion, i.e. the model with ln($B$)=0. For each companion, the adopted model is in bold, which is by default the clear model unless one of the cloudy models is preferred over the clear model by $>3\sigma$, which corresponds to $>3$ in ln($B$). Negative values of ln($B$) for an alternative model means the clear model is preferred over this model. For GQ~Lup~b, HIP~79098~b, and DH~Tau~b, we ran retrievals without $^{13}$CO opacities to quantify the detection significance of $^{13}$CO.
The companion RVs from our retrievals are provided in Appendix~\ref{app:rvs}.
}
\tablenotetext{a}{Based on \citet{Trotta2008}, ln($B$) of $3$ and $11$ correspond to $3\sigma$ and $5\sigma$ significance, respectively.}
\end{deluxetable*}

Using this $P$--$T$ prescription, we are able to set meaningful priors on the fitted $\Delta T/ \Delta P$ values by considering atmospheric profiles from self-consistent models. After defining the pressure extent and pressure points used for a given companion, we fit SPHINX \citep{Iyer2023} and Sonora \citep{marley_Sonora_2021} profiles with \Teff~and \logg~similar to the companion's \Teff and \logg using our $P$--$T$ function. The expected \Teff and \logg for each companion are determined from evolutionary models (\S~\ref{sec:m_r_priors}). We save the best-fit $P$--$T$ values of each self-consistent profile, noting that the $\Delta T/ \Delta P$ values are very similar between profiles with different \Teff and \logg as the slope is set by radiative-convective equilibrium. In the retrieval, we then set uniform $P$--$T$ priors that bracket the mean of the best-fit values with a wide prior range of $\pm~50-100\%$ of the mean value. While the self-consistent models we considered are cloudless, the wide uniform priors we impose allow sufficient flexibility in the profile shapes, as we discuss in \S~\ref{sec:pt_teff}. We note that our method of setting physically-motivated priors on the $P$--$T$ profile is similar in spirit to the approach from \cite{Zhang2023}, who imposed Gaussian priors on the temperature gradient $d\ln{T}/d\ln{P}$ informed by self-consistent models in their retrievals. Both approaches leverage information from self-consistent $P$--$T$ profiles to prevent the retrievals from returning overly isothermal or unphysical profiles. 

\subsubsection{Mass and Radius Priors from Evolutionary Models} \label{sec:m_r_priors_retr}
The KPIC spectra are not flux-calibrated and cover a very small wavelength range. Hence, these data provide little information on a companion's radius and mass. We find that preliminary retrievals occasionally yield unphysical radii and \logg for our companions (e.g. $\logg>5$ for a young, low-gravity companion). Therefore, we adopt mass and radius priors from evolutionary models in our retrievals. These priors are described in \S~\ref{sec:m_r_priors} and listed in Table~\ref{tab:sys_prop}.

\subsubsection{Model Fitting with Nested Sampling} \label{sec:ns_bayes}
We use nested sampling as implemented by \texttt{dynesty} \citep{speagle_DYNESTY_2020} to find the posterior distributions for the model parameters. Specifically, we use between 500-800 live points and adopt the stopping criterion that the estimated contribution of the remaining prior volume to the total evidence is less than 1\%. 

One advantage of adopting nested sampling is that we can use the Bayesian evidence from each fit to calculate the Bayes factor $B$, which quantifies the relative probability of model $M_2$ compared to $M_1$. We use the Bayes factor to compare different models throughout this paper to determine whether a given $M_2$ is justified over $M_1$.

\begin{figure*}[t!]
    \centering
    \includegraphics[width=.44\linewidth]{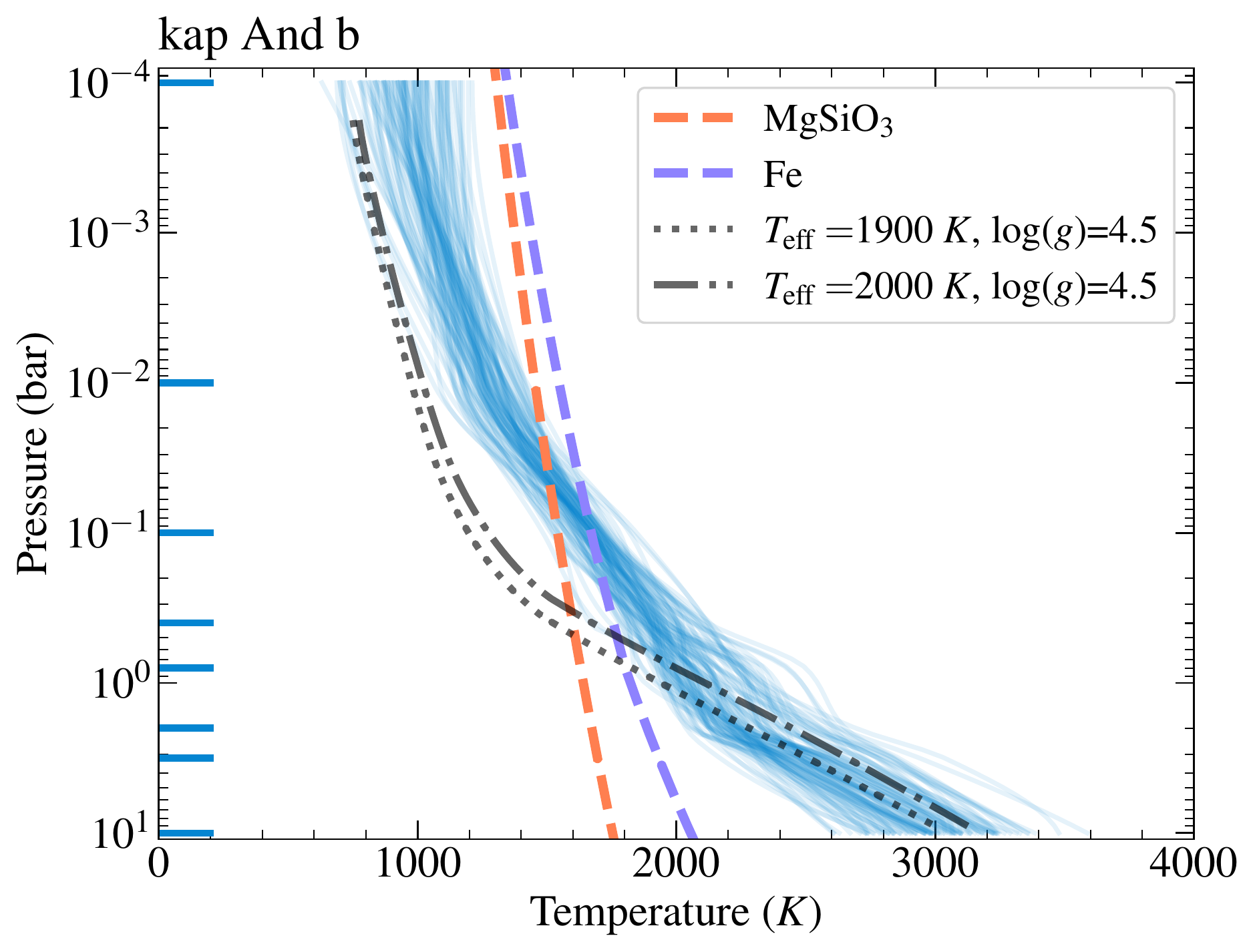}
    \centering
    \includegraphics[width=.42\linewidth]{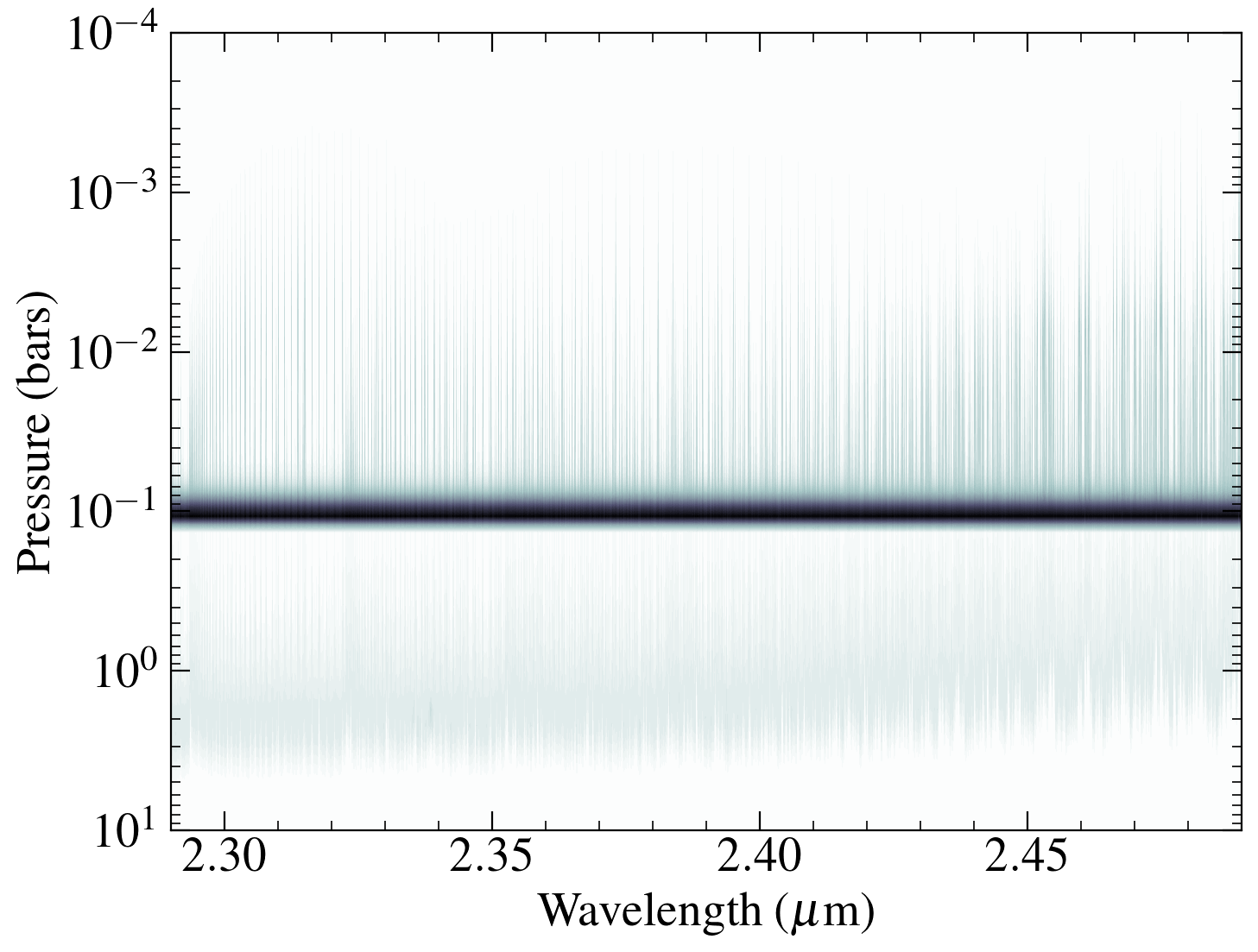}
    \centering
    \includegraphics[width=.44\linewidth]{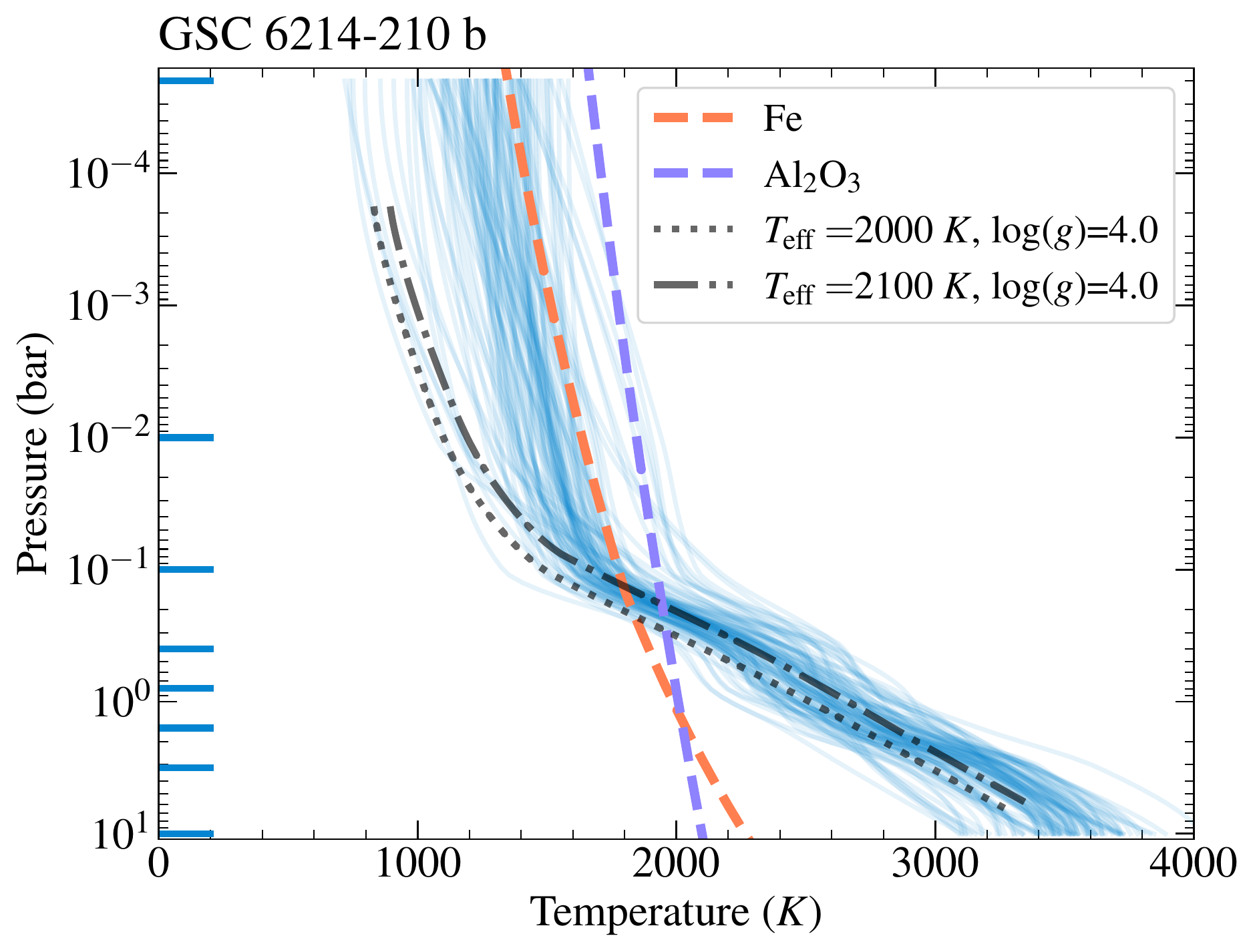}
    \centering
    \includegraphics[width=.42\linewidth]{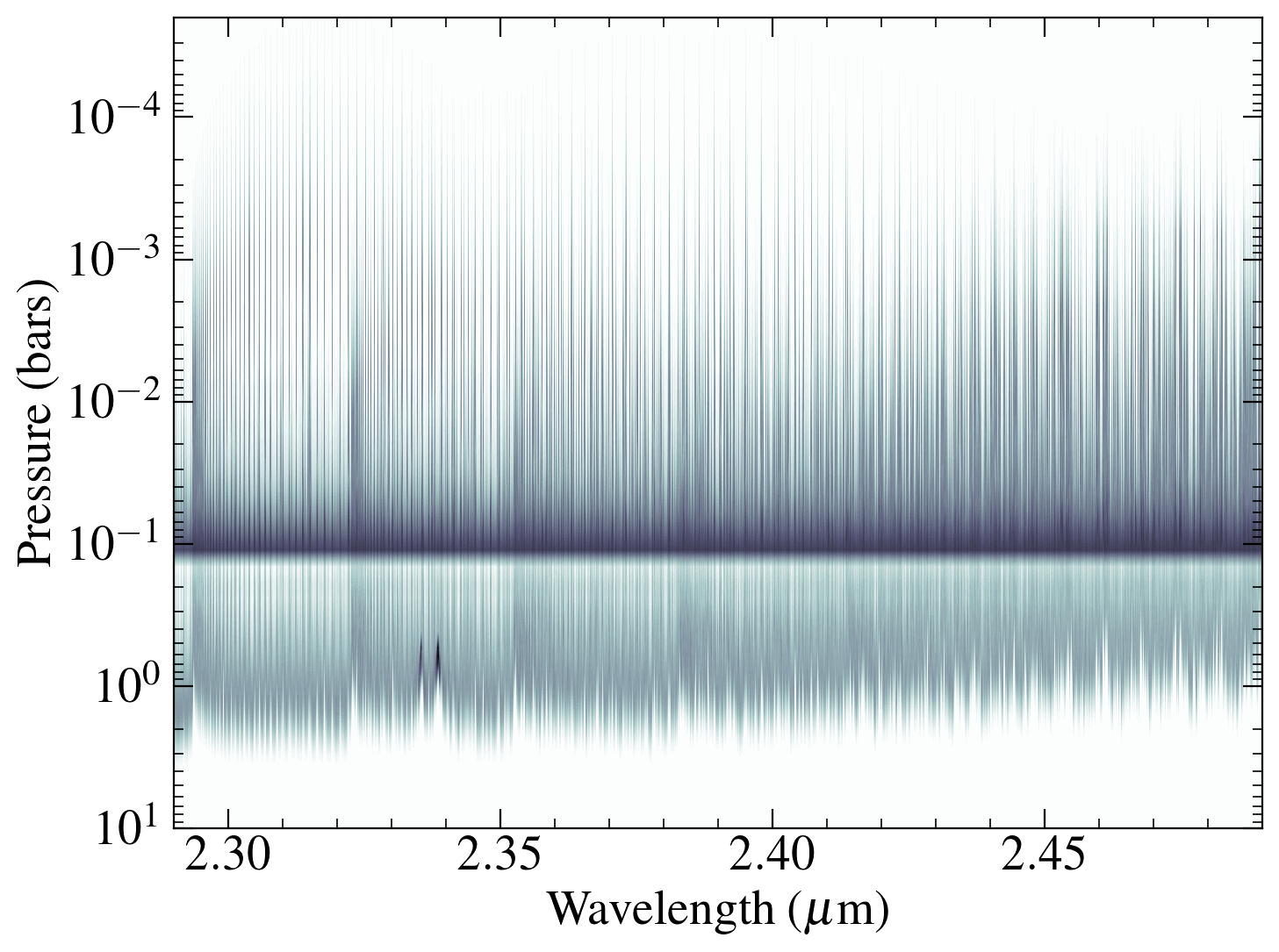}
    \centering
    \includegraphics[width=.44\linewidth]{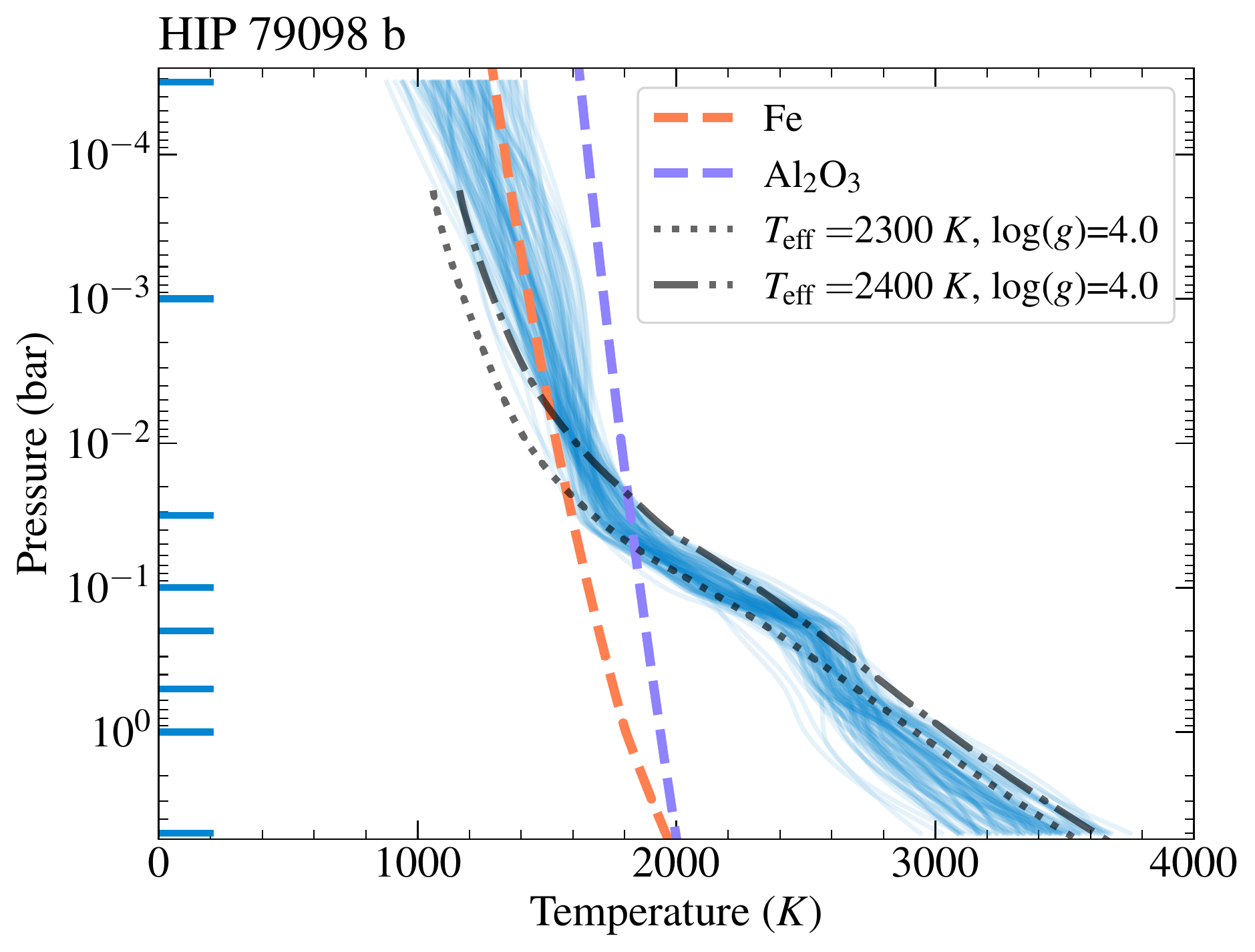}
    \centering
    \includegraphics[width=.42\linewidth]{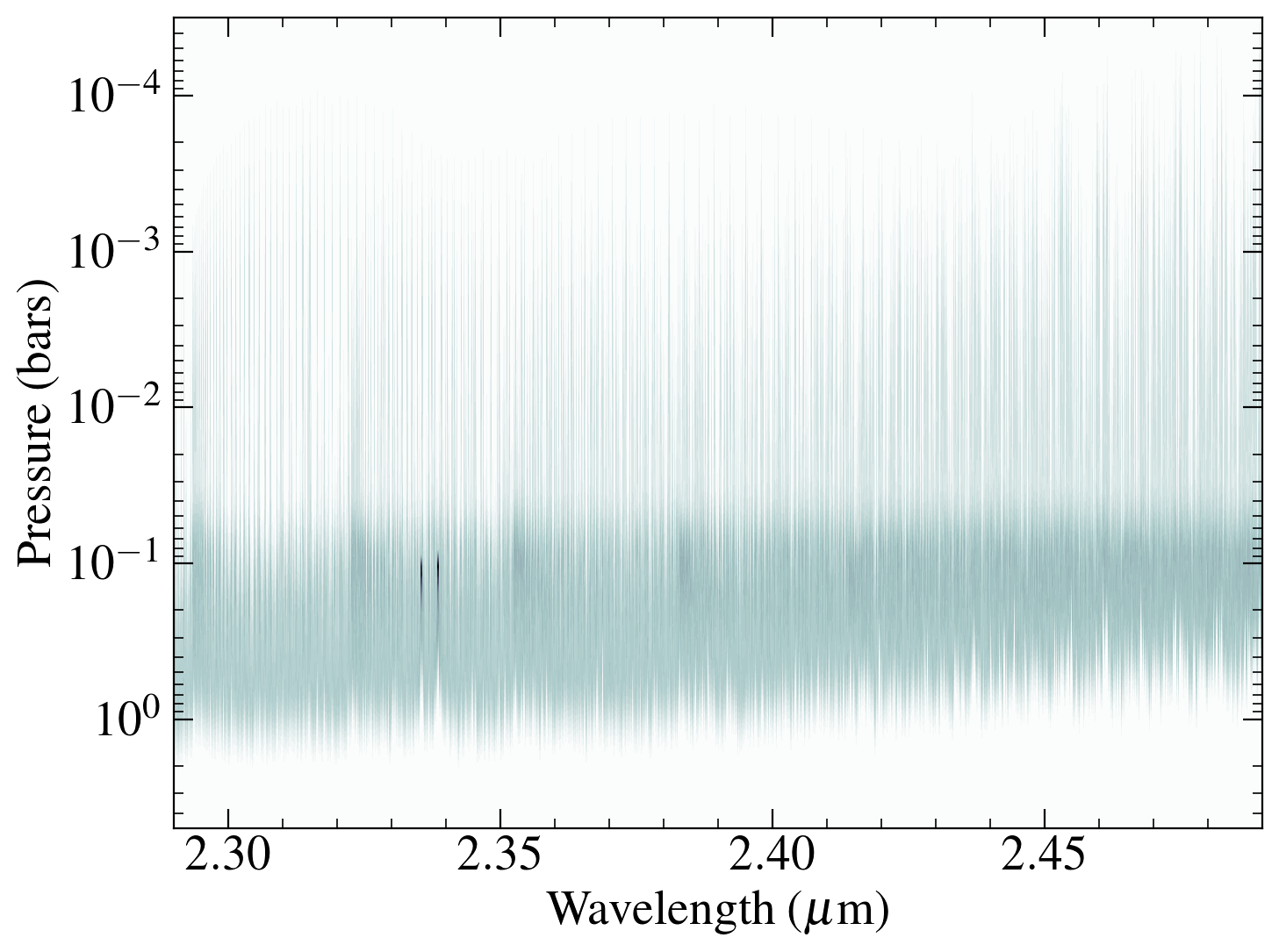}
    \caption{Retrieved $P$--$T$ profiles and emission contribution functions for three example companions (one per row). Left panels: We plot random draws of the retrieved $P$--$T$ profiles in blue. The gray lines show Sonora cloudless models \citep{marley_Sonora_2021} with similar bulk properties as the companions. Different cloud condensation curves are plotted as colored dashed lines. The horizontal blue lines mark pressure points between which we fit $\Delta T$ values in our $P$--$T$ parameterization. Right panels: The emission contribution functions of the best-fit models. For a given wavelength, darker colors mean that a larger fraction of emission originates from that pressure level. The darker, bar-like structures seen for \kapandb~and GSC~6214-210~b coincide with the cloud base locations. In our best-fit EddySed models, we find that the total cloud optical depth ($\tau$) is about 1.4/0.6 for~\kapandb/GSC~6214-210~b at their retrieved cloud bases, so a non-negligible part of the flux beneath the clouds can propagate through the cloud base. These two companions prefer the EddySed cloud model over the clear model to $>3\sigma$ significance (see \S~\ref{sec:cloud_insensitive}). Plots for the other companions are included in Appendix~\ref{app:pt}.}\label{fig:ptprofile}
\end{figure*}

\begin{figure*}[t!]
    \centering
\includegraphics[width=0.9\linewidth]{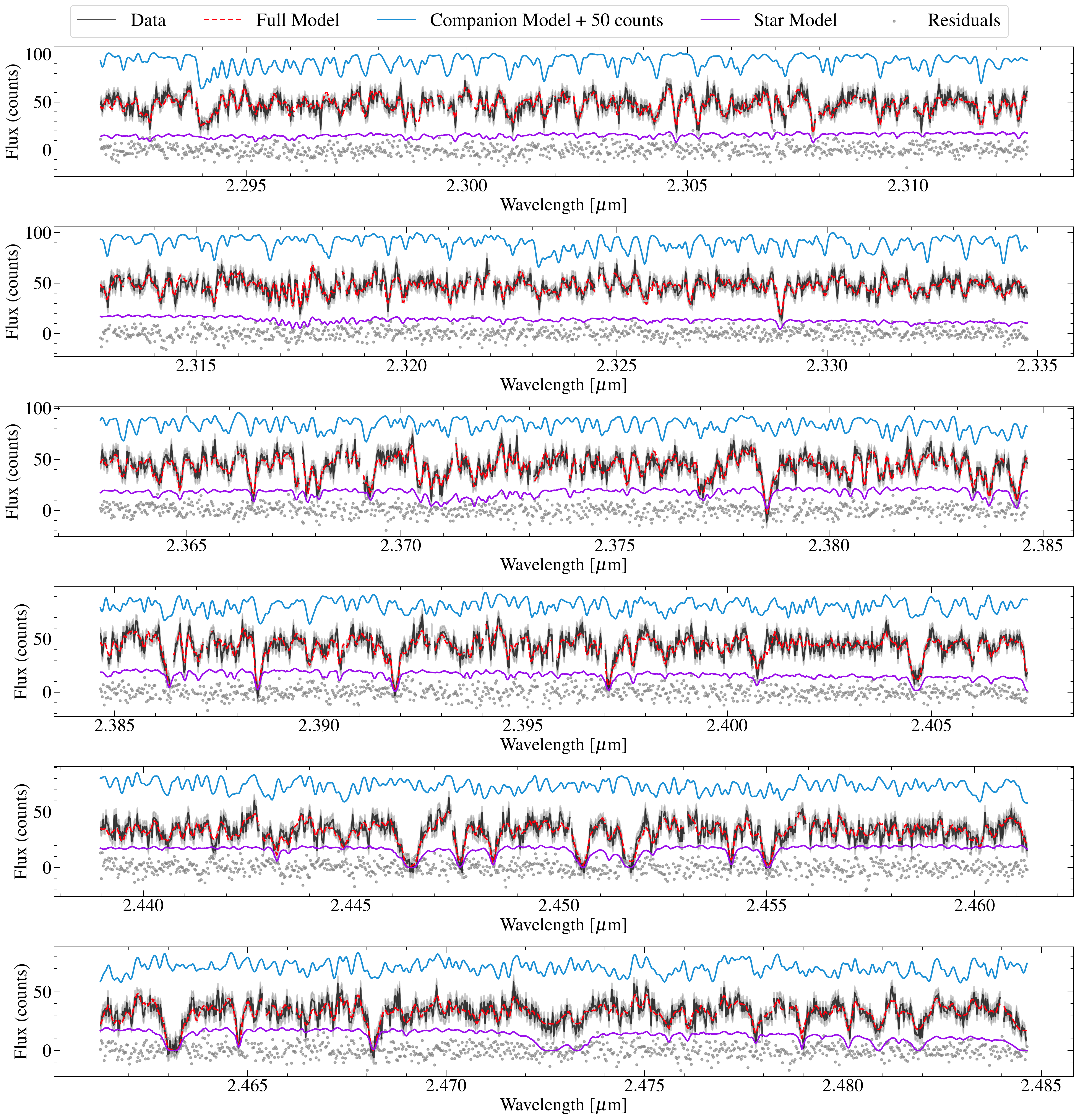}
    \caption{KPIC data of for GQ~Lup~b from one science fiber are shown in black, with error bars in gray. This represents half the data, since we nodded between two fibers. Each spectral order used in the retrieval is broken into two panels. The full model is shown in red dashed lines and consists of the RV-shifted and broadened companion model in blue ($M_b$ in Eq.~\ref{eqn:comp}), the stellar model in purple ($M_s$) to model the speckle contribution, and the telluric and instrumental response ($T$). The residuals are shown as gray points. For clarity, an offset of +50 counts was added to the companion model. Plots for the other companions can be found in Appendix~\ref{app:model_bestfit}.}
    \label{fig:kpic_model_gqlup}
\end{figure*}

\section{Atmospheric Retrieval Results} \label{sec:results}
We summarize the retrieval results in this section. Given the large number of objects, we focus on representative examples here. In Fig.~\ref{fig:kpic_model_gqlup}, we plot the data and best-fit model for GQ~Lup~b, our highest S/N dataset. Plots for other companions are shown in Appendix~\ref{app:model_bestfit}. Key parameters from the retrievals for all companions are listed in Table~\ref{table:bayes_factors}. In \S~\ref{sec:mr_prior_effect}, we show the insensitivity of our results to the mass and radius priors used in the retrievals. We discuss the retrieved $P$--$T$ profiles and compare the retrieved \Teff~with predicted \Teff~from evolutionary models in \S~\ref{sec:pt_teff}. In \S~\ref{sec:cloud_insensitive}, we discuss whether cloudy models are preferred over clear models for our companions, and whether clouds impact the retrieved abundances. Finally, we discuss our C/O, [C/H] measurements in \S~\ref{sec:c_and_o} and \cratio~constraints for three companions in \S~\ref{sec:13co_result}.

\subsection{Effect of Mass and Radius Priors}\label{sec:mr_prior_effect}
We adopt mass and radius priors in our retrievals to prevent the fits from yielding unphysical radii and \logg. In principle, jointly fitting photometry or low-resolution data with the KPIC HRS could provide the flux information needed to anchor the \lbol~and radius. For example, \citet{Stolker2021} fit the 0.6-5 $\mu$m spectro-photometry for GQ~Lup~b accounting for dust extinction and emission from a circumplanetary disk and find $3.77\pm0.10~\Rj$, which is consistent with our priors derived from evolutionary models. However, as \citet{Stolker2021} illustrates, for broadband spectro-photometry it is necessary to model both extinction from dust and potential circumplanetary disk emission for the youngest companions. Several companions in our study likely possess circumplanetary disks as well \citep[e.g.][]{Bowler2011, vanHolstein2021, Stolker2021}. Doing such modeling for all the companions in our sample is beyond the scope of this paper, which is focused on measuring atmospheric abundances from high-resolution spectra. Hence, we choose to adopt mass and radius priors to achieve the same purpose. We verified that these mass and radius priors do not affect the retrieved abundances by comparing against retrievals without the mass and radius priors, and found that the measured abundances vary by much less than $1\sigma$ when the priors are imposed. 

\subsection{$P$--$T$ Profile and Effective Temperature}\label{sec:pt_teff}
Our retrieved $P$--$T$ profiles mostly follow self-consistent (cloudless) profiles in the deeper atmosphere (see Fig.~\ref{fig:ptprofile} and Appendix~\ref{app:pt}), but they can be hotter and more isothermal than the corresponding self-consistent models above the cloud bases. This could be due to several reasons. First, the cloudless model $P$--$T$ profiles do not provide the best point of comparison for cloudy atmospheres, and cloud formation is expected to heat up the atmosphere above the cloud base resulting in a detached radiative zone \citep{Tsuji2002, Burrows2006}. This effect can be seen as a kink the $P$--$T$ profile, which is most obvious for \kapandb~at pressures lower than $\sim1$ bar (see Fig.~\ref{fig:ptprofile}). Upcoming self-consistent models that include cloud radiative feedback \citep{Morley2024} would provide a better comparison. Second, our retrievals could be showing evidence of the well-known cloud and isothermal $P$--$T$ correlation \citep[e.g.][]{burningham_retrieval_2017, molliere_Retrieving_2020}, whereby a cloudier atmosphere can be re-produced by one with a more isothermal temperature gradient. While we attempt to address this behavior with our informed $P$--$T$ profile priors, our priors are uniform and wide enough that isothermal behavior is still allowed. A stricter, Gaussian prior such as implemented by \citet{Zhang2023} could prevent this behavior at the expense of limiting the parameter space for the retrieval to explore. 

To assess whether our priors on the $P$--$T$ profile affect the retrieved parameters, we repeat the baseline HIP~79098~b retrieval with wide priors from $0-2000~K$ for each $\Delta T$ value. This second retrieval yielded the same median value and error bars for the abundances, RV, and \vsini as our baseline retrieval, indicating that our default $P$--$T$ priors are conservative enough and not biasing the results.

From our retrievals, we compute \Teff~by sampling the posteriors to generate low-resolution models from 0.1 to 30~$\mu$m, calculating the integrated flux, and applying the Stefan-Boltzmann law with the retrieved radii. To see how our results compare with evolutionary models, we compare our retrieved \Teff posteriors with the evolutionary-model predicted \Teff that were estimated in \S~\ref{sec:m_r_priors}. As with the mass and radius priors, we estimate a range of evolutionary \Teff that encompass all the models, which are listed in Table~\ref{table:bayes_factors}. 

\begin{figure*}
    \centering
    \includegraphics[width=0.5\linewidth]{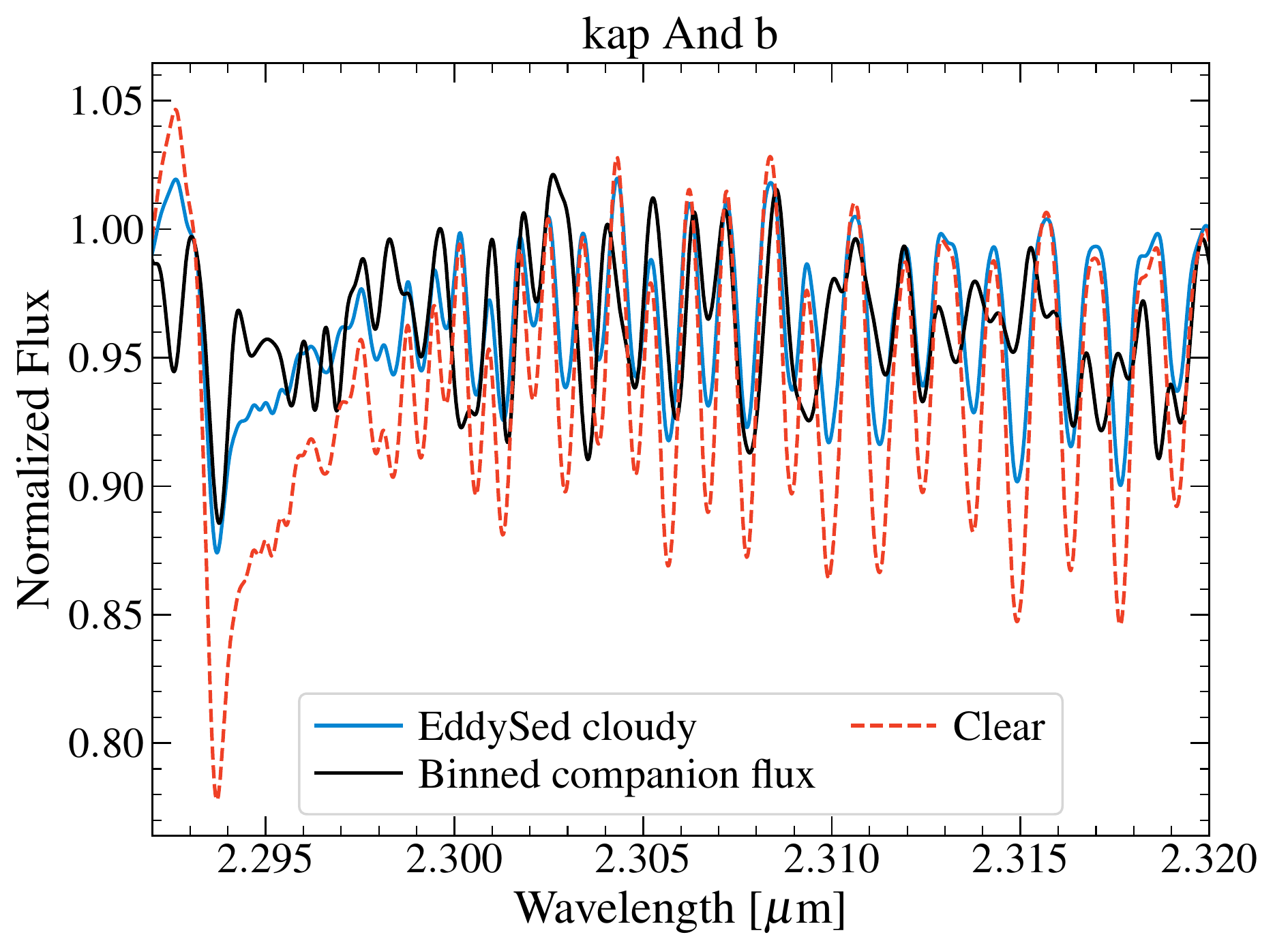}
    \centering
    \includegraphics[width=0.4\linewidth]{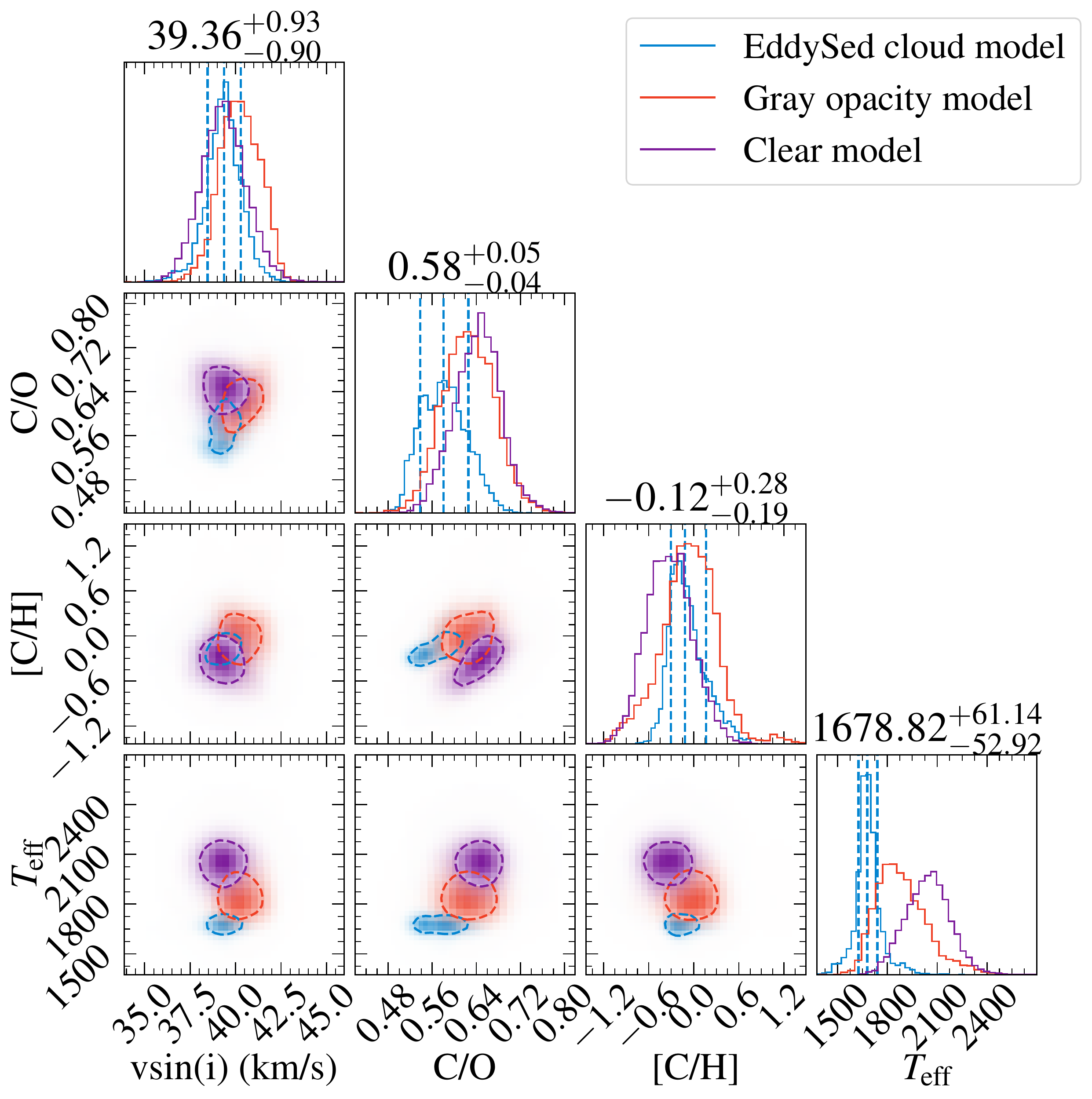}
    \centering
    \includegraphics[width=0.5\linewidth]{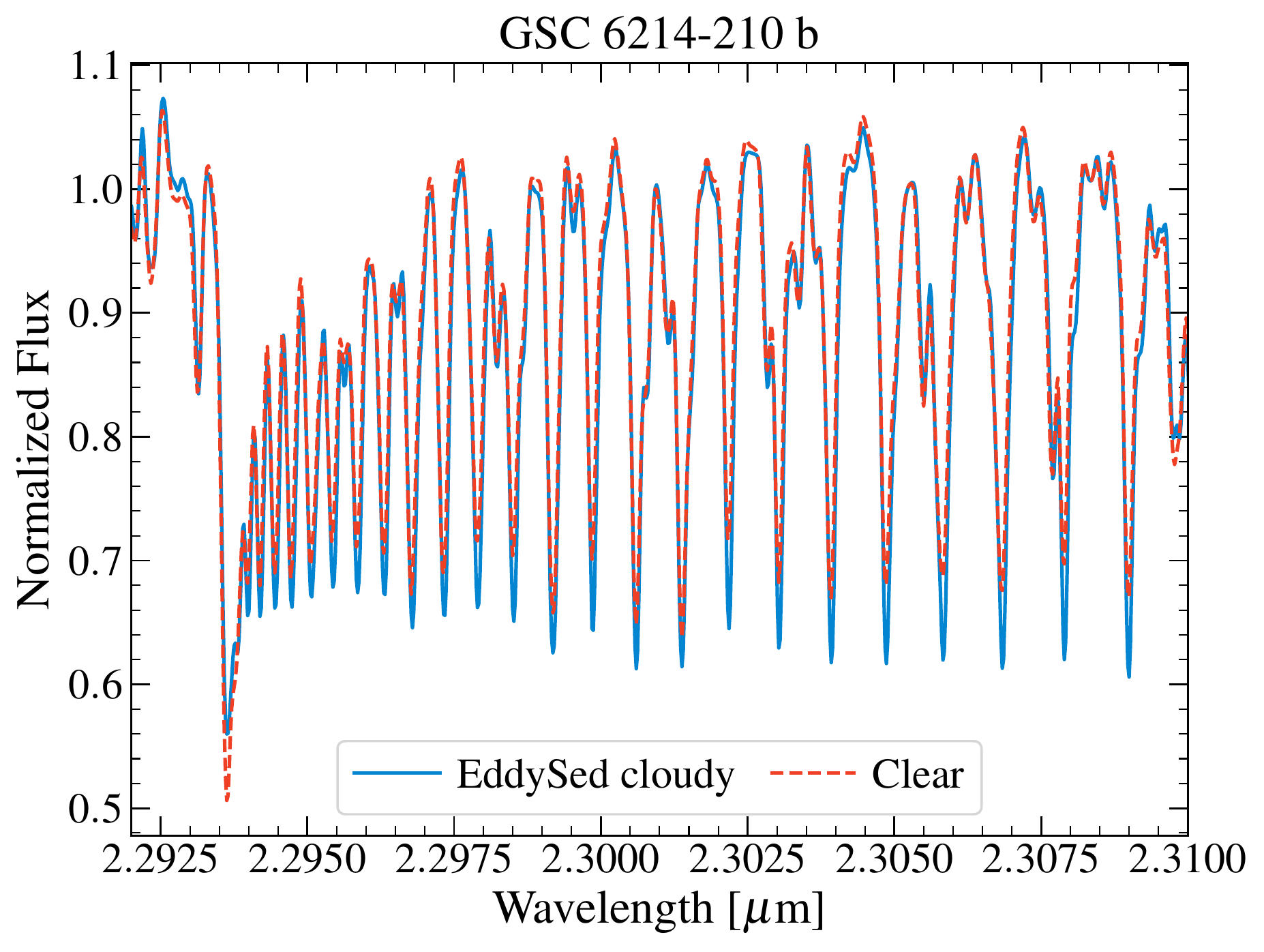}
    \centering
    \includegraphics[width=0.4\linewidth]{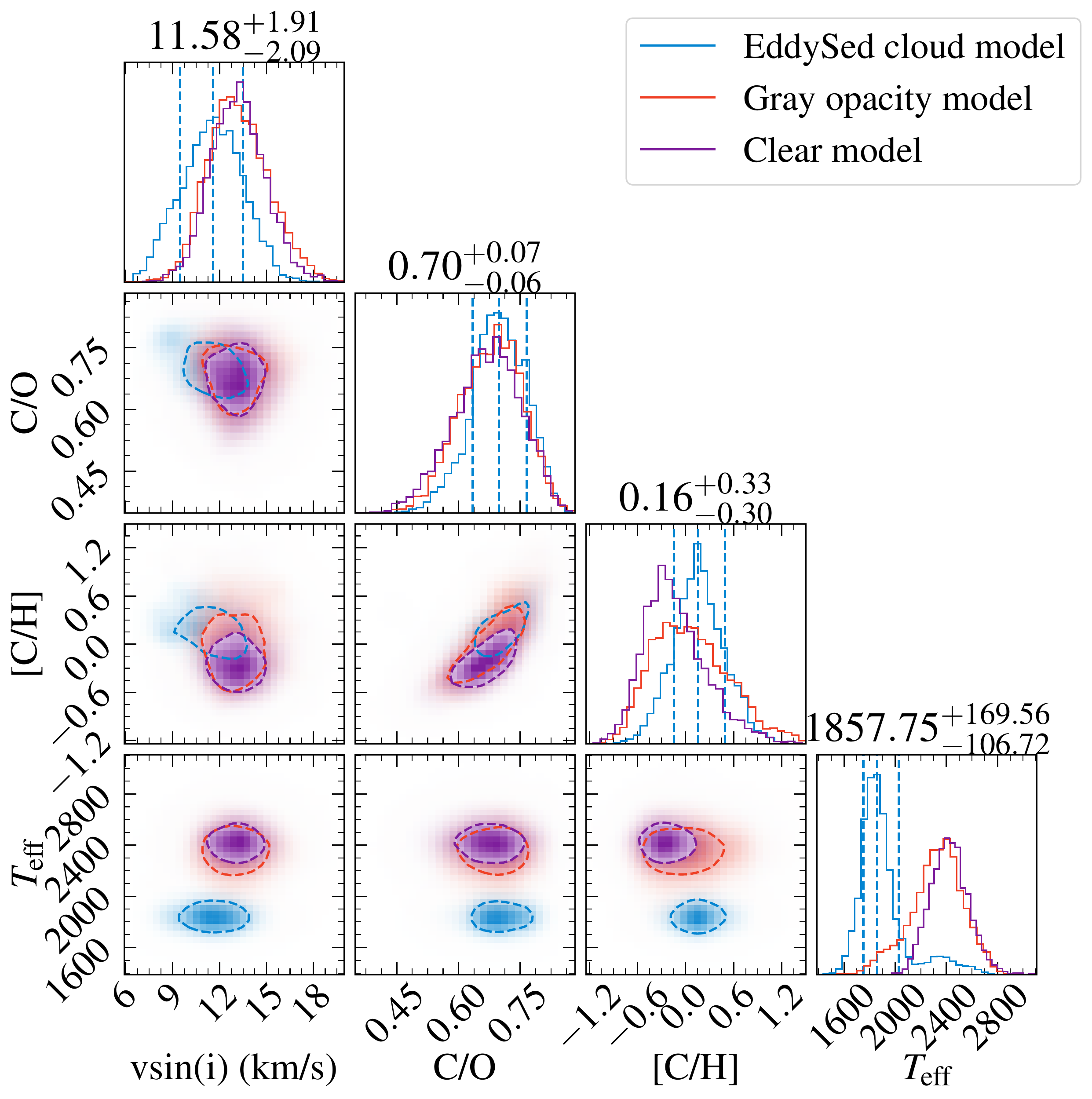}
    \caption{Top left: The best-fit Eddysed cloud model (blue) and clear model (red dashdot) for \kapandb. These models are normalized companion templates from \texttt{petitRADTRANS}, and show non-gray variations in line depths and continuum location. We also overplot the normalized companion spectra (data - best-fit speckle model), which has been binned down to $R=5000$ to better visualize the companion lines. Top right: the joint posterior distributions of \vsini, C/O, [C/H], and \Teff from the different cloud models. There are $\approx1\sigma$ variations in the abundances and \vsini, while the \Teff is $\approx 350~K$ lower for the EddySed model compared to the clear model.
    Bottom panel: same but for GSC~6214-210~b. Note the different x-axis scales for the spectral plots, which are chosen for visualization purposes. The retrieved \Teff for GSC~6214-210~b is $\approx500~K$ lower in the EddySed model compared to the clear model. Since the continuum-to-line contrast increases for lower \Teff atmospheres, this could cause absorption lines in the cloudy model to be deeper than those in the clear model despite the added cloud opacity, which is evident in the case of GSC 6214-210 b.}
    \label{fig:cloud_kap}
\end{figure*}

For the two companions that showed $>3\sigma$ preference for the EddySed cloud model, GSC~6214-210~b and \kapandb, our retrieved \Teff from the EddySed model is lower compared to those from the clear and gray opacity models (see Fig.~\ref{fig:cloud_kap}). In addition, the AMES-Dusty and SM08 models which include clouds generally predict lower \Teff for the same object compared to the cloudless ATMO 2020 and AMES-COND models (Appendix~\ref{app:evol_model}). The EddySed-retrieved \Teff for these two companions are closer to the predictions of the cloudy evolutionary models. For example, the EddySed-retrieved $\Teff$ is $1680^{+60}_{-50}~K$ for \kapandb. For this companion, the cloudy SM08 evolutionary model predicts $\Teff\approx1760~K$ while the cloudless ATMO 2020 evolutionary model predicts $\Teff\approx1860~K$ (see Fig.~\ref{fig:evol_all}). Thus, the overlap between the EddySed-retrieved \Teff and predicted \Teff from cloudy evolutionary models for GSC~6214-210~b and \kapandb is consistent with the fact that we find evidence of clouds in these objects (discussed further in \S~\ref{sec:cloud_discuss}).

Next, we discuss the six companions that did not show strong preference for cloudy models; these companions have similar retrieved \Teff posteriors between different cloud models. First, for ROXs~12~b, ROXs~42~Bb, and 2M0122~b, we find good agreement ($<1\sigma$) between their retrieved and evolutionary \Teff. For the remaining three, GQ~Lup~b, DH~Tau~b, and HIP~79098~b, which are among our hottest companions with late M spectral types, the retrieved \Teff from each cloud model is lower by $\sim300-400~K$ compared to the evolutionary model predictions (a $\sim2\sigma$ discrepancy). We find that the retrieved \lbollsun~of these companions are also slightly lower than those predicted from the models by $\sim 0.1-0.2$ dex (or by $1-2\sigma$). On the other hand, the median-retrieved radius is slightly higher by $\sim5-10\%$ compared to the median of the radius prior used in the retrieval. 

\citet{Xuan2024} also noted \Teff~and radius discrepancies from retrievals of a M7.5 stellar companion with similar \Teff, where the \Teff/radius was slightly lower/higher than evolutionary models. One potential explanation is that high-resolution spectra after high-pass filtering are more sensitive to the slope of the $P$--$T$ profile than the absolute temperature value \citep{Landman2023}, and therefore may not provide accurate \Teff. In our case, the retrieved \Teff~is also influenced by the mass and radius priors we placed. However, this alone does not explain why only these companions show disagreement with the evolutionary models. Another reason for the \Teff discrepancy could be shortcomings in the treatment of clouds in evolutionary models, especially at the M to L transition. As shown by a systematic study of brown dwarfs in \citet{Sanghi2023}, clouds may be highly inaccurate at the M to L transition. This means that the evolutionary \Teff are not necessarily correct for late M objects. 

For purposes of this paper, we simply check that our retrieved abundances are not affected by the retrieved \Teff~by repeating retrievals with the $P$--$T$ profile fixed to a self-consistent profile that has higher \Teff. Specifically, for HIP~79098~b, we tried fixing the $P$--$T$ profile to a SPHINX model profile \citep{Iyer2023} with $\Teff=2600~K$ and $\logg=4.0$ (overplotted in Fig.~\ref{fig:ptprofile}) and repeated the retrieval. We find that all retrieved parameters shift by $<1\sigma$ compared to the baseline retrieval. Therefore, we conclude that discrepancies in \Teff and radius have negligible impact on the retrieved abundances from our high-resolution data. 

\begin{figure*}
    \centering
  \includegraphics[width=.32\linewidth]{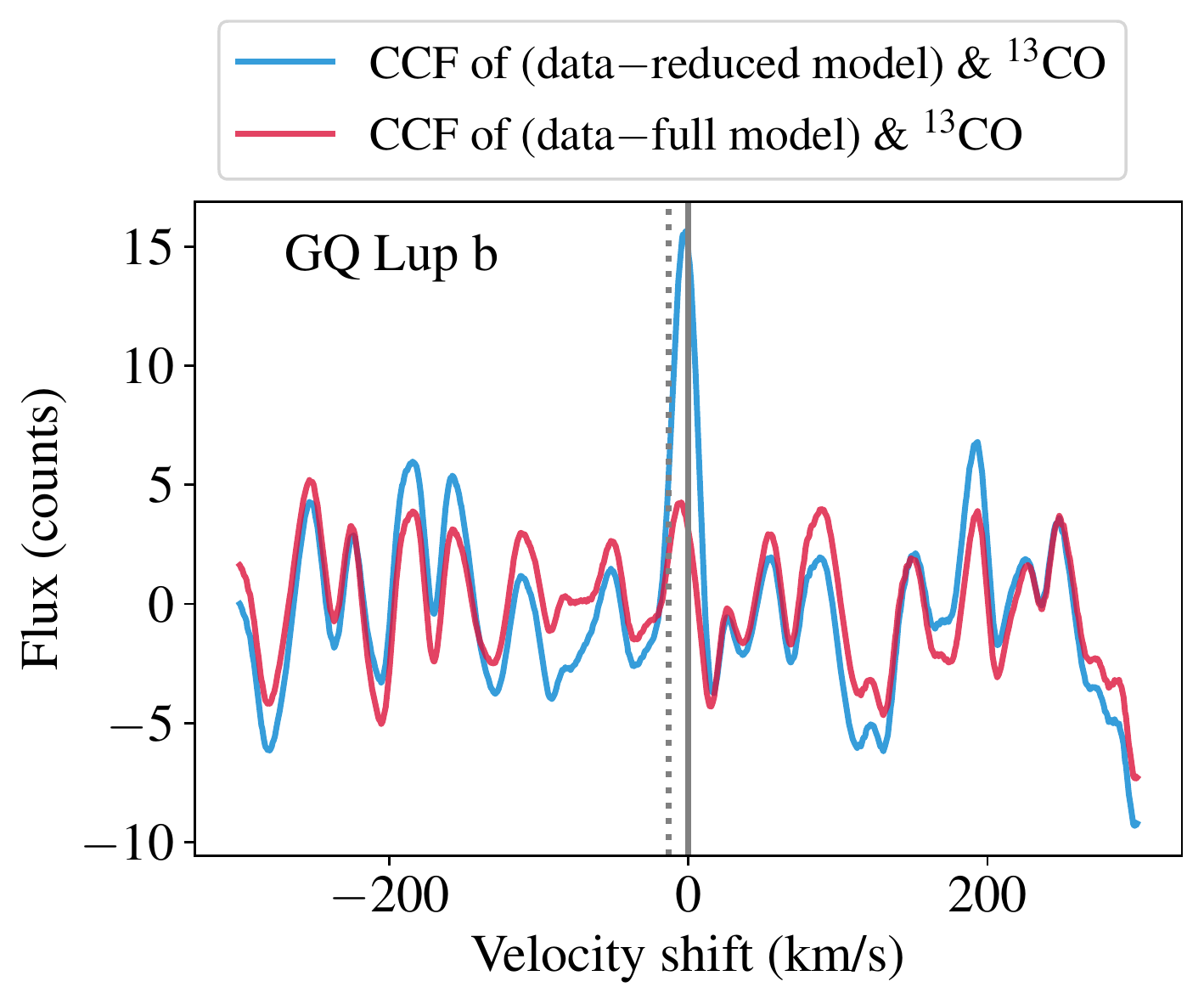}
    \centering
  \includegraphics[width=.32\linewidth]{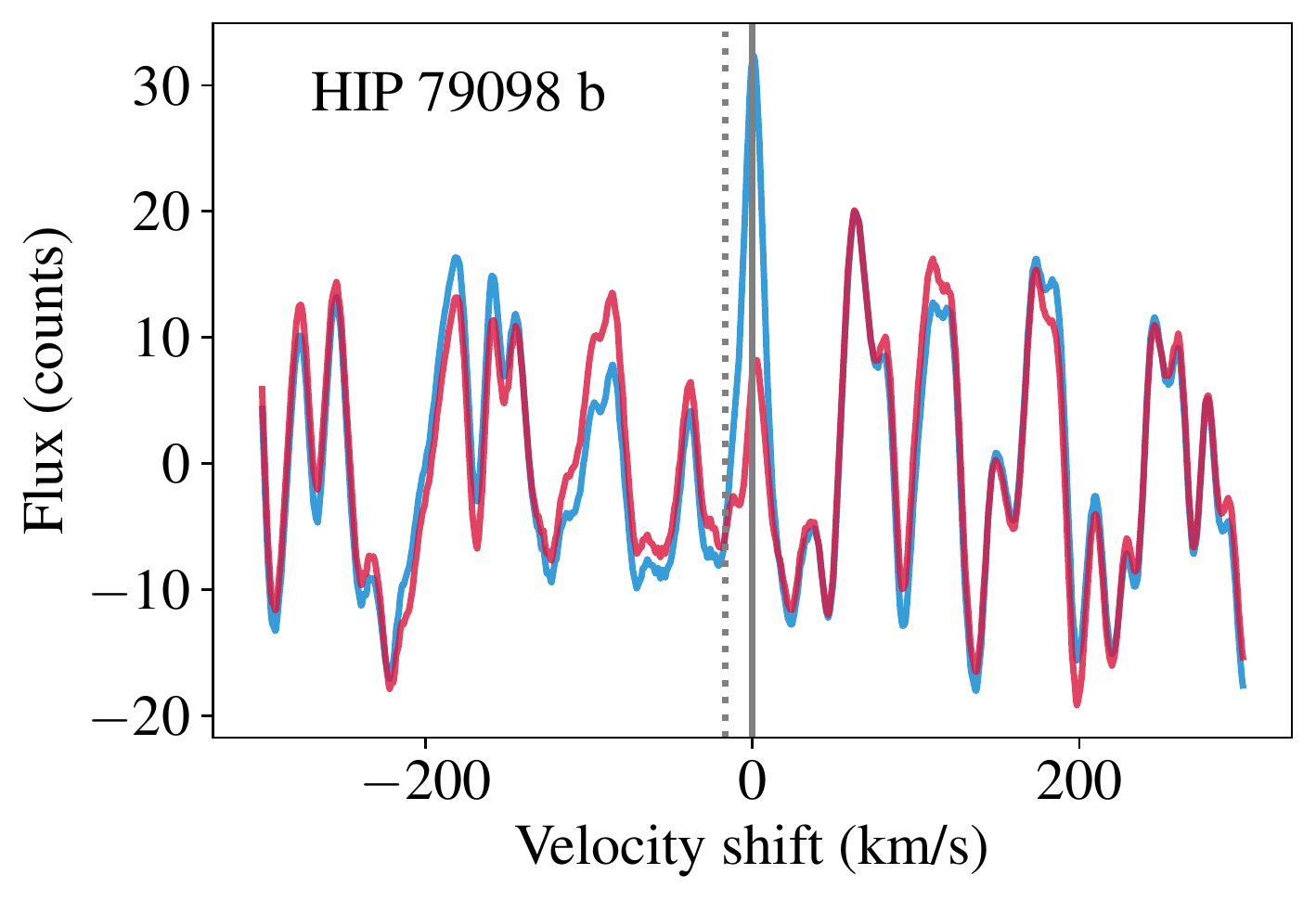}
    \includegraphics[width=.32\linewidth]{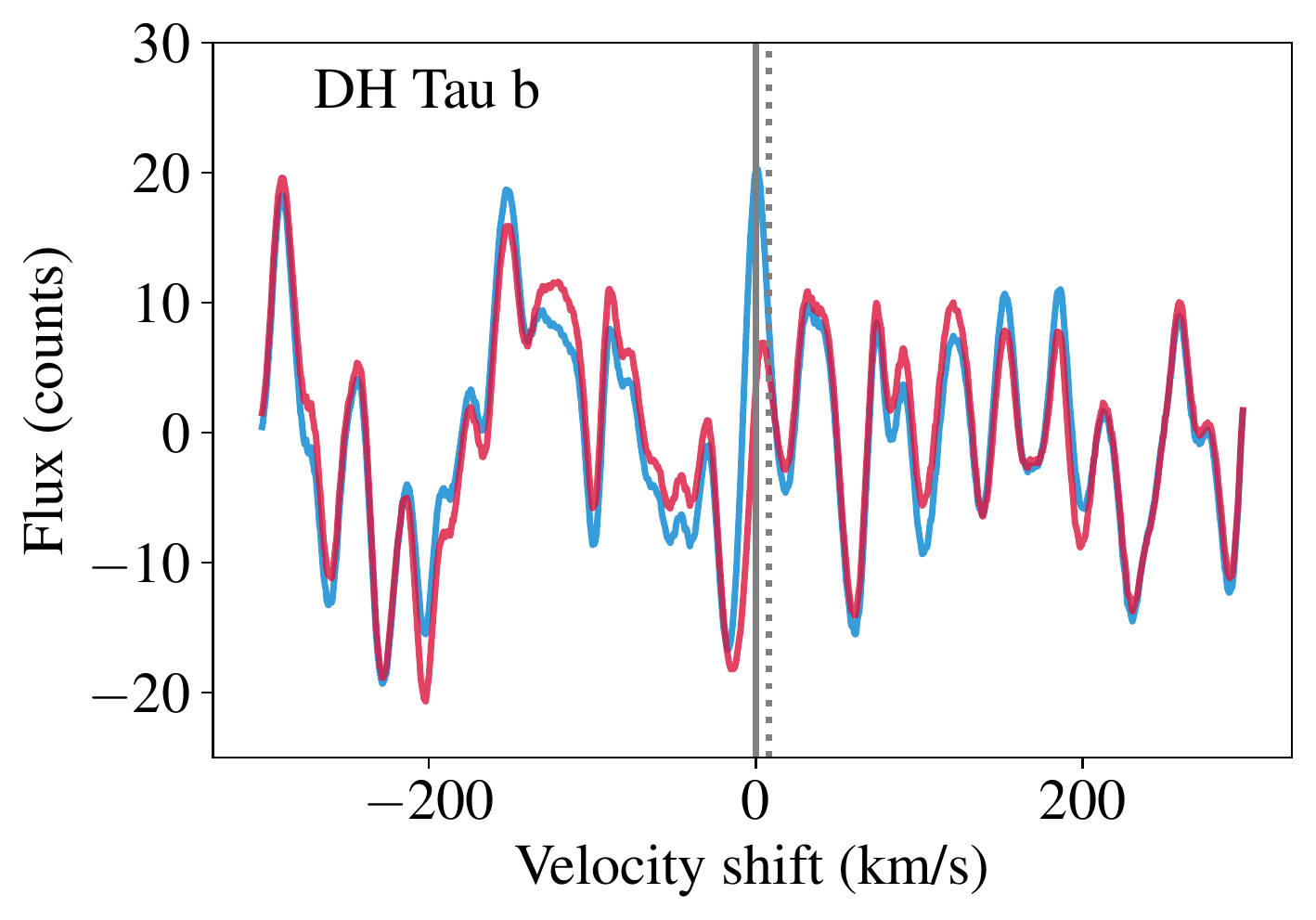}
    \caption{Left panel: The CCF between the $\rm{^{13}CO}$-only template and (data - best-fit reduced model) in blue for GQ~Lup~b. The reduced model retrieval does not include $\rm{^{13}CO}$ opacities. The CCF between the $\rm{^{13}CO}$-only template and (data - best-fit full model) is shown in red. The full model retrieval includes $\rm{^{13}CO}$ opacities. The fact that the blue CCF shows a peak at the companion's rest frame (gray solid line) indicates a real $\rm{^{13}CO}$ detection. For comparison, the gray dotted line is the telluric rest frame. In the red CCF, we do not expect a peak since $\rm{^{13}CO}$ is fitted for in this model, so the residuals should be free of $\rm{^{13}CO}$. Middle and right panels: same but for HIP~79098~b and DH Tau b. The $\rm{^{13}CO}$ detection for DH Tau b is tentative.}\label{fig:13co_ccfs}
\end{figure*}

\subsection{Impact of Clouds} \label{sec:cloud_insensitive}
For most of our companions, the posteriors from our clear and cloudy retrievals overlap significantly; the median or best-fit retrieved abundance parameters are identical within $\ll$1$\sigma$. We compare the cloudy models and clear models statistically with the log Bayes factor, or ln($B$), which are listed in Table~\ref{table:bayes_factors}. The cloudy models are compared against the clear model, so a positive ln($B$) for a cloudy model means it is preferred over the clear model. While all companions prefer cloudy models, the preference is only significant for \kapandb\, and GSC~6214-210~b, which show $3.7\sigma$ and $3.4\sigma$ preferences for the EddySed cloud model. For these two companions, our median retrieved abundances from the EddySed models can differ by $\approx1\sigma$ compared to the clear model, as we show in Fig.~\ref{fig:cloud_kap}. For all other companions, the cloudy models are preferred by $0-2.6\sigma$, and the addition of clouds negligibly impact their retrieved abundances and other parameters.

We illustrate the impact of clouds on the retrieved spectra of \kapandb\, and GSC~6214-210~b in Fig~\ref{fig:cloud_kap}, which shows that both the continuum and line depths are varying in a distinctly non-gray behavior. This explains why the spectra could not be fit as well with the gray opacity model. As the data have relatively low S/N ($\sim5$ and $\sim2$ per pixel for \kapandb~and GSC 6214-210~b, respectively), the residual noise makes it difficult to clearly see the companion lines. Thus, we try to visualize the effect of clouds in the data after binning down the companion flux contribution of in data. Since the \kapandb~spectra have a non-zero speckle component (see Eq.~\ref{eqn:comp}), we subtract the best-fit speckle model from its KPIC spectra and then bin them down to $R=5000$. As shown in Fig~\ref{fig:cloud_kap}, we see that the companion line depths do match better with the cloudy, EddySed model compared to the clear model for \kapandb. However, the same effect is not as clear when we plot the GSC 6214-210 b companion flux, which is more noisy due to its lower S/N. In addition, the relative difference between the clear and cloudy model is smaller for this object. Thus, we conclude that a clear confirmation of the effect of clouds in high-resolution spectra would require higher S/N data. This could be achieved with longer integration times on our targets, or existing high S/N spectra of isolated brown dwarfs.

Why do only \kapandb\, and GSC~6214-210~b prefer EddySed cloud models in our retrievals? Based on Table~\ref{table:bayes_factors}, we see that they are among the colder companions. Except for 2M0122~b, all other companions are predicted to be hotter. For \kapandb, we retrieve $\Teff=1680^{+60}_{-50}~K$ from the EddySed model and $\Teff=2050^{+130}_{-140}~K$ for a clear atmosphere. For GSC~6214-210~b, we retrieve $\Teff=1860^{+170}_{-110}~K$ and $\Teff=2420\pm150~K$ from the EddySed and clear retrievals, respectively. As noted in \S~\ref{sec:pt_teff}, the EddySed-retrieved \Teff for these two companions matches predictions from cloudy substellar evolutionary models (SM08, AMES-dusty) better than the \Teff retrieved from the clear model, which are overly high. The fact that we see evidence of clouds for only \kapandb~and GSC 6214-210~b suggests that the impact of clouds on the spectra is smaller for hotter companions and therefore harder to detect despite their higher S/N data (e.g. GQ~Lup~b has S/N per pixel 12 v.s. 5 for \kapandb). When examining the $P$--$T$ profiles for \kapandb~and GSC~6214-210~b, we see that their cloud bases are predicted to be closer to the $K$ band continuum due to their lower \Teff (see Fig.~\ref{fig:ptprofile}). In contrast, for hotter companions, the cloud bases are at much lower pressures than the continuum. Physically, this means that there is more condensable cloud material for \kapandb~and GSC~6214-210~b at pressures closer to their photospheres, thereby causing them to appear cloudier. 

Therefore, our results match the trend of decreasing cloud opacity with rising temperature that is observed across the L to M transition, for instance from the decreasing silicate cloud absorption strength at $\approx$8--11~$\mu$m seen in mid-IR Spitzer/IRS spectra of brown dwarfs \citep[e.g.][]{cushing_Spitzer_2006, Suarez2022_L2silicate}. We note that the coldest of our companions, 2M0122~b, shows a weak preference ($2\sigma$) for the EddySed model, but its S/N is the lowest among the sample ($\sim1$ per pixel), which may prevent a stronger constraint of its clouds properties. Alternatively, viewing geometry also affects the strength of the silicate cloud feature of brown dwarfs, whose equators tend to be cloudier than their poles \citep{Vos2017, Suarez2023_viewgeo}. Given the measured companion spin axis inclination of $33^{+17}_{-9}\deg$ \citep{Bryan_obliquity_2020}, 2M0122~b is close to a pole-on geometry, which may reduce the effect of clouds on its measured emission spectrum relative to companions with more equatorial viewing geometries. Finally, because spectral features become deeper compared to the continuum at lower \Teff for brown dwarfs \citep[e.g.][]{Ruffio2024arXiv}, this causes CO lines to be deeper in the EddySed model for GSC 6214-210 b, despite the impact of clouds (Fig.~\ref{fig:cloud_kap}). This suggests a potential trade-off between \Teff (set by the P-T profile) and clouds. Future work using broader wavelength coverage or flux-calibrated data should be able to constrain \Teff better in order to confirm our findings. We further discuss the effect of clouds on high-resolution spectroscopy in \S~\ref{sec:cloud_discuss}.

\subsection{Carbon and Oxygen
Abundances}\label{sec:c_and_o}
Our retrieved C/O and [C/H] values for the companions are listed in Table~\ref{table:bayes_factors}. As with previous high-resolution studies \citep[e.g.][]{Finnerty2023, Xuan2024}, our constraints on the relative abundances of different molecular species as indicated by the ratio of their relative line depths, which are used to calculate C/O, are tighter than our constraints on the absolute abundances of individual molecular species, which are used to determine the overall atmospheric [C/H]. Specifically, our uncertainties range between $0.02-0.08$ for C/O and $0.2-0.5$ dex for [C/H] depending on the object and S/N. Both the C/O and [C/H] of our companions are consistent at the 1--2$\sigma$ level with the solar composition from \citet{Asplund2021}. The implications of our measured abundances for formation pathways are discussed in \S~\ref{sec:discuss_abunds}.

\subsection{\co~Measurements}\label{sec:13co_result}
We obtain bounded constraints on \co~for GQ~Lup~b, HIP~79098~b, and DH~Tau~b, which are the three companions with the highest S/N detections as shown by the CCFs (Fig.~\ref{fig:ccf_detect}). To quantify the detection significance of $^{13}$CO, we run additional retrievals where we leave out $^{13}$CO opacities, and compute the Bayes factor between these `reduced models' without $^{13}$CO and the `full model' that includes $^{13}$CO and fits for \co. The resulting log Bayes factors are listed in Table~\ref{table:bayes_factors}, and correspond to $5.7\sigma$, $4.3\sigma$, and $3.4\sigma$ detections of the $^{13}$CO isotopologue for GQ~Lup~b, HIP~79098~b, and DH~Tau~b, respectively. We note that ROXs~12~b also has a \co~posterior which peaks at $\sim100$, but the corresponding detection significance is $<3\sigma$ so we do not consider it in the following discussion. The other four companions show unbounded \co~posteriors from our tests, so their baseline models do not include \co. 

Following \citet{Xuan2022}, we perform a cross-correlation analysis to obtain a complementary perspective on the robustness of these $^{13}$CO detections. The goal of this analysis is to assess whether the full models prefer $\rm{^{13}CO}$ independent of the Bayes factor calculation. First, we compute the CCF between a $\rm{^{13}CO}$-only model and the (data - model without $\rm{^{13}CO}$). The latter is equivalent to the residuals of the reduced model, and will contain residual $\rm{^{13}CO}$ lines if the data contain $\rm{^{13}CO}$. Then, we compute the CCF between the $\rm{^{13}CO}$-only model and the (data - model with $\rm{^{13}CO}$). This second CCF should not show a detection, as $\rm{^{13}CO}$ is already fitted for in the full model. We generate $\rm{^{13}CO}$-only models by manually zeroing the opacities of all other line species except $\rm{^{13}CO}$ when computing the full model. The CCF calculations follow the framework described in \S~\ref{sec:ccf}.

From the CCF analysis, we find that the CCFs for GQ~Lup~b and HIP~79098~b show convincing peaks at the companion's rest frame that indicate a strong $\rm{^{13}CO}$ detection (see Fig.~\ref{fig:13co_ccfs}). However, the $\rm{^{13}CO}$ CCF for DH~Tau~b is much noisier, showing residual structure comparable to the peak. Therefore, we consider the $\rm{^{13}CO}$ detection in DH~Tau~b to be tentative, since it is possible that $\rm{^{13}CO}$ is being used to fit for systematics that are unaccounted for in the model. This is consistent with the larger uncertainties in the \cratio~of DH~Tau~b, and is expected given the lower S/N of the DH~Tau~b data compared to the data for the other two companions. We further discuss our \cratio~measurements in the context of previous results in \S~\ref{sec:discuss_13co}.

\section{Discussion} \label{sec:discuss}

\begin{deluxetable*}{lcccc}
\tablecaption{Comparison of Our Measurements with Previous Work \label{table:compare_res}}
\tabletypesize{\small}
\tablehead{
Target name  & C/O & Metallicity ($\times$ solar) & \vsini (\kms) & Source}
\startdata
GQ~Lup~b & $0.70^{+0.01}_{-0.02}$ & $2.5^{+1.5}_{-1.0}$ & $6.4^{+0.3}_{-0.4}$ & TP \\
  & $0.44^{+0.13}_{-0.11}$ & $1.7\pm0.3$ & $5.3^{+0.9}_{-1.0}$ & 1, 7 \\
\hline
 \kapandb  & $0.58^{+0.05}_{-0.04}$  & $0.8^{+0.8}_{-0.3}$ & $39.4\pm0.9$ & TP \\
  & $0.70^{+0.09}_{-0.24}$ & 1.0--1.6 & $38.42\pm0.05$ &  6, 9 \\
\hline
GSC~6214-210~b  & $0.70^{+0.07}_{-0.06}$ & $1.4^{+1.6}_{-0.7}$ & $11.6^{+1.9}_{-2.1}$ & TP \\
 & $0.48^{+0.16}_{-0.12}$ & $0.7^{+0.3}_{-0.2}$ & $6.1^{+4.9}_{-3.8}$ & 2, 7 \\
\hline
2M0122~b  & $0.37\pm0.08$ & $0.5^{+0.6}_{-0.2}$ & $19.6^{+3.0}_{-2.5}$ & TP \\ 
& ... & ... & $13.4^{+1.4}_{-1.2}$ & 3 \\ 
\hline
ROXs 42B b  & $0.48\pm0.08$ & $1.0^{+2.8}_{-0.7}$ & $4.4^{+1.6}_{-2.1}$\tablenotemark{a} &  TP \\
& $0.50\pm0.05$ & $0.2^{+0.3}_{-0.1}$ & $10.5\pm0.9$ & 8 \\
\hline
DH Tau b & $0.54^{+0.06}_{-0.05}$ & $0.5^{+0.6}_{-0.2}$ & ${5.7 ^{+0.8}_{-1.0}}$ & TP \\
& ... & ... & $9.6\pm0.7$ & 5 \\
\hline
ROXs~12~b & $0.54\pm0.05$ & $0.5^{+0.4}_{-0.2}$ & $3.6^{+1.2}_{-1.6}$\tablenotemark{a} & TP \\
& ... & ... & $8.4^{+2.1}_{-1.4}$ & 4 \\
\enddata
\tablecomments{TP refers to this paper.}
\tablenotetext{a}{These companions have a weak detection of spin with $\sim1\sigma$ significance in the KPIC data. Future higher S/N or resolution spectra is required to definitively measure their \vsini.}
\tablerefs{(1) \citet{Schwarz2016}, (2) \citet{bryan_constraints_2018}, (3) \citet{Bryan_obliquity_2020}, (4) \citet{bryan_Worlds_2020}, (5) \citet{xuan_Rotation_2020} (6) \citet{Hoch2020}, (7) \citet{Demars2023}, (8) \citet{Inglis2024}, (9) Morris et al. 2024, accepted}
\end{deluxetable*}

\subsection{Comparison with Previous Work}\label{sec:discuss_previous}
As noted in \S~\ref{sec:system_prop}, many our companions have previous medium- or high-resolution studies which reported on their \vsini~and/or atmospheric abundances. Table~\ref{table:compare_res} compares our baseline KPIC measurements with previous results.

\citet{Hoch2020} used Keck/OSIRIS medium-resolution spectra ($R\sim4000$, $2.22-2.4~\mu$m) to measure the abundances of \kapandb~using a custom grid of PHOENIX models \citep{Barman2011, barman_simultaneous_2015} that vary in metallicity and C/O. They report $\rm C/O=0.70^{+0.09}_{-0.24}$ and $\rm [M/H]=0.0-0.2$, which are consistent with our measured $\rm C/O=0.58^{+0.05}_{-0.04}$ and $\rm [C/H]=-0.12^{+0.28}_{-0.19}$. Our measured $\vsini=39.4\pm0.9~$\kms for \kapandb~also agrees with the recent study by Morris et al. 2024, accepted, who find $38.42\pm0.05~$\kms using KPIC data from a different observing night.  

\citet{Schwarz2016} used VLT/CRIRES spectra ($R\sim100,000$, $2.30-2.33~\mu$m) to measure $\vsini=5.3^{+0.9}_{-1.0}$ \kms for GQ~Lup~b, which is consistent with our $\vsini=6.4^{+0.3}_{-0.4}$ \kms to within $\approx1.1\sigma$. This agreement is reassuring given the $\sim3\times$ lower spectral resolution of Keck/KPIC compared to VLT/CRIRES, and confirms the ability of KPIC to measure \vsini~values below the resolution limit of $\sim7.5$ \kms, as demonstrated by \citet{Xuan2024} as well. 

\citet{Demars2023} used VLT/SINFONI medium-resolution spectra and the self-consistent model ATMO \citep{Tremblin2015, Tremblin2016} to fit the abundances of GQ~Lup~b and GSC~6214-210~b. They adopted $\rm C/O=0.44^{+0.13}_{-0.11}$ and $\rm [M/H]=0.23\pm0.06$ for GQ~Lup~b and $\rm C/O=0.48^{+0.16}_{-0.12}$ and $\rm [M/H]=-0.16\pm0.17$ for GSC~6214-210~b. Overall, our measured abundances agree with their results at the $1-2~\sigma$ level. However, we note that the \citet{Demars2023} results are discrepant between different observing epochs at the $\sim60\%$ level in C/O and $>0.4$ dex in [M/H], and sometimes hit the limits of their grid. In contrast, as we show in Appendix~\ref{app:spline}, our KPIC results between different observing epochs of GQ~Lup~b are consistent at the $<3\%$ level in C/O and $\sim0.1$ dex in [C/H]. This demonstrates the reliability of high-resolution spectra and our forward model + retrieval approach in measuring atmospheric abundances. 

Five of our targets (ROXs~42~Bb, ROXs~12~b, DH~Tau~b, 2M0122~b, GSC 6214–210 b) have reported \vsini measurements with pre-upgrade Keck/NIRSPEC spectra ($R\sim25,000$) as part of a spin survey for low-mass substellar companions \citep{bryan_constraints_2018, xuan_Rotation_2020, bryan_Worlds_2020}. We list the literature values along with our new values in Table~\ref{table:compare_res}, and discuss them below. 

For the faster rotators GSC-6214-610 b and 2M0122~b, our measured \vsini are consistent with the pre-upgrade NIRSPEC values from \citet{bryan_constraints_2018} and \citet{Bryan_obliquity_2020} at the $1\sigma$ and $2\sigma$ levels, respectively. For the slower rotators (ROXs 12 b, ROXs 42 Bb, and DH Tau b), we find $\gtrsim3\sigma$ discrepancies in \vsini, with a trend that KPIC measured \vsini values are lower than those from pre-upgrade NIRSPEC studies. However, we note that for ROXs~12~b and ROXs 42 Bb our KPIC data only prefer a non-zero spin with $\approx1\sigma$ significance, so future data is required to definitively measure their \vsini. For ROXs 42 Bb, \citet{Inglis2024} perform atmospheric retrievals with pre-upgrade NIRSPEC data and report $\vsini=10.5\pm0.9~$\kms, which agrees with the $\vsini=9.5^{+2.1}_{-2.3}~$\kms found by \cite{bryan_constraints_2018} for the same data. However, we find $\vsini=4.2^{+1.8}_{-2.2}~$\kms for ROXs~42~Bb. Despite these differences in \vsini, our retrieved abundances for ROXs~42~Bb agree with those from \citet{Inglis2024}. Finally, in \citet{xuan_Rotation_2020} we found $\vsini=9.6\pm0.7~$\kms for DH~Tau~b using pre-upgrade NIRSPEC data, while we measure $\vsini={5.7^{+0.8}_{-1.0}}~$\kms with KPIC. 

Low S/N spectra can preclude confident spin measurements, especially for very slow rotators. However, in all three cases above, the KPIC data have higher CCF S/N compared to the pre-upgrade NIRSPEC data. For ROXs~12~b, ROXs~42~Bb, and DH~Tau~b, we obtain $\sim$25$\sigma$, $\sim$15$\sigma$, and $\sim$20$\sigma$ detections (Fig.~\ref{fig:ccf_detect}), respectively, while the previous NIRSPEC detections have CCF S/N $\sim10\sigma$ \citep{xuan_Rotation_2020, Inglis2024}. If the companions are in fact rotating faster, our KPIC data should have been able to reveal this. 

Despite the higher CCF S/N of the KPIC data, the uncertainties of our new \vsini measurements are comparable with the older measurements. This suggests that there may be additional sources of uncertainty that the older NIRSPEC studies did not consider. For example, the retrieval approach allows us to vary the P-T profile and therefore adjust the pressure-broadened line shapes in each iteration. In contrast, \citet{xuan_Rotation_2020} and \citet{bryan_Worlds_2020} used fixed atmospheric templates to compute \vsini, and may under-estimate the uncertainty introduced by the uncertain atmospheric properties of the companion.

We note that for a high S/N dataset of the isolated brown dwarf 2MASS J03552337+1133437 (hereafter 2M0355), \citet{zhang_12CO_2021} found similar discrepancies in \vsini~compared to \citet{bryan_constraints_2018}. Using the same NIRSPEC data for the brown dwarf, \citet{zhang_12CO_2021} found an upper limit of 4~\kms for \vsini, consistent with a non-detection of spin, but \citet{bryan_constraints_2018} reported $\vsini=14.7\pm1~$\kms. Using earlier $R\sim25,000$ NIRSPEC data, \citet{Blake2010} also found a similarly large spin for this brown dwarf which agrees with the \citet{bryan_constraints_2018} value. Recently, however, \citet{Zhang2022} used much higher resolution VLT/CRIRES+ spectra ($R$$\sim$80,000) to confirm the slow spin of 2M0355, finding $\vsini=2.5-3.0~$\kms. The fact that \citet{zhang_12CO_2021} and \citet{bryan_constraints_2018} used the same NIRSPEC data and found different \vsini for 2M0355 point to details in the data reduction and spectral extraction process as possible sources of discrepancy.

For example, the higher spin values from older NIRSPEC data may be the result of systematic underestimation of the instrumental line spread function, which is degenerate with \vsini. For KPIC, the LSF is estimated using the spectral trace widths in the spatial direction, and conservatively allowed to vary between 1.0 to 1.2 times this width following \citet{wang_Detection_2021}, who found that the LSF is $\approx1.1$ times broader in the spectral direction than the spatial direction (see \S~\ref{sec:forward_m}). A major advantage of the single-mode fiber injection used by KPIC is the stability of the LSF, whose shape does not vary with the adaptive optics (AO) correction quality (only the intensity varies). Furthermore, we confirm that the KPIC LSF is constant over the course of an observing night by re-fitting the LSF every 20 minutes over a 3.5-hour period. The LSF weights in each refitting vary by $<5\%$, with most of the variation in the line wings, suggesting we are limited by the signal-to-noise of each substack and that the flux-weighted LSF is even more stable. In addition, the LSF in KPIC is measured to vary as a function of wavelength by $\sim10-20\%$ within and across different spectral orders. In regular NIRSPEC, the resolution should also vary with wavelength at a similar level, but this wavelength-dependence is neglected in pre-upgrade NIRSPEC spin studies since it is hard to measure \citep{xuan_Rotation_2020}.

In contrast, for the pre-upgrade NIRSPEC analysis, the LSF was determined by fitting telluric lines in the host star spectra. For example, \citet{xuan_Rotation_2020} found $R=24,800\pm1000$ from the DH Tau dataset, which was adopted as a Gaussian instrumental broadening kernel and assumed to be constant with wavelength. We note that NIRSPAO (NIRSPEC in AO mode) was used for these observations, and the size of the PSF was comparable or smaller than the slit width of 0.041\arcsec \citep{xuan_Rotation_2020}. In this case, variations in the AO correction quality cause varying PSF sizes, and when the PSF becomes smaller than the slit this would cause a narrower LSF. However, the effect of a varying LSF should be captured by measuring the instrumental resolution from telluric lines in the data, which are varying in the same way. However, as shown in \citet{xuan_Rotation_2020}, the approach of determining the LSF width from telluric lines can sometimes produce poor fits and discrepant results between different spectral orders, which may bias the adopted LSF estimate. 

We conclude that future work is required to explain the \vsini discrepancies in more detail. Given the higher spectral resolution, higher S/N, and more stable LSF of KPIC, and the fact that our KPIC-measured spin of GQ~Lup~b agrees with $R$$\sim$100,000 CRIRES data \citep{Schwarz2016}, we conclude that our measurements are likely more reliable than pre-upgrade NIRSPEC observations of the same objects. Despite some differences, our new \vsini~results are in good agreement with the findings from \citet{bryan_Worlds_2020} that low-mass substellar companions rotate much slower than their breakup velocities, pointing to mechanisms such as magnetic breaking that can efficiently reduce the companion's angular momentum on short timescales \citep{bryan_constraints_2018, batygin_terminal_2018}. Finally, our measured abundances for \kapandb~and ROXs~42B~b both agree with previous work using medium- and high-resolution spectra \citep{Hoch2020, Inglis2024}, which is a validation of the power of these data to constrain chemical abundances in substellar companions and exoplanets. Except for ROXs 42B b, our KPIC measurements of the other seven companions are the first high-resolution retrievals of these objects.

\subsection{Rotation Rates and RVs}\label{sec:rotation}
Here, we place our measured \vsini~into context with literature spin measurements for low-mass companions and exoplanets. 
Given the discussion above, we update the \vsini's~for DH~Tau~b, ROXs 42B b, ROXs~12~b, 2M0122~b, and GSC 6214-210~b. 
For bound companions, we compile literature measurements from \citet{Bryan2020}, and recent KPIC studies from \citet{wang_Detection_2021, Xuan2022, wang_Retrieving_2022, Xuan2024, Hsu2024_submitted}.
For field brown dwarfs, we compile results from \cite{crossfield_global_2014, Tannock:2021aa, Hsu:2021aa, Hsu:2024aa, vos_Let_2022}. The spin measurements come from both \vsini and rotational period measurements. For the \vsini values, we assume isotropically distributed inclinations to remove the unknown $i$.

\begin{figure}[t!]
\centering
\includegraphics[width=1.0\linewidth]{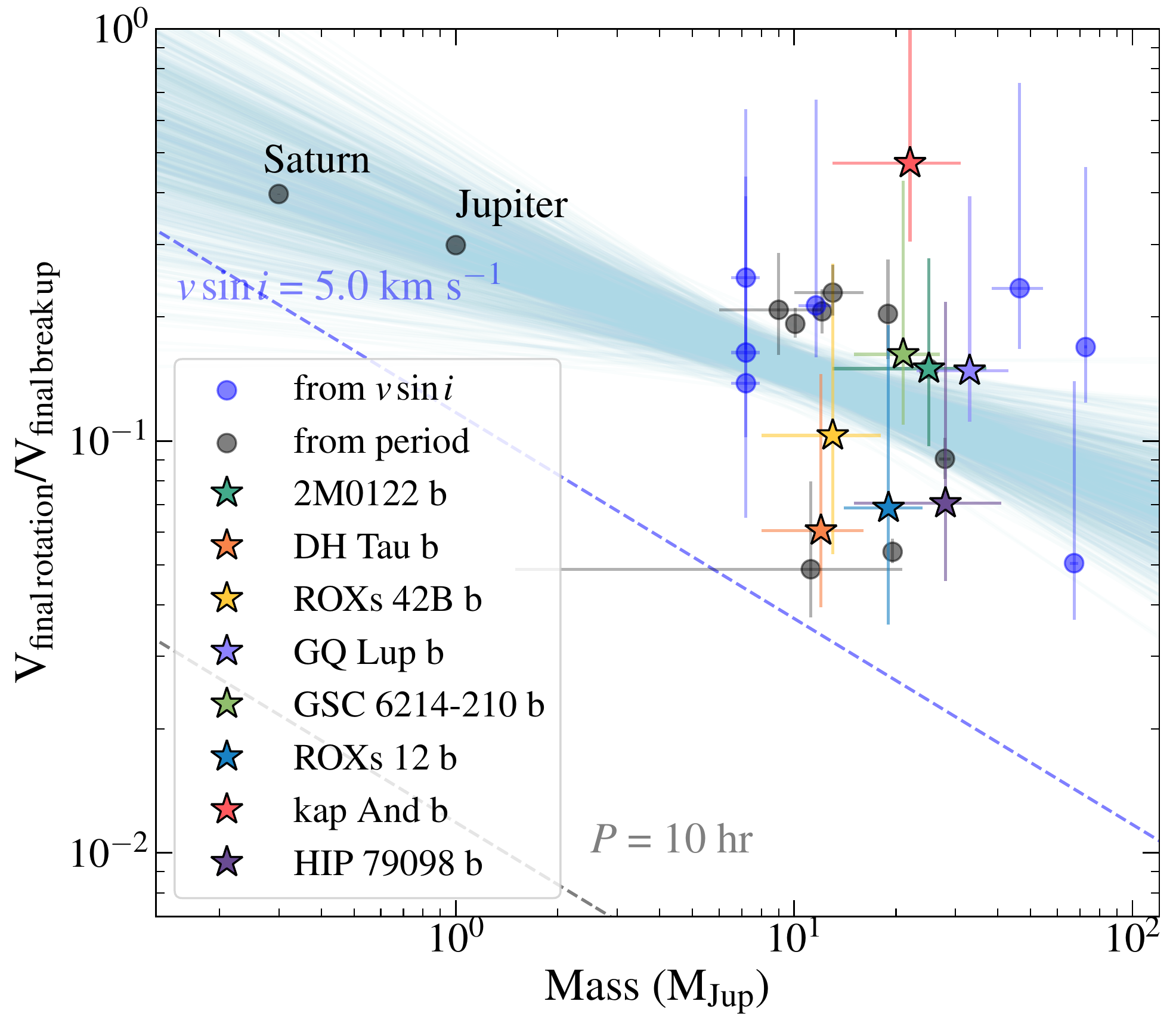}
\includegraphics[width=1.0\linewidth]{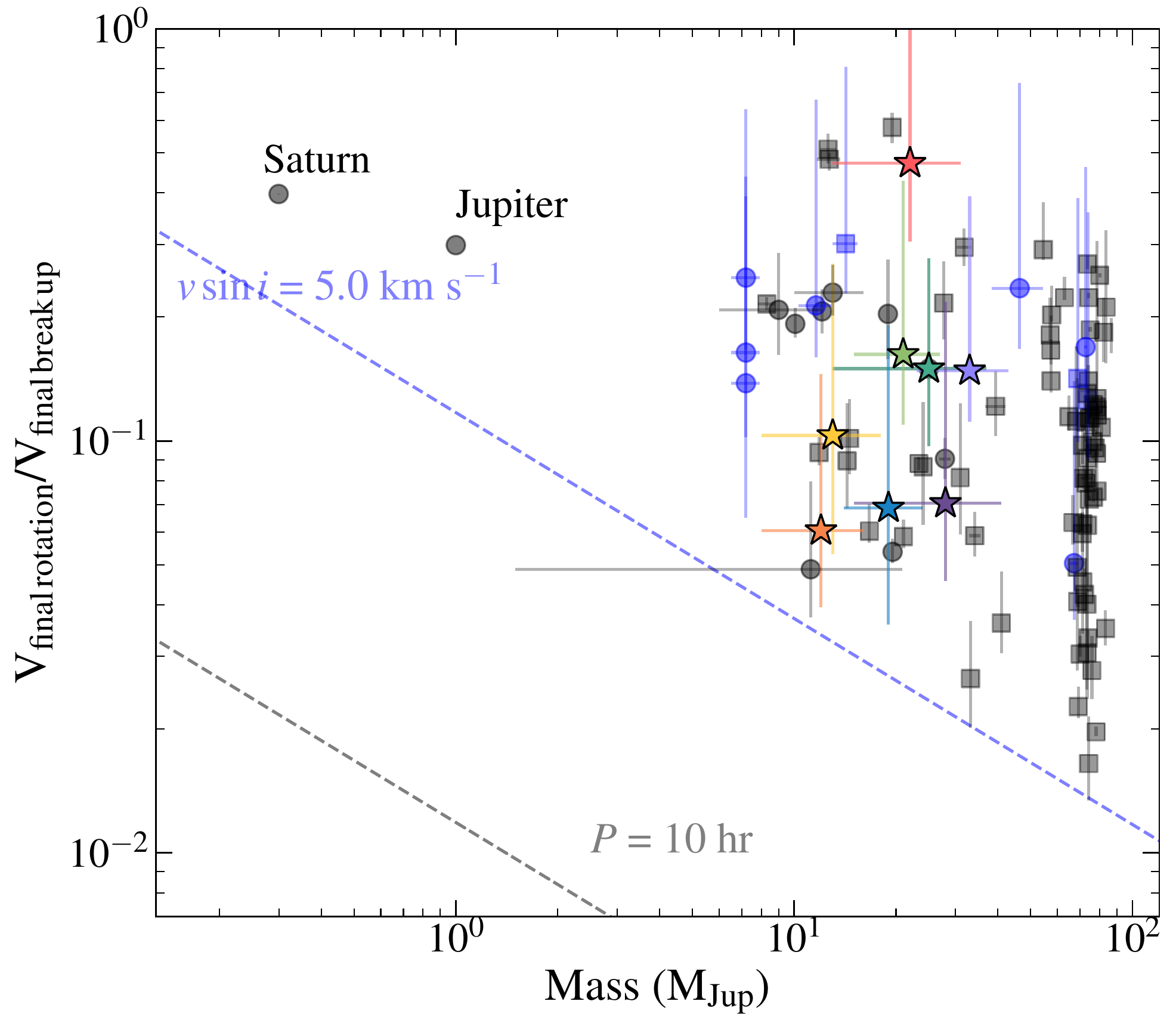}
\caption{Top: The fractional final rotational velocities versus companion masses for sources in our sample (labeled) compared to companions with spin measurements from the literature, as well as Jupiter and Saturn.
The literature values from $v\sin{i}$ and rotational periods are shown in blue and gray circles, respectively.
The best-fit rotational trends from \cite{wang_Detection_2021} are plotted as light blue lines. The blue and gray dashed lines indicate typical measurement limits of $v\sin{i}$ (5.0~\kms) and period (10~hr) from the instruments used \citep{Hsu2024_submitted}. Bottom: Same as the top panel but with the addition of field brown dwarfs as squares.}
\label{fig:spins}
\end{figure}

We compute how close the objects' final rotation speeds compare to their final break-up velocities by evolving their ages, radii, and spin velocities to 5 Gyr assuming constant angular momentum evolution, following the methodology detailed in \citet{wang_Retrieving_2022} and \citet{Hsu2024_submitted}. We adopt the evolutionary model from \cite{baraffe_evolutionary_2003} for this.
Evolving the rotation to the same age allows us to remove age-dependent effects from literature measurements of \vsini or photometric rotational periods.
Fig.~\ref{fig:spins} shows that the spin measurements in our sample fall within the overall trend compared to the literature measurements, and the best-fit rotational trend from \cite{wang_Detection_2021}, who identified a tentative anti-correlation between fractional rotation speed and object mass. 
Five companions in our sample have ages $\lesssim10$~Myr (ROXs 12 b, ROXs 42 Bb, DH Tau b, GQ Lup b, HIP 79098 b). These companions also have slower \vsini values below 10~\kms and are likely still undergoing gravitational contraction and gradually spinning-up. 
This fits the findings of \citet{bryan_Worlds_2020}, who showed that younger substellar objects generally have lower rotation speeds than older objects, and that their rotation speeds increase as their radii contract with age following constant angular momentum evolution. On the other hand, the fractional rotation velocities of isolated brown dwarfs (squares) are much more scattered, and these field objects do not exhibit a clear trend for rotation. However, we note that field brown dwarfs with masses $\lesssim$30~M$_\text{Jup}$ fall into the Y dwarf regime, for which rotation rates are extremely challenging to measure due to their faintness. To confirm the tentative trend between rotation rate and mass, it would be useful to extend \vsini or rotational period measurements to young, directly imaged planets with masses $\sim1-10~\Mj$.
A detailed analysis of all KPIC rotation measurements will be presented in a future study.

In addition to measuring \vsini, we also measure the RV of the companions at the observed epochs. We provide the companion RVs in Appendix~\ref{app:rvs}, which could be used to refine their orbits \citep{Ruffio2019, DoO2023, Xuan2024}.
 
\begin{figure*}[t!]
    \centering
    \begin{subfigure}
      \centering
        \includegraphics[width=0.485\linewidth]{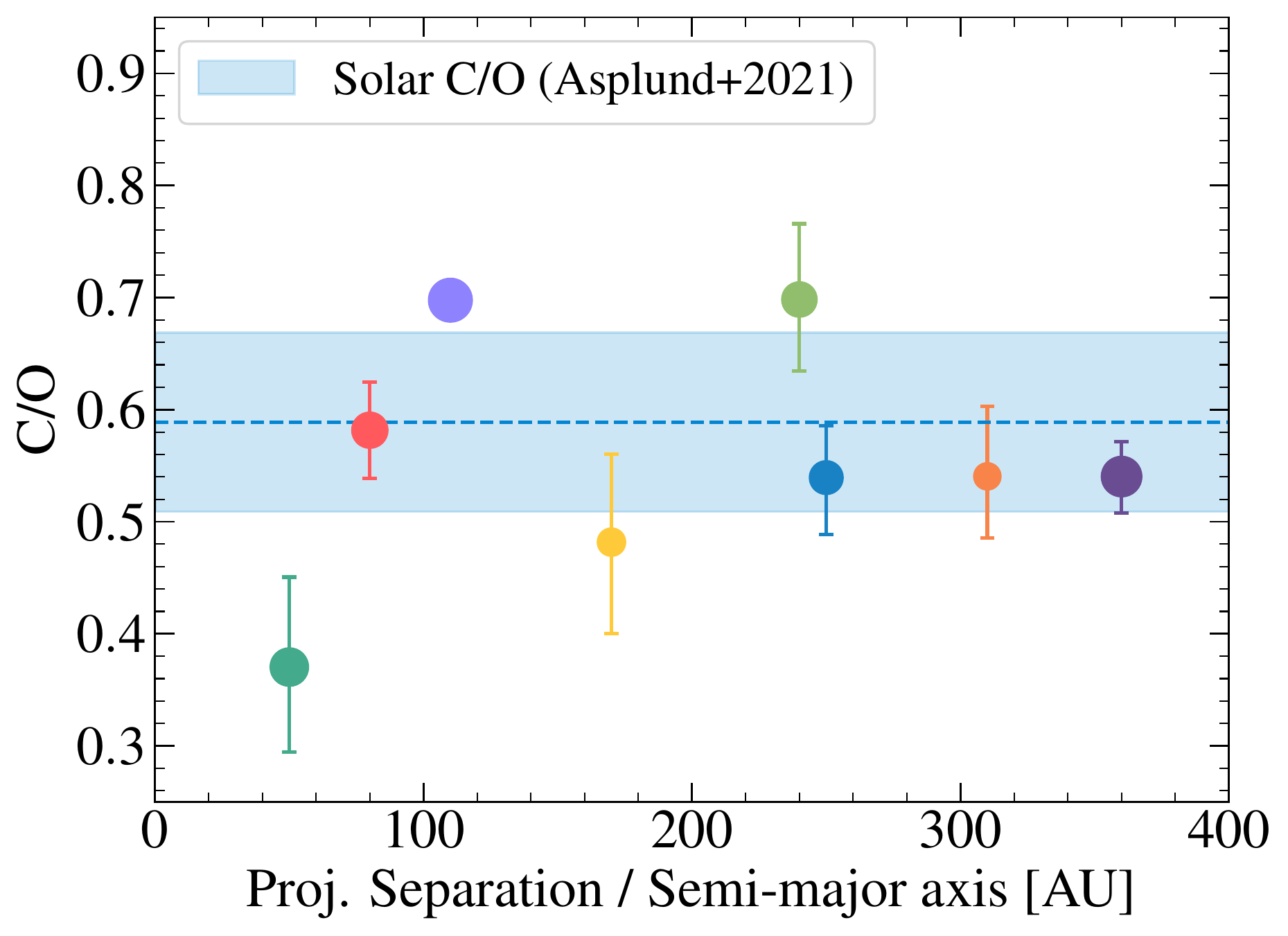}
    \end{subfigure}
    \begin{subfigure}
      \centering
    \includegraphics[width=0.495\linewidth]{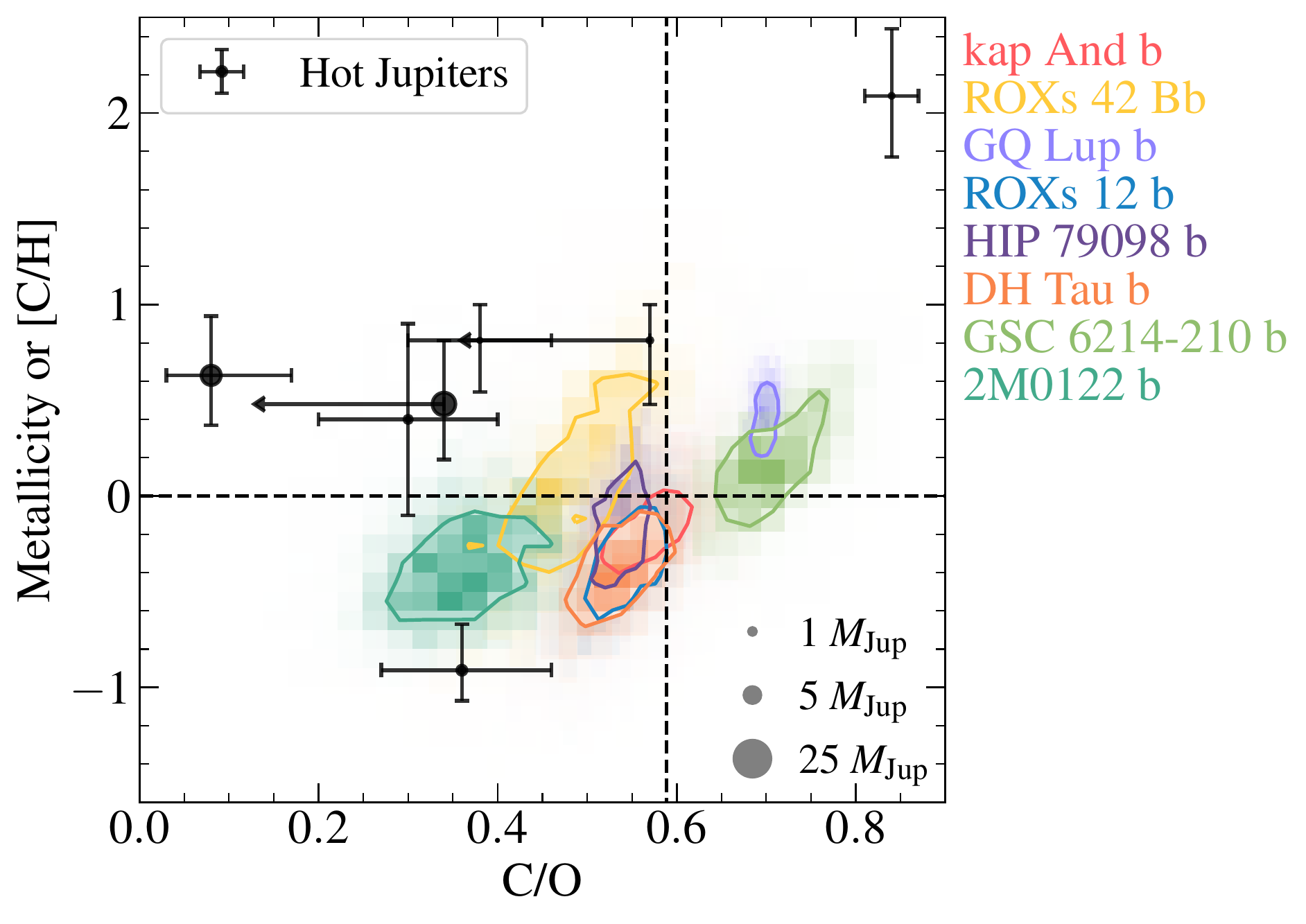}
    \end{subfigure}
    \caption{Left: C/O as a function of projected separation or semi-major axis for the eight companions in this study. The points are sized by companion mass, using the mass legend in the right panel. The blue line and region show the solar C/O value from \citet{Asplund2021}. Right: In color: $1\sigma$ contours of C/O and [C/H] for the eight companions, which are all consistent with solar composition (intersection of dashed gray lines) to the $2\sigma$ level. In \S~\ref{sec:host_abunds}, we justify that their host stars likely have solar composition as well. We overplot hot Jupiters which have abundance measurements for at least one oxygen-bearing and one carbon-bearing species from JWST or high-resolution spectroscopy as black points \citep{Alderson2023, August2023, Bean2023, Bell2023, Brogi2023, Finnerty2024, Xue2024}, which are also sized by mass. The hot Jupiters show considerable scatter and super-solar (and super-stellar) metallicities except for one planet, unlike our companions which cluster around solar (and stellar) composition.}
    \label{fig:abund_summary}
\end{figure*}

\subsection{The Effect of Clouds on Narrowband High-Resolution Spectroscopy}\label{sec:cloud_discuss}
A rather unexpected finding of this paper that narrow-band high-resolution spectra can be sensitive to clouds for $\Teff\sim1700-2000~K$ objects, even after high-pass filtering the data and models. We illustrate the impact of clouds in Fig.~\ref{fig:cloud_kap} for \kapandb~and GSC~6214-210~b, the two companions which show $>3\sigma$ preferences for the EddySed cloud model \citep{ackerman_Precipitating_2001}. In our retrievals with \texttt{petitRADTRANS}, the EddySed model accounts for both absorption and scattering from clouds, whose opacities are computed from optical constants of real condensates \citep{molliere_petitRADTRANS_2019}. We find that a simple, gray opacity cloud model is not sufficient to explain their spectra. The sensitivity to clouds appears to arise from line depth variations with wavelength, which we can visualize in the \kapandb~spectra after binning it down to average out the residual noise. While our data does not contain absolute flux information, the line depth variations are preserved after the HPF continuum removal procedure.

Despite their preference for clouds, we note that the retrieved C/O and metallicity for \kapandb~and GSC~6214-210~b are only weakly affected by clouds ($\lesssim 1\sigma$ level shifts). However, compared to the clear models, the EddySed cloudy models retrieve a much lower \Teff. Specifically, the \Teff from the EddySed models are more consistent with cloudy evolutionary models (see \S~\ref{sec:pt_teff}). The fact that \kapandb~and GSC 6214-210~b prefer clouds is consistent with their $\Teff$ of $1700-2000~K$ and early L spectral types. Specifically, silicate clouds cause a broad absorption feature around $\approx8-11~\mu$m which has been directly observed in L dwarfs using mid-IR spectra from Spitzer and JWST \citep[e.g.][]{cushing_Spitzer_2006, Suarez2022_L2silicate, Miles2023}. For brown dwarfs with $\Teff\gtrsim2000~K$, the silicate absorption band starts to disappear \citep{Suarez2022_L2silicate}, which is consistent with the fact that the hotter $\Teff>2200~K$ companions in our sample show zero or weak preference clouds ($0-2.6\sigma$).

2M0122~b, with $\Teff\sim1500-1700~K$, presents an exception to this pattern. This companion should also be cloudy, but its low S/N data or nearly pole-on viewing angle (see \S~\ref{sec:cloud_insensitive}) may prevent a stronger constraint on clouds. In addition, due to our limited S/N, it is difficult to directly visualize the effect of clouds in the GSC 6214-210 b spectra. We emphasize that future work with flux-calibrated data and wider wavelength coverage is required to confirm our findings of clouds in \kapandb~and GSC 6214-210 b. Retrieval studies of high S/N, high-resolution spectra for cloudy L dwarfs would also provide a good test of our findings. 

Finally, we note that our results represent an update to \citet{Xuan2022}, who found that the $K$ band spectrum of brown dwarf companion HD~4747~B ($\Teff$$\approx$1400~K and $\logg$$\approx$5.3) is insensitive to clouds because the MgSiO$_3$ and Fe cloud bases are expected to lie below its $K$ band photosphere. Our new findings indicate that when the cloud bases do intersect with the $P$--$T$ profile near the photosphere, as is the case for $\Teff$$\sim$1700-2000~K companions, $K$ band high-resolution spectroscopy could be sensitive to cloud opacities despite continuum removal. There remain significant challenges in retrieving accurate abundances in the presence of clouds from low-resolution data \citep[e.g.][]{burningham_retrieval_2017, Xuan2022, Lueber2022, Inglis2024}, so high-resolution retrievals of L dwarfs may provide a complementary way forward.

\subsection{Towards Atmospheric Abundance Trends for Directly Imaged Companions} \label{sec:discuss_abunds}

\subsubsection{Host Star Abundances}\label{sec:host_abunds}
Knowledge of the host star abundances is important for formation inferences, since they represent proxies for the natal elemental abundances in the protoplanetary disk or molecular cloud. For the young stars ($\sim1-100$ Myr) in our sample, however, it is difficult to measure their C and O abundances due to complicating factors such as rotation, magnetic fields, veiling, and stellar activity. Even for field stars, C and O abundance calibrations are only recently being worked out for M and K spectral types \citep[e.g.][]{Souto2022, Hejazi2023}; six of the eight stars in our sample fall into this category. The remaining two B9 stars, HIP 79098 and \kapand, have extremely rapid rotation ($\vsini>100$ \kms) that leads to significant spectral line broadening. This causes neighboring individual lines to blend with each other, most of which cannot then be distinguished even with high-resolution spectroscopy.

To circumvent these challenges, studies have focused on more favorable targets: early K to late F young stars with lower rotation rates and no apparent veiling. The C and O abundances of these more solar-like stars can be estimated from atomic C and O lines in the visible using established abundance calibrations \citep{Reggiani2024}. Alternatively, we can utilize abundance measurements for other species, including Ca, Mg, Si, and Fe, which also have strong, isolated lines in the visible \citep[e.g.][]{Santos2008, Biazzo2012}. Importantly, Ca, Mg, and Si are alpha elements along with C and O. Using the elemental abundances of $\approx6000$ stars from the Hypatia catalog \citep{Hinkel2014}, we confirm that the abundances of Ca, Mg, and Si scale together with those of C and O to the $\lesssim0.2$ dex level. Below, we summarize previous abundance measurements for stars in the same star-forming associations as our host stars. Since stars in open clusters are found to be chemically homogeneous at the $<0.03$ dex level \citep[e.g.][]{DeSilva2006, Bovy2016, Ting2012, Poovelil2020}, stars in the same star-forming associations should also be chemically homogeneous at a similar level \citep{Reggiani2024}.

In terms of C and O, \citet{Reggiani2024} showed that the F7 star HD~181327 in the $\beta$ Pic moving group has [C/H]$=-0.08\pm0.06$, [O/H]$=-0.10\pm0.06$, and C/O$=0.62\pm0.08$, which is very close to solar. Kinematic studies have shown that the $\beta$ Pic moving group likely originated from the  Scorpius–Centaurus (Sco-Cen) association \citep{Mamajek2001, Ortega2002}. This means that to first order, stars in the Sco-Cen association should also have roughly solar C and O abundances. Five of our stars belong to sub-regions within Sco-Cen (HIP 79098, GSC 6214-610, GQ Lup, ROXS 12, and ROXs 42B). \citet{wang_Chemical_2020a} measured [C/H]$=0.04\pm0.12$, [O/H]$=0.08\pm0.14$, and C/O$=0.54^{+0.12}_{-0.09}$ for HR~8799, which is an early F $\lambda$ Boo star. The kinematic association of HR~8799 is unclear, but a recent study posits that HR~8799 may have formed either alone or in a since-dispersed small stellar group within a larger star-forming complex that gave birth to the Columba and Carina groups \citep{Faramaz2021}. If \kapand~is a member of Columba as suggested by the age measurement from \citet{Jones2016}, then it likely has a solar composition as well. 

In terms of other elements, \citet{Santos2008} measured Fe and Si abundances for six star-forming regions including Lupus, $\rho$ Oph, and Taurus and found solar [Fe/H] and [Si/H] values across their sample. Four of our stars belong to these regions (GQ Lup, ROXS 12, ROXs 42B, DH Tau). More recently, \citet{Biazzo2017} analyzed the spectra of six other stars in Lupus and also found $\rm [Fe/H]\approx0.03$ on average. We note that the Lupus and $\rho$ Oph regions are both embedded in the larger Sco-Cen association. In Taurus, \citet{DOrazi2011} found an average $\rm [Fe/H]=-0.01\pm0.05$ and $\rm [Si/H]$ consistent with zero from seven stars. In the older AB Dor moving group, where 2M0122 is, \citet{Biazzo2012} find average values of $\rm [Fe/H]=0.10\pm0.03$, $\rm [Mg/Fe]=-0.03\pm0.03$, and $\rm [Ca/Fe]=-0.01\pm0.05$, again consistent with solar abundances to within 0.1 dex. Together, these studies paint a broad picture of solar abundances in young star-forming regions \citep[e.g. see also][]{Biazzo_orion_2011, Spina2014}.

Therefore, to first order, all our host stars are expected to have nearly solar C and O abundances to the $\sim0.1-0.2$ dex level. We therefore proceed under the assumption that our host stars have solar compositions, but emphasize that future work should attempt to perform precise C and O abundance measurements for young stars in various star-forming regions in order to provide context for abundance measurements of directly imaged planets and brown dwarfs, a majority of which are young. Specifically, high-resolution, near-infrared spectroscopy has proven vitally important in characterizing the fundamental parameters and measuring the abundances of lower-mass host stars with spectral types later than mid-K \citep[e.g.][]{LopezValdivia2021, Souto2022, Hejazi2023, Cristofari2023}. In these latter type stars, the atomic lines of C and O in the optical are too weak for abundance measurements, but instead, C- and O-bearing molecules such as OH and CO can be utilized to infer the C and O abundances.  As mentioned, performing abundance measurements for young, late-type stars will involve a detailed accounting of the effects of magnetic fields, line veiling, and rotation, which we are working on including in a follow-up study in order to directly determine the stellar abundances.

\subsubsection{Formation Pathways of Widely-separated, 10--30~\Mj Companions} \label{sec:formation_discuss}
The C/O and [C/H] measurements for our eight companions are summarized in Fig.~\ref{fig:abund_summary}. The companions all have C/O consistent to solar within $1\sigma$ level with the exception of our lowest S/N target, 2M0122~b, which has a lower C/O of $0.37\pm0.08$ but is still consistent with solar at the $2\sigma$ level. Their [C/H] values are likewise all consistent with solar at the $\approx1-2\sigma$ level. Since we established that their host stars most likely possess solar C and O abundances, our measurements suggest chemical homogeneity between these low-mass companions and their stars. The trend of solar C/O ratios for widely-separated companions with $m\approx10-30~\Mj$ was first noted by \citet{Hoch2023}, who compiled literature measurements and their own Keck/OSIRIS results. Our only overlap with the \citet{Hoch2023} sample is \kapandb, as discussed in \S~\ref{sec:discuss_previous}. While \citet{Hoch2023} only studied C/O, our measurements of solar [C/H] provides a complementary piece of information. High-resolution studies of high-mass ($m\approx60-70~\Mj$) brown dwarf companions indicate that their atmospheric abundances are consistent with those of their host stars at the $\lesssim2\sigma$ level \citep{wang_Retrieving_2022, Xuan2022}. Therefore, our results, along with \citet{Hoch2023}, indicate that widely-separated ($\gtrsim50$ au) 10--30~\Mj companions, which have masses in between those of directly imaged planets and high-mass brown dwarfs, have abundance pattern that more closely resemble those of brown dwarfs.

On a population level, we can also compare our abundance measurements with those made for close-in hot Jupiters (HJs; $m\sim0.4-10~\Mj$, $a\sim0.01-0.1$ au), which are found to have a large scatter in C/O and generally super-stellar atmospheric metallicities \citep[e.g.][]{Brogi2023, Alderson2023, August2023, Bell2023, Bean2023, Finnerty2024, Xue2024}. These hot Jupiters almost certainly form via core accretion (within the CO snowline), though a variety of post-formation evolution processes are possible \citep[see review by][]{fortney_Hot_2021}. The strong correlation between total heavy element mass and the total planet mass suggests these planets have solid cores consistent with core accretion \citep{Thorngren2016}, and the correlation between host star metallicity and their occurrence \citep{Petigura2018, Osborn2020} also indicates that metals are needed for their formation. As shown in Fig.~\ref{fig:abund_summary}, our widely separated, 10--30~\Mj companions occupy a region of the C/O and metallicity space that is clustered around solar (and likely stellar) composition, which is distinct than the the HJs. We carry out a weighted two-sample t-test to quantify the statistical likelihood that the compositions of HJs and those of our sample are drawn from the same population. We find p-values of $4.2\times10^{-4}$ for C/O (3.5$\sigma$) and $7.5\times10^{-6}$ for [C/H] ($4.5\sigma$), indicating that our measurements are inconsistent with the null hypothesis that the atmospheric compositions of HJs and our sample belong to the same population. Therefore, this serves as empirical evidence that our companions likely did not form via core accretion inside the CO snowline.

Our $\sim$10--30~$\Mj$ companions orbit a diverse group of host stars, with stellar masses spanning an order of magnitude between $\approx0.35-0.45~\Msun$ for DH Tau and 2M0122 \citep{Sebastian2021, Yu2023} to $\approx3-4~\Msun$ for kap~And~b and HIP~79098 \citep{Janson2019, Gaia2022_DR3}. This translates to a range of mass ratios ($q$) from $\approx0.007$ to $0.07$ (with $\sim30-60\%$ uncertainties on $q$ mostly due to the uncertain companion masses). Previous studies find that multiple star formation operates in a largely scale-invariant manner, with mass ratio being a more informative quantity than masses of either component \citep[e.g.][]{Goodwin2013, Duchene2023}. For example, \citet{Duchene2023} found a dearth of low-mass stellar companions around intermediate-mass stars ($M=1.75-4.5~\Msun$) with $q\approx0.05-0.10$, which matches the `brown dwarf desert' around solar-mass stars \citep[e.g.][]{Grether2006, Metchev2009, Sahlmann2011}. They postulate that this mass ratio desert delineates two distinct formation regimes with systems with $q\lesssim0.02$ forming in disks (either disk instability or core accretion), and those with $q\gtrsim0.07$ forming through cloud fragmentation. Seven of our companions have $q\lesssim0.03$, while 2M0122 b has $q\approx0.07$, so a mix of formation mechanisms are possible.

How do our results fare in terms of the different formation mechanisms? Perhaps most directly, our findings are consistent with birth via direct gravitational collapse in a massive protostellar disk or cloud fragmentation, which should produce broadly stellar compositions as is seen for stellar binaries \citep[e.g.][]{hawkins_Identical_2020}. While some studies predict that low-mass objects which form via disk instability can be subsequently enriched by the accretion of solids post-formation \citep{Boley2011}, it would be difficult to significantly alter the metallicity of the 10--30~$\Mj$ objects studied here \citep[see also][]{Inglis2024}. For example, a $10\times$ increase in metallicity for a $20~\Mj$ object would require the addition of $\sim$2~$\Mj$ of solids. Assuming a solid-to-gas mass ratio of 0.01 for the disk, accretion of $\sim$2~$\Mj$ of solids would require a disk-to-star ratio exceeding 0.2 for a solar-mass star and that all the solids in the disk accrete onto the companion. Therefore, significant metal enrichment for our companions is challenging, especially as most of our stars are less massive the Sun. By a similar logic, significantly sub-solar metallicities are also unexpected for 10--30~$\Mj$ companions that form via disk instability.

Alternatively, our composition measurements could also be compatible with core accretion outside the CO snowline (i.e. in situ). Outside the CO snowline, nearly all the metals are condensed into solids as there is very little gas. As a result, the solids inherit a stellar composition. This means that unless the planet accreted more gas than solids and became extremely metal poor, accretion of solids at these locations would typically produce stellar C/O and metallicities \citep[e.g.][]{Chachan2023}. Since our companions have nearly solar and stellar metallicities, they are chemically consistent with forming via core accretion outside the CO snowline. We note that planetesimal accretion is less efficient at wider orbital distances, but pebble accretion may be able to produce a sufficiently massive core on a reasonable timescale \citep{lambrechts_rapid_2012}. For example, \citet{Gurrutxaga2024} demonstrated pebble accretion pathways to form the giant planets PDS~70 b ($m\approx2-6~\Mj$, $a\approx21$ au) and c ($m\approx3-10~\Mj$, $a\approx34$ au; planet values from \citealt{wang_Constraining_2021}). However, it remains to be seen whether realistic pebble accretion scenarios can produce our higher mass and wider separation (up to $360$ au) companions. For example, half of our companions have $q\gtrsim0.03$, which means that companion formation must begin early in a disk with disk-to-star mass ratio $>0.03$ and finish before the bulk of the gas reservoir accretes onto the star. By the protoplanetary disk phase, typical disk-to-star mass ratios of $\sim0.01$ have insufficient mass to produce $>10$~\Mj companions around solar-mass stars even if all the disk mass goes into the companion (let alone around the M dwarfs in our sample). Finally, early, massive circumstellar disks can be prone to gravitational instability \citep{Kratter2016}, so an additional challenge of forming companions via core accretion at large distances is the need to prevent disk fragmentation from occurring.

\begin{figure}[t!]
    \centering
      \includegraphics[width=\linewidth]{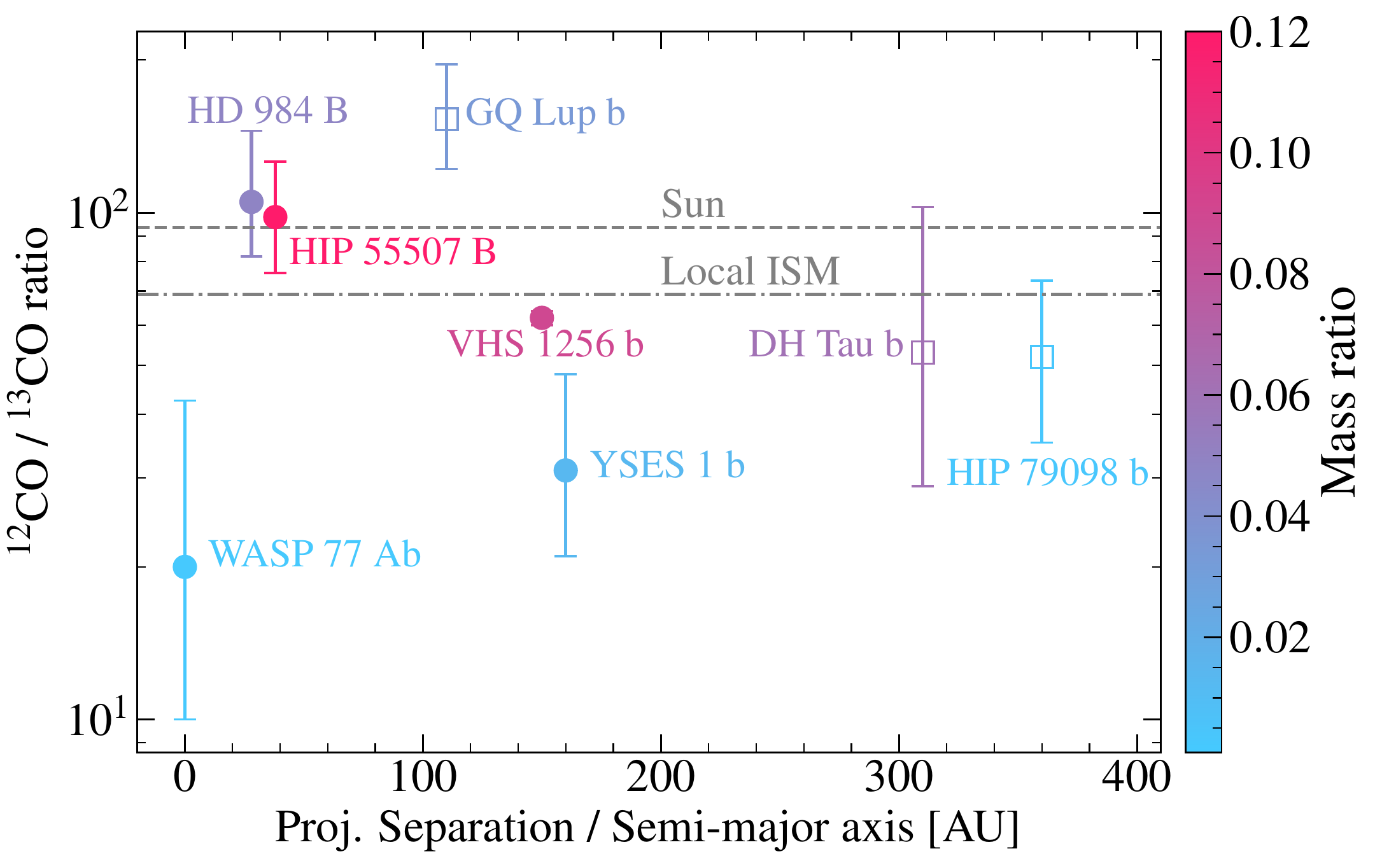}
    \caption{\cratio~measured from CO as a function of projected separation or semi-major axis. The points are color-coded by companion-star mass ratio, and the three open squares are from this paper. The dashed and dashdot lines denote the solar value \citep{Lyons2018} and ISM value \citep{Wilson1999}, respectively. The other points are from \citet{line_solar_2021a} for WASP-77 Ab, \citet{zhang_13COrich_2021} for YSES 1 b, \citet{Xuan2024} for HIP~55507~B, \citet{Costes2024} for HD 984 B, and \citet{Gandhi2023} for VHS 1256 b.}
    \label{fig:13co_summary}
\end{figure}

\subsection{The \cratio~Isotopic Ratio} \label{sec:discuss_13co}
Minor isotopologues of carbon, oxygen, and nitrogen have recently been detected in exoplanets and brown dwarfs \citep{zhang_13COrich_2021, zhang_12CO_2021, line_solar_2021a, Barrado2023, Gandhi2023}, allowing constraints on isotopic ratios such as $\rm ^{12}C/^{13}C$, $\rm ^{12}O/^{18}O$ and $\rm ^{14}N/^{15}N$. \citet{Xuan2024} also demonstrated carbon and oxygen isotopic homogeneity between a late-M dwarf companion and its late-K primary star with Keck/KPIC spectroscopy. In this paper, we find bounded constraints on \co~for three companions, GQ~Lup~b, HIP~79098~b, and DH~Tau~b, though the $^{13}$CO detection for DH Tau b is tentative. 

$\cratio$~is proposed to be a formation diagnostic as CO ice outside the CO snowline is expected to be enriched in $^{13}$C due to isotopic selective fractionation processes such as CO self-shielding \citep{zhang_13COrich_2021}. Planets forming via core/pebble accretion outside the CO snowline may therefore show a lower $\cratio$~due to this effect. Our three companions show a range of $\cratio$~values between $\approx$50--150. The values for DH~Tau~b ($53^{+50}_{-24}$) and HIP~79098~b (${52^{+22}_{-17}}$) are broadly consistent with the local ISM value of $\cratio=69\pm6$ \citep{Wilson1999}, while GQ~Lup~b shows a higher \co~of $153^{+43}_{-31}$. In Appendix~\ref{app:spline}, we compare retrieval results from several different nights of KPIC data for GQ~Lup~b, and find $\approx$0.1 dex differences in the retrieved \logco~ between different nights, which is consistent with the findings of \citet{Xuan2024}. When including the night-to-night scatter of $0.10$ dex as a systematic error, GQ~Lup~b's measurement is consistent with the solar value of $\cratio=93.5\pm3.1$ \citep{Lyons2018} at the $1\sigma$ level. Overall, these findings do not provide evidence that our companions accreted a significant amount of ice with low \cratio, consistent with their likely star-like formation pathways. 

In Fig.~\ref{fig:13co_summary}, we plot our new measurements along with previous constraints on the \cratio~for exoplanets and low-mass companions. There is considerable scatter in the measurements, and no clear trend with orbital distance or mass ratio. Some of these results should be re-visited with better quality or higher-resolution data to reduce the error bars. For directly imaged companions, the exquisite S/N from JWST is poised to provide more precise measurements of isotopologue ratios, as demonstrated by \citet{Gandhi2023}. Another direction is to obtain higher S/N data to detect additional isotopologues such as $\rm C^{18}O$ and $\rm H_2^{18}O$ \citep{Zhang2022, Xuan2024} Finally, to better interpret isotopic ratio data, more work is required on the modeling side to understand the details of isotopic variability and fractionation chemistry in disks \citep{Oberg2023}.

\subsection{System and Orbital Architectures} \label{sec:architecture}
The formation of substellar companions should also be considered in the context of the host stellar system. As noted in \S~\ref{sec:system_prop}, five out of eight of our systems are either confirmed or likely multiple-star systems. ROXs 42B is a resolved binary with projected separation $\sim$12 au \citep{kraus_three_2013} while HIP~79098 is a suspected binary \citep{Janson2019}, making the b components around them `circumbinary.' One way to explain these configurations is via disk fragmentation, which generally produces multiple fragments that interact gravitationally \citep[e.g.][]{Kratter2016}. For example, the protostellar disk around ROXs 42B could first fragment and form the secondary star, which continues to grow given high infall rates from the surrounding nebula. At a later stage when the disk mass is lower, but still gravitationally unstable, a second fragmentation could produce the $\sim$13~\Mj companion. Dynamical interactions may cause the substellar companion to move away from the star while the secondary star moves inward. 

On the other hand, DH Tau, ROXs 12, and GQ Lup have wide stellar companions at thousands of au \citep{Kraus_Hillenbrand_2009, bowler_young_2017, Alcala2020}. \citet{Alcala2020} hypothesized that GQ Lup A and its wide ($\approx2400$ au) stellar companion may have formed via turbulent fragmentation of a molecular cloud core, while GQ~Lup~b fragmented out of the circumprimary disk around GQ Lup A. Similar formation pathways may be responsible for the DH Tau and ROXs 12 systems as well. While disk fragmentation simulations do produce low-mass substellar companions \citep[e.g.][]{stamatellos_properties_2009}, it is challenging to prevent these companions from further accretion to become low-mass stars or be tidally destroyed during a rapid inward migration phase powered by infall \citep{Zhu2012, Forgan2013}. Indeed, to produce the 10--30~\Mj companions we observe today, fragmentation must occur in a narrow window after the period of high infall rates, but before the disk mass becomes too small for Toomre instability \citep{kratter_runts_2010}. Therefore, if disk fragmentation is a common formation mechanism, we should observe more low-mass stars around the same stellar types compared to substellar or planetary companions. Published observations are in qualitative agreement with this prediction. For example, \citet{Duchene2023} find a significantly higher number of low-mass stellar companions around intermediate-mass stars compared to substellar companions.

Orbital architectures provide another piece of the puzzle. One informative probe is the relative alignment between different angular momentum vectors in the system. Recently, \citet{Bowler2023} presented a summary of stellar obliquity constraints for directly imaged substellar companions, finding that misalignment between the companion orbits and stellar spin axes are common. Specifically, GQ~Lup~b may be on a nearly perpendicular orbit with respect to the circumstellar disk around GQ Lup A, and is also misaligned with the star's spin axis \citep{Stolker2021}. Similarly, the orbit of ROXs~12~b is also likely misaligned compared to the stellar spin axis \citep{bowler_young_2017, Bowler2023}. \citet{Bryan_obliquity_2020} provided the first measurement of companion obliquity for 2M0122~b, finding that the companion's spin axis is tentatively misaligned with respect to the stellar spin.\footnote{We note that our higher KPIC \vsini measurement of $19.6^{+3.0}_{-2.5}~$\kms for 2M0122~b (see Table~\ref{table:compare_res}) would increase the line-of-sight inclination of the companion's spin axis, making it more aligned with the stellar spin axis. Due to the low S/N of our 2M0122~b spectra however, we emphasize that future data will be required to confirm the exact \vsini of this companion.} Since stellar binaries that form via turbulent fragmentation tend to have misaligned spin and orbital orientations \citep[e.g.][]{Jensen2014, Lee2016, Offner2016}, random spin-orbit distributions should also be prevalent for systems with substellar companions if they form in this manner, which is consistent with current findings. Alternatively, wide, misaligned stellar companions in these systems could torque the disk around the primary, causing the companions that fragment in the disk to inherit such misalignments \citep[e.g.][]{bowler_young_2017}. Finally, disk turbulence itself could result in random spin-orbit misalignments  \citep{Jennings2021}. Regardless of the detailed mechanism, the system and orbital architectures of our systems are broadly consistent with star-like formation. This is consistent with the trend of chemical homogeneity we infer between these companions and their stars. 

\section{Conclusion}\label{sec:conclude}
We have carried out a uniform atmospheric retrieval study of eight widely-separated substellar companions ($\sim$50--360 au, $\sim$10--30~\Mj) with Keck/KPIC high-resolution $K$ band spectroscopy. From these retrievals, we measure the companion's C/O, metallicity (denoted [C/H]), isotopic abundances, in addition to their cloud properties, spins, radial velocities, and temperature profiles. To complement the continuum-removed high-resolution data, we adopt mass and radius priors from evolutionary models in the retrievals.

First, we find that these companions have broadly solar composition (to within $2\sigma$ level), and likely stellar composition given the trend of solar abundances seen for stars in the same star-forming associations. Their abundance pattern is similar to systems with high-mass brown dwarf companions and stellar binaries, which show chemically homogeneity \citep[e.g.][]{hawkins_Identical_2020, wang_Retrieving_2022, Xuan2022}. On the other hand, their abundances are distinct from hot Jupiters, which show a range in C/O and generally super-stellar metallicities (see Fig.~\ref{fig:abund_summary}), and directly imaged giant planets with $m\sim3-10~\Mj$, which show tentative metal enrichment \citep{Wang2023b_arxiv}. Thus, the population of low-mass substellar companions from direct imaging likely traces the tail-end of star formation processes such as gravitational disk instability and cloud fragmentation, making them low-mass brown dwarfs instead of `super-Jupiter' planets or `planetary-mass companions.' Alternatively, we note that our composition measurements are also consistent with core accretion outside the CO snowline where these companions are observed today, since accretion at these locations would also yield stellar C/O and metallicities. However, such a scenario requires core accretion to proceed early and rapidly in a massive, protostellar disk in order to explain the accretion of 10--30~\Mj of material, especially for the systems with lower mass host stars.

Second, we find evidence of clouds in two of the colder companions (Fig.~\ref{fig:cloud_kap}), kap~And~b ($\Teff=1640^{+220}_{-170}~K$) and GSC 6214-210~b ($\Teff=1860^{+170}_{-110}~K$), with the EddySed cloud model being preferred by $\gtrsim3\sigma$ compared to both the clear and gray cloud models. This indicates that narrow-band, high-resolution spectra can be sensitive to non-gray, scattering clouds when the cloud opacity is high near the photosphere, as is the case for these companions. This result highlights the potential of high-resolution spectroscopy in constraining both abundances and clouds for brown dwarfs and exoplanets. 

Third, we present three new measurements of \co~for GQ~Lup~b ($153^{+43}_{-31}$), HIP~79098~b ($52^{+22}_{-17}$), and DH~Tau~b ($53^{+50}_{-24}$). A cross-correlation analysis shows solid detections for the first two companions, and a tentative detection of $^{13}$CO for DH~Tau~b. From retrievals of six independent KPIC datasets of GQ~Lup~b data (Appendix~\ref{app:spline}), we find $\approx0.1$ dex systematic errors in \co, which is consistent with the findings of \citet{Xuan2024}. After accounting for systematics, our measurements agree with either the ISM or solar \cratio~to within $1\sigma$. We place these measurements in context of previous work and do not identify any clear trends between \cratio~and mass, mass ratio, or orbital distance at this stage (Fig.~\ref{fig:13co_summary}). More precise and accurate measurements of \cratio~and complementary measurements of the isotopic ratio in the host stars are necessary to further interpret these results. 

Finally, we present radial velocity and spin measurements for the companions. We find some discrepancies between our $\vsini$ values and previous studies using pre-upgrade Keck/NIRSPEC data \citep{bryan_Worlds_2020}, while our $\vsini$ for GQ~Lup~b agrees with previous VLT/CRIRES measurements at a higher resolution of $R\sim100,000$ \citep{Schwarz2016}. Due to the higher spectral resolution ($R\sim35,000$), higher S/N, and more stable line spread function of KPIC (which uses post-upgrade NIRSPEC; \citealt{martin_overview_2018}) compared to pre-upgrade NIRSPEC ($R\sim25,000$), we adopt our new spin measurements in this paper. Seven out of eight of our companions have relatively slow $\vsini\approx4-20~$\kms, with kap~And~b being an outlier with $\vsini=39.4\pm0.9~$\kms (see also Morris et al. 2024, accepted). Our objects follow the overall trend of literature spin measurements, and display a large scatter in their fractional rotational velocities, $v/v_{\rm breakup}$, as shown in Fig.~\ref{fig:spins}. 

Looking forward, it would be useful to obtain higher S/N spectra for some of the companions in order to obtain higher precision measurements. The results for 2M0122 b in particular should be re-visited, as this is our lowest S/N dataset and we also retrieve a much lower C/O for this companion than the rest of the sample. Higher S/N data would also help increase the quantity and quality of \co~measurements. For example, \citet{Gandhi2023} showed that JWST/NIRSpec enables isotopologue measurements with unprecedented precision  for widely-separated substellar companions. In addition, detailed forward modeling of the residual starlight could also open the door to studying close-in, high-contrast companions with the JWST/NIRSpec integral field unit \citep{Ruffio2023}. The wide wavelength coverage of JWST would be useful for constraining cloud properties and measuring more elemental abundances besides C and O in order to break degeneracies in formation inferences as well \citep[e.g.][]{Cridland2020, Turrini2021, Ohno2023, Chachan2023}.

Many of the companions presented in this work have relatively large separations and low star-to-companion contrasts, making them accessible with traditional NIRSPEC in AO mode \citep{bryan_constraints_2018}. A natural next step would be to extend our measurements to companions with lower masses and smaller orbital distances. Keck/KPIC and similar instruments like VLT/HiRISE \citep{Vigan2024} are ideal for these targets, as shown by \citet{wang_Detection_2021}. By targeting more directly imaged planets with high-resolution spectroscopy, we could further test the tentative trend of metal enrichment suggested by \citet{Wang2023b_arxiv} for these planets. Such measurements are especially important given literature discrepancies in the retrieved abundances for the same planet, either due to using data with different spectral resolution \citep[e.g. $\beta$ Pic b;][]{nowak_Peering_2020, Landman2023}, or using different retrieval methods on similar low-resolution data \citep[e.g. 51 Eri b;][]{BrownSevilla2023, Whiteford2023}. As demonstrated by this paper and previous work, atmospheric abundance measurements from medium-to-high resolution spectroscopy are providing results that are reliable to uncertain assumptions about clouds \citep{zhang_12CO_2021, Xuan2022, Inglis2024}, and consistent across different observing nights \citep{ruffio_Deep_2021, Xuan2024, Landman2023a}. Therefore, medium-to-high resolution spectroscopy is poised to improve our understanding of the nature and formation of directly imaged companions in the near future, bringing more clarity in the delineation between giant planets and their brown dwarf counterparts.

\acknowledgments
J.X. thanks Melanie Rowland, Yayaati Chachan, Michael Liu, Kazumasa Ohno, Douglas Lin, Ricardo López-Valdivia, Henrique Reggiani, and Jingwen Zhang for helpful discussions. 
J.X. is supported by the NASA Future Investigators in NASA Earth and Space Science and Technology (FINESST) award \#80NSSC23K1434. J.X. also acknowledges support from the Keck Visiting Scholars Program (KVSP) to commission KPIC Phase II capabilities. D.E. is supported by the NASA FINESST award \#80NSSC19K1423. D.E. also acknowledges support from the Keck Visiting Scholars Program (KVSP) to install the Phase II upgrades.
Funding for KPIC has been provided by the California Institute of Technology, the Jet Propulsion Laboratory, the Heising-Simons Foundation (grants \#2015-129, \#2017-318, \#2019-1312, \#2023-4598), the Simons Foundation, and the NSF under grant AST-1611623.
The computations presented here were conducted in the Resnick High Performance Center, a facility supported by Resnick Sustainability Institute at the California Institute of Technology.
W. M. Keck Observatory access was supported by Northwestern University and the Center for Interdisciplinary Exploration and Research in Astrophysics (CIERA).
The data presented herein were obtained at the W. M. Keck Observatory, which is operated as a scientific partnership among the California Institute of Technology, the University of California and the National Aeronautics and Space Administration. The Observatory was made possible by the generous financial support of the W. M. Keck Foundation. The authors wish to recognize and acknowledge the very significant cultural role and reverence that the summit of Mauna Kea has always had within the indigenous Hawaiian community.  We are most fortunate to have the opportunity to conduct observations from this mountain. 
This work benefited from the 2023 Exoplanet Summer Program in the Other Worlds Laboratory (OWL) at the University of California, Santa Cruz, a program funded by the Heising-Simons Foundation and NASA. 
The research here acknowledges use of the Hypatia Catalog Database, an online compilation of stellar abundance data as described in \citet{Hinkel2014}, which was supported by NASA's Nexus for Exoplanet System Science (NExSS) research coordination network and the Vanderbilt Initiative in Data-Intensive Astrophysics (VIDA).

\facilities{Keck (KPIC)}
\software{\texttt{petitRADTRANS}~\citep{molliere_petitRADTRANS_2019}, \texttt{dynesty}~\citep{speagle_DYNESTY_2020}}

\clearpage

\appendix
\onecolumngrid

\FloatBarrier
\section{Interpolated properties from evolutionary models}\label{app:evol_model}
\restartappendixnumbering

Here, we show the evolutionary-model predicted mass, radius, \logg, and \Teff distributions for the seven other companions. Different colors indicate different evolutionary models, with ATMO 2020 in blue \citep{Phillips2020, Chabrier2023}, AMES-Dusty in red \citep{Allard2001}, BHAC15 in purple \citep{Baraffe2015}, and SM08 in gray \citep{Saumon_2008}. The distributions for GQ Lup b are shown in Fig.~\ref{fig:mr_prior}.

\begin{figure*}
    \centering
  \includegraphics[width=0.9\linewidth]{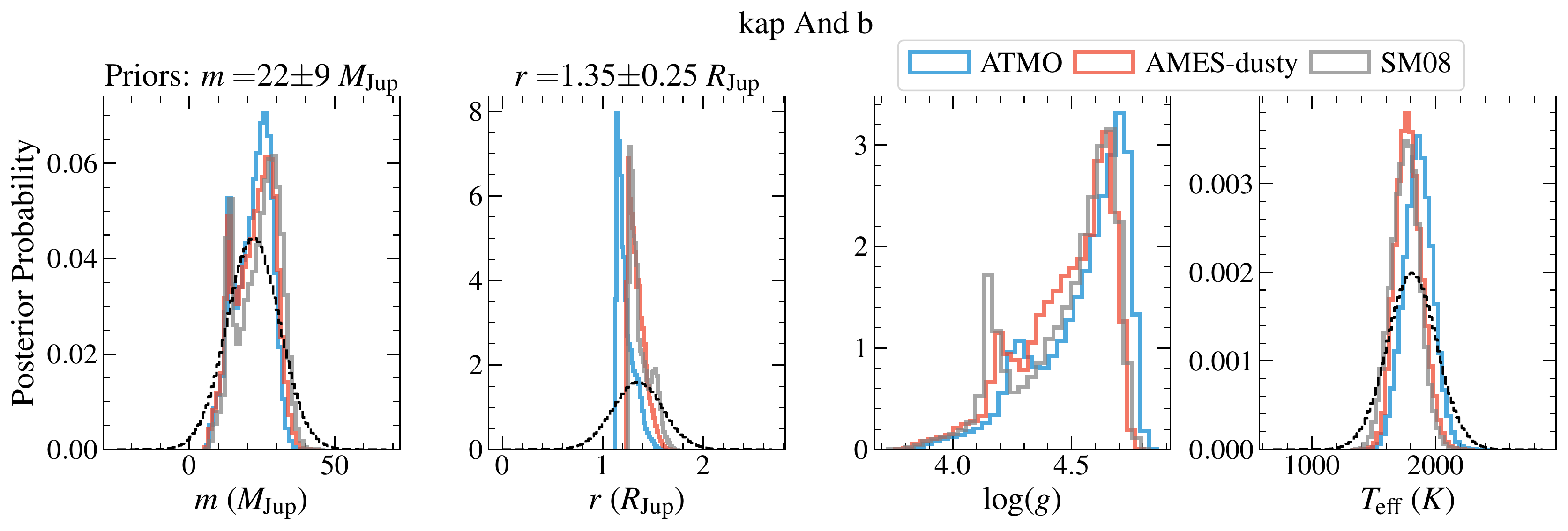}
  \centering
  \includegraphics[width=0.9\linewidth]{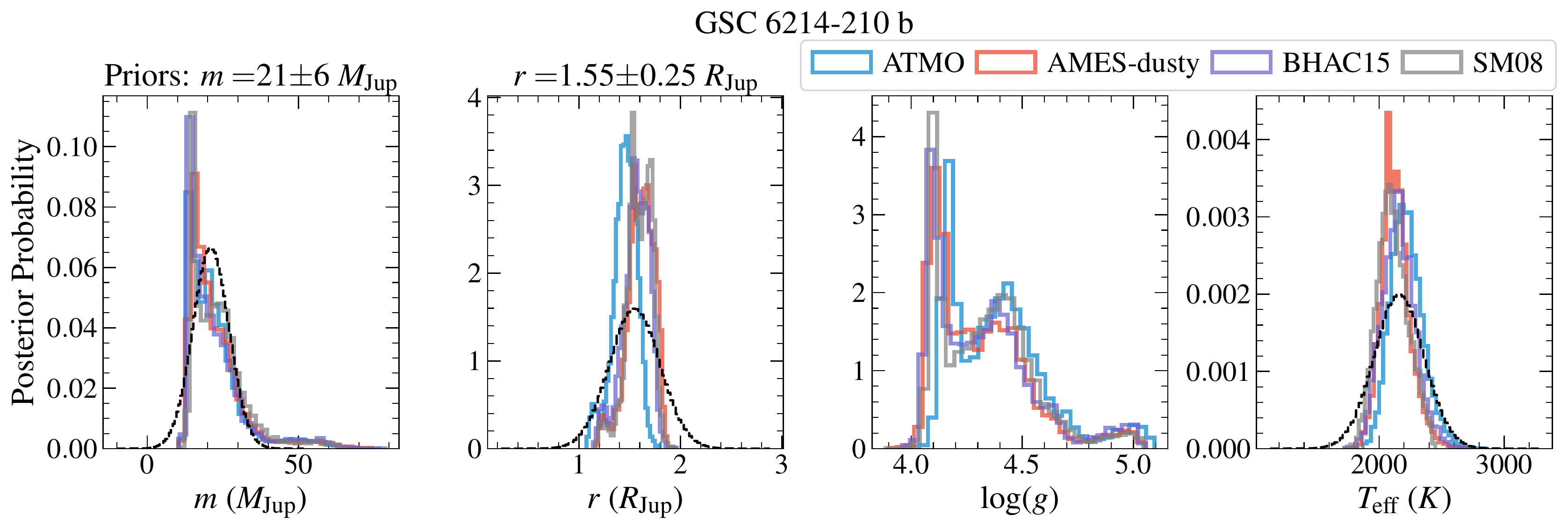}
  \centering
  \includegraphics[width=0.9\linewidth]{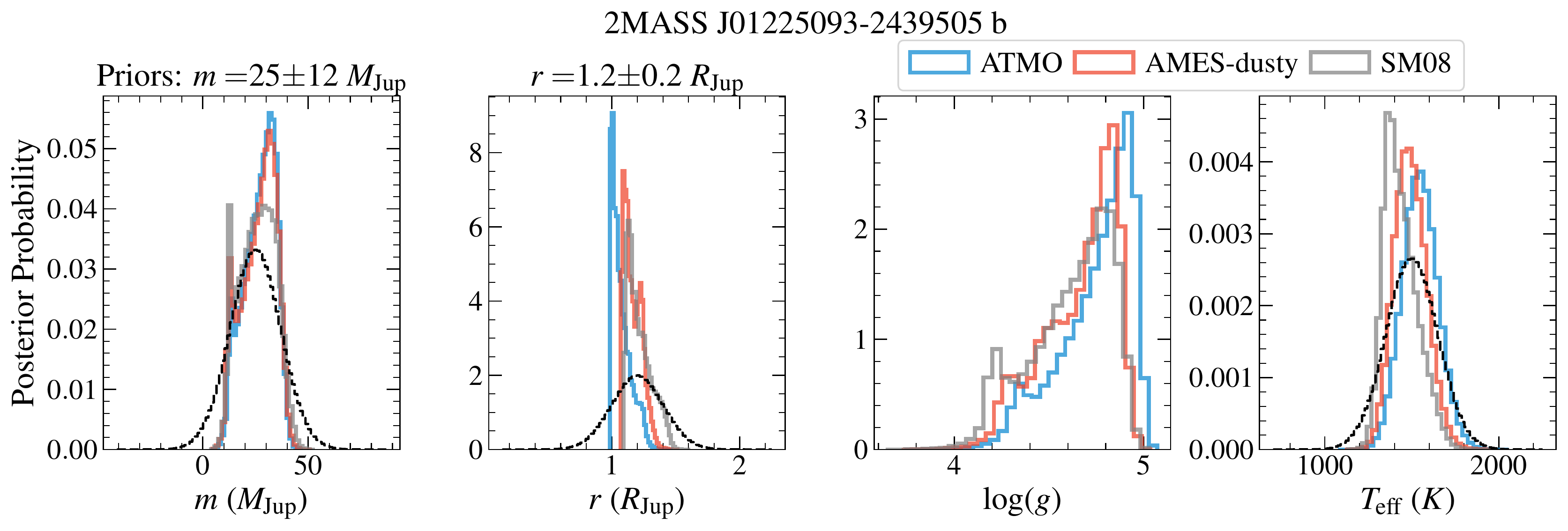}
\end{figure*}

\begin{figure*}
  \centering
  \includegraphics[width=0.9\linewidth]{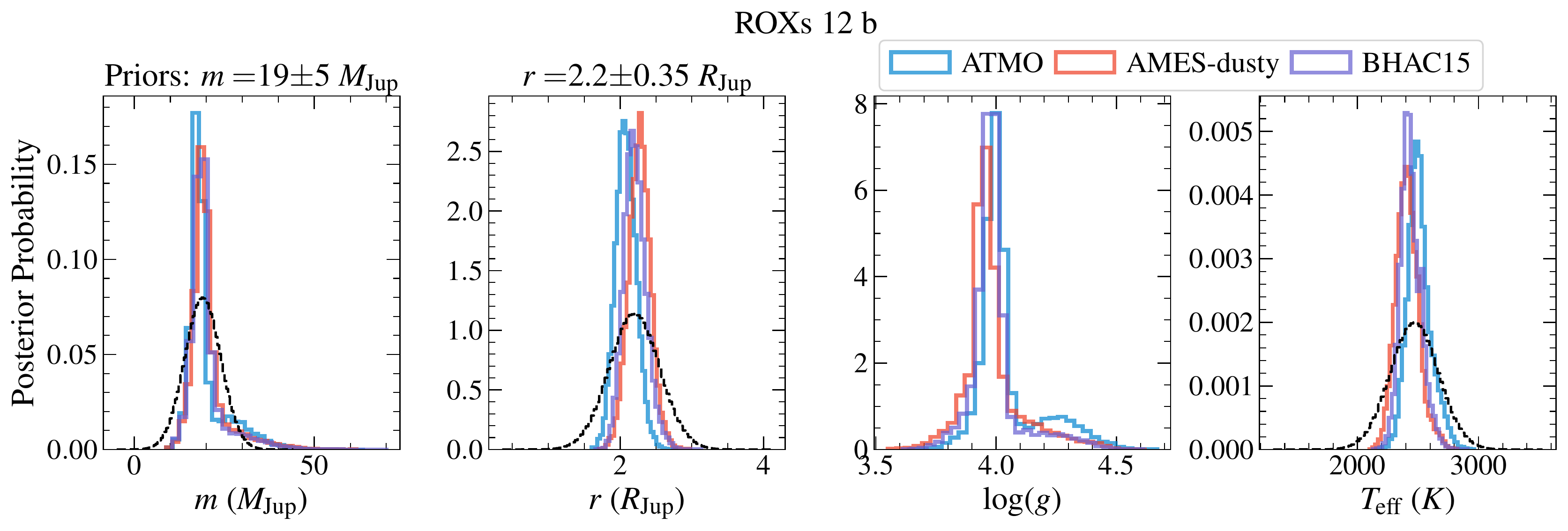}
  \centering
  \includegraphics[width=0.9\linewidth]{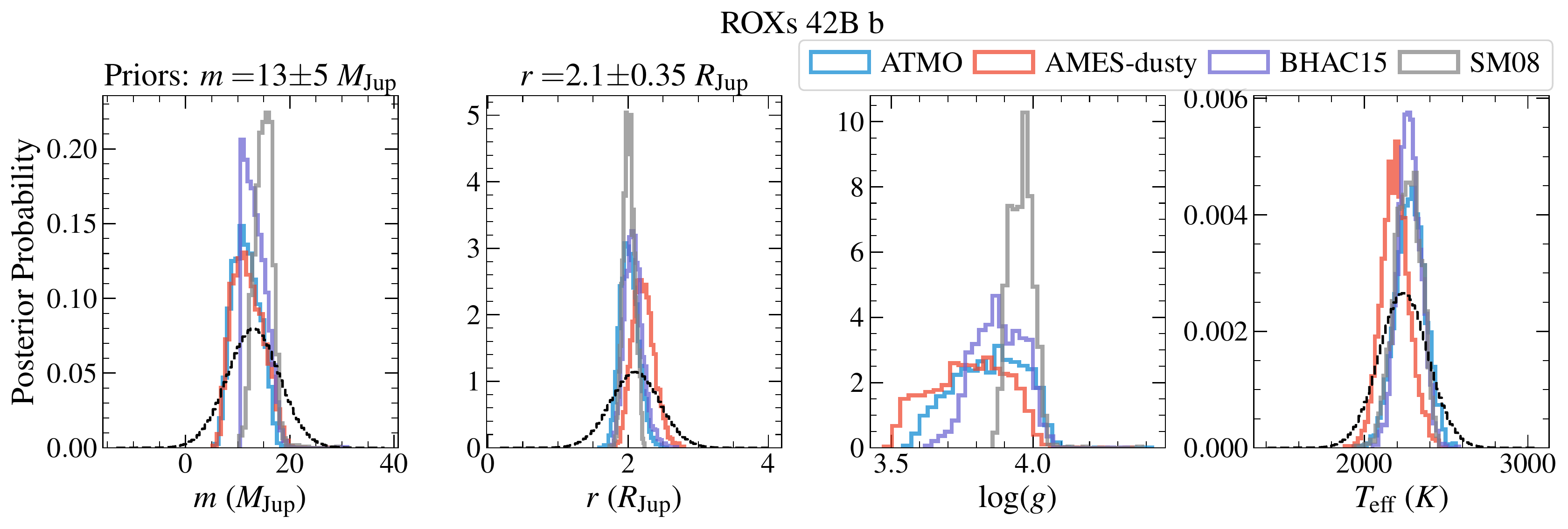}
  \centering
  \includegraphics[width=0.9\linewidth]{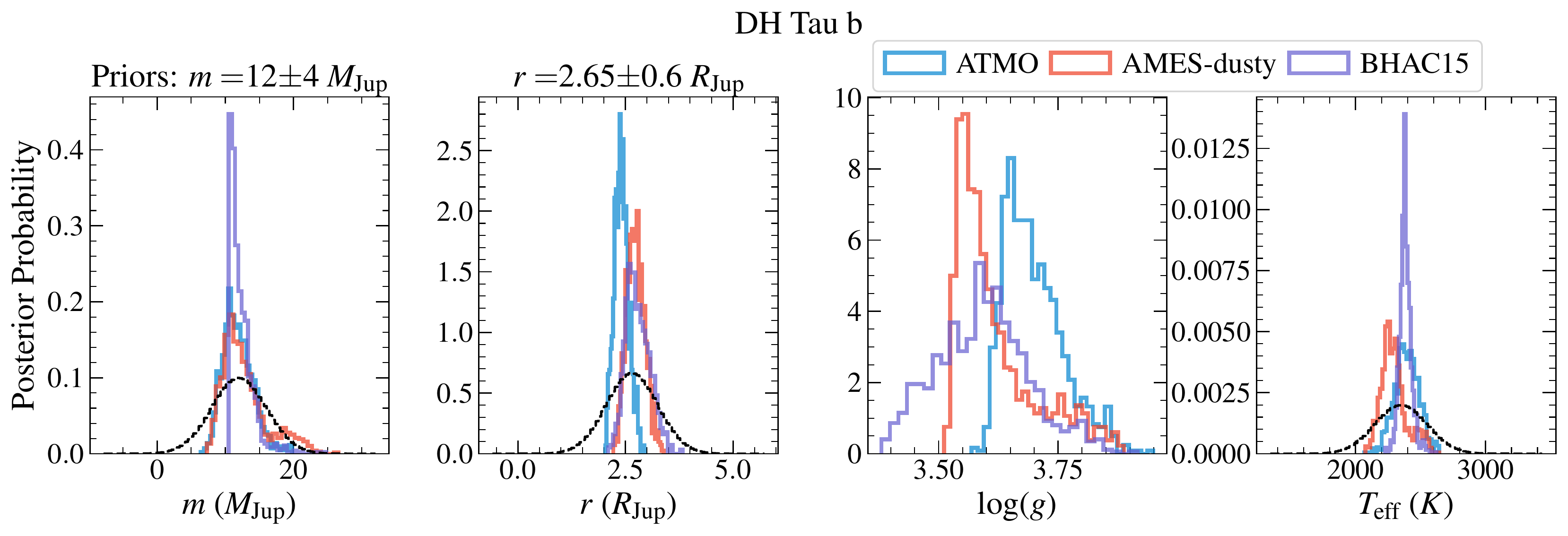}
  \centering
  \includegraphics[width=0.9\linewidth]{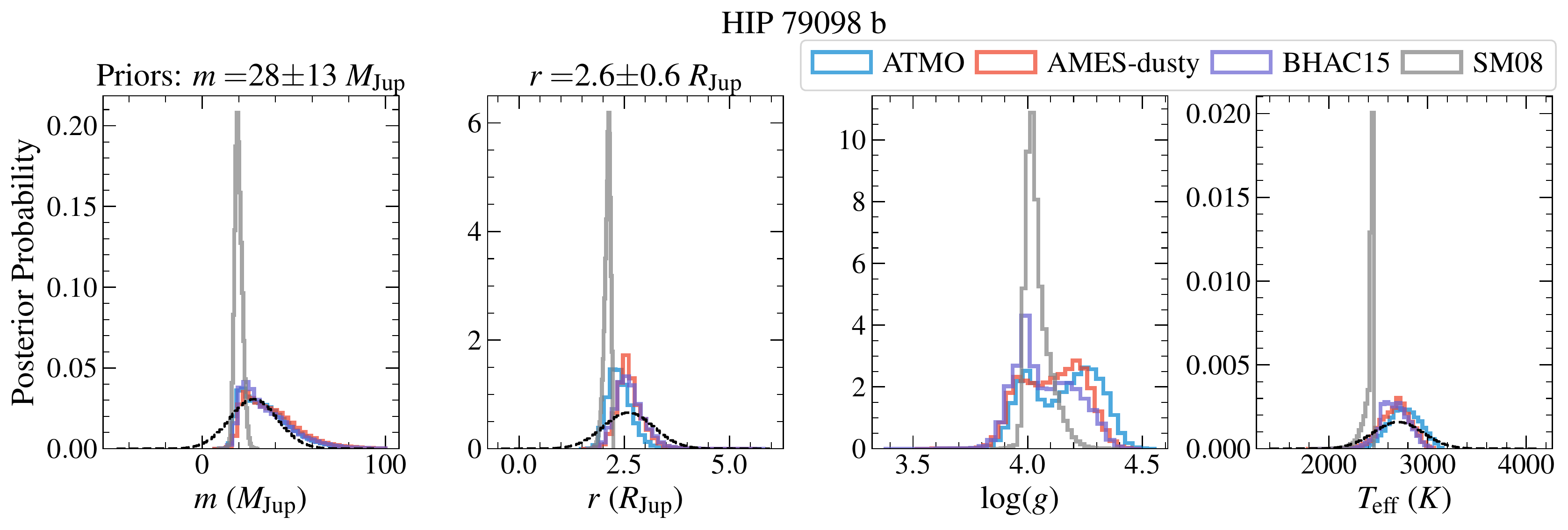}
  \caption{Same as Fig.~\ref{fig:mr_prior}, but for the other seven companions.}
\label{fig:evol_all}
\end{figure*}

\FloatBarrier
\section{Comparing continuum removal methods with multiple datasets of GQ~Lup~b}\label{app:spline}
\restartappendixnumbering
We compare the retrieved abundances and \vsini~between different observing epochs for GQ~Lup~b, our highest S/N target. In addition, we compare two methods of removing the continuum: the high-pass filter (HPF) method used in the paper and a spline model. As described in \S~\ref{sec:forward_m}, a median-filter with a size of 100 pixels ($\approx0.002~\mu$m) is adopted as the HPF. The choice of the filter size is motivated by injection-recovery tests in \citet{Xuan2022}, while the median-filter is found to perform better compared to Fourier-based filters or Gaussian filters \citep{wang_Detection_2021}. 

For the alternative, spline model, we follow the method in \citet{Ruffio2023} using the open-source package \texttt{breads} \citep{Agrawal2023}.\footnote{\url{https://github.com/jruffio/breads}} Since the spline model allows us to modulate the companion continuum and the speckle continuum separately, we experimented with a few different choices. Our default spline model modulates the continuum of both speckle and companion with a third-order spline model. Specifically, we use three spline nodes per spectral order. We note that such a spline operates over a much larger scale ($\approx$ 1000 pixels) compared to our 100-pixel median filter, so is not an exact comparison to the HPF method. When we instead used 10 nodes per spectral order for the speckle continuum, the results were consistent with the 3-node model but the continuum oscillations appear more stochastic and less smooth, leading us to disfavor these higher number of nodes. Finally, we also experimented with a simpler model of not modulating the companion continuum and only modulating the speckle continuum. We find that this did not change the results significantly either. 

To compare the two continuum removal methods, we perform retrievals on six independent datasets for GQ~Lup~b across three different nights (two fibers were used each night), UT 2023 June 23, June 24, and June 29. The same retrieval setup and priors are used. For the HPF, we jointly fit for flux scales for the companion and the speckle contribution, $\alpha_b$ and $\alpha_s$ (see Eq.~\ref{eqn:comp}). Similarly, the spline model performs a linear optimization to determine the spline parameters for each proposed model. 

\begin{figure*}
     \centering
  \includegraphics[width=0.4\linewidth]
  {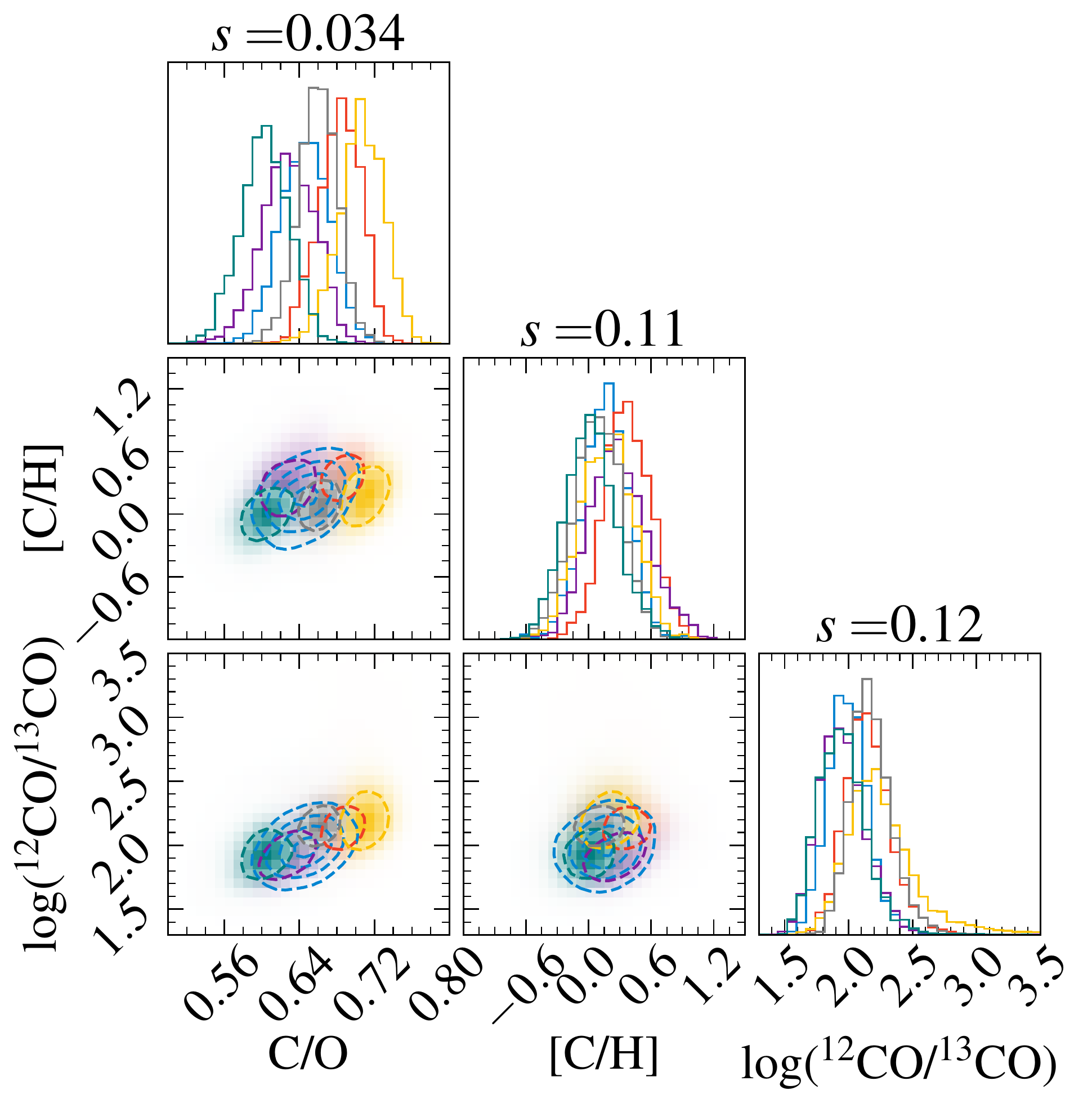}
  \hspace{12mm}
  \centering
  \includegraphics[width=0.4\linewidth]
  {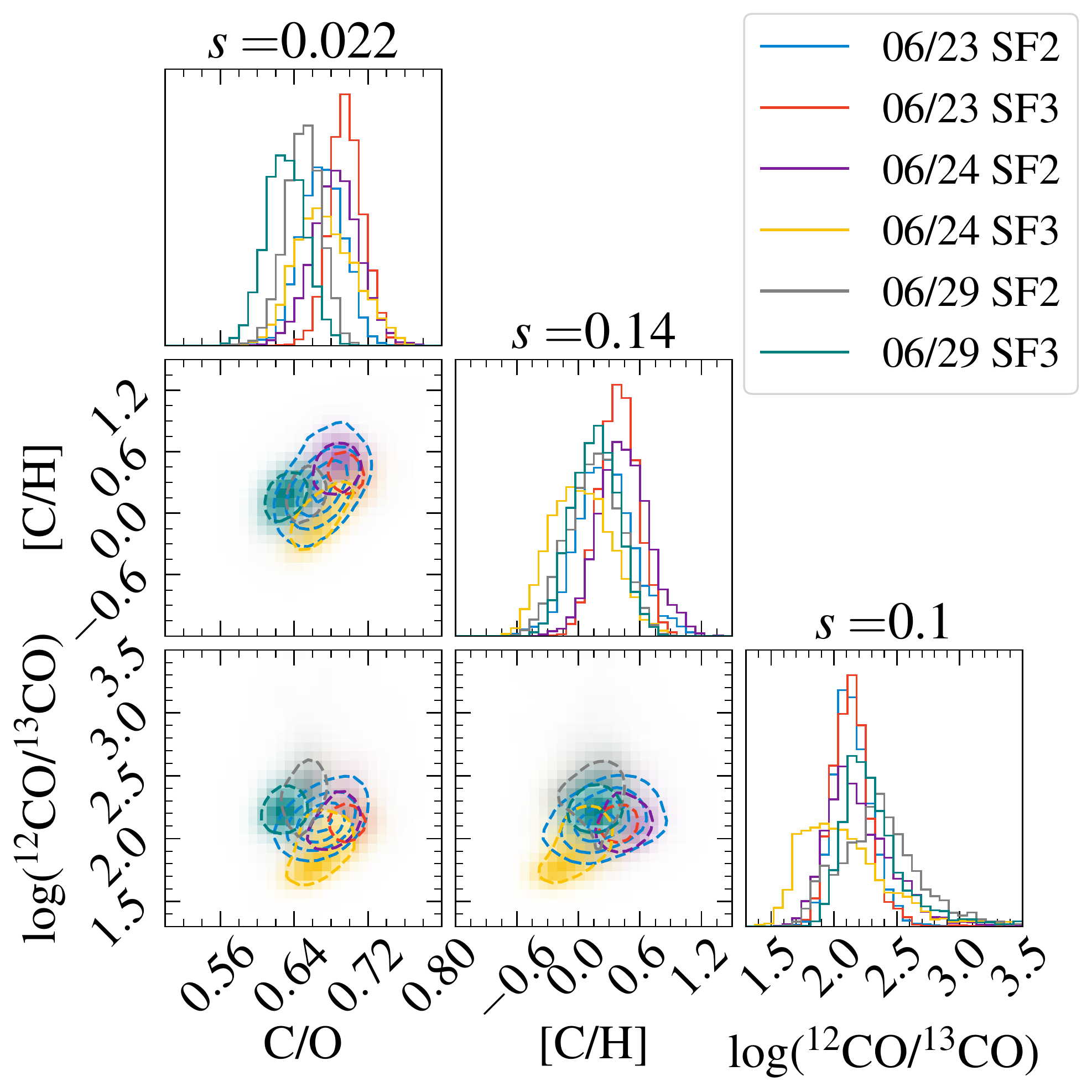}
    \caption{Left: posterior distributions of three key parameters from using the spline model. Six different datasets are compared (3 nights $\times$ 2 fibers per night). Right: the same but using the HPF (median-filter of 100 pixels). The titles on each histogram show the standard deviation ($s$) of the median between the six different posteriors. The legend is shared between both plots.}
\label{fig:hpf_spline}
\end{figure*}

In Fig.~\ref{fig:hpf_spline}, we compare the retrieved C/O, [C/H], and \logco~between the different datasets for each continuum removal method. First, we note that the retrieved values between the two methods agree well; taking the median across the results from six datasets, we obtain $\rm C/O=0.65$, $\rm [C/H]=0.17$, $\logco=2.07$ from the HPF method, and $\rm C/O=0.67$, $\rm [C/H]=0.21$, $\logco=2.15$ from the spline method.
Next, we quantify the relative agreement between different datasets with the standard deviation ($s$) of the median retrieved values. We find that the HPF yields a slightly lower spread in C/O by $\approx35\%$ compared to the spline method. The spread in [C/H] and \logco~is comparable between the two methods. Therefore, we conclude that the performances of the two continuum removal methods are comparable for our data. We adopt the HPF as the continuum removal method in this paper to be consistent with previous KPIC retrieval papers \citep[e.g.][]{wang_Detection_2021, Xuan2022, Xuan2024}, but note that our results will not be affected if we chose the spline model instead.

By comparing results from different datasets, we can also assess the level of systematic error in the retrieved parameters. Here, we focus on the HPF results. Approximating the standard deviation as the systematic error, we find night-to-night systematics of $0.022$, $0.14$ dex, and $0.10$ dex for C/O, [C/H], and \logco. These systematic errors are comparable in size to those found in \citet{Xuan2024} by comparing two independent KPIC datasets for a M7.5 companion.

\onecolumngrid
\FloatBarrier
\section{Priors for Retrievals}\label{app:priors}
\restartappendixnumbering
Here, we list the fitted parameters and adopted priors for retrievals of the seven other companions. The corresponding information for \kapandb~is shown in Table~\ref{tab:param_prior}.

\begin{deluxetable*}{llll}
\tablecaption{Fitted Parameters and Priors for Retrievals}
\tabletypesize{\footnotesize}
\tablehead{Parameter & Prior & Parameter & Prior}
\startdata
GQ Lup b\\
\hline 
Mass ($\Mj$) & $\mathcal{N}(33.0, 10.0)$ & Radius ($\Rj$) & $\mathcal{N}(3.7, 0.7)$  \\
$T_{\rm anchor}$ [log($P$)=-1.1)] (K) & $\mathcal{U}(2000, 3000)$ & RV (\kms) & $\mathcal{U}(-50 , 50)$ \\
$\Delta T_1$ [0.5 to -0.5] (K) & $\mathcal{U}(400, 1100)$ & $v\sin{i}$ (\kms) & $\mathcal{U}(0, 80)$ \\
$\Delta T_2$ [-0.5 to -0.8] (K) & $\mathcal{U}(0, 400)$ & $\rm C/O$ & $\mathcal{U}(0.1,1.0)$  \\
$\Delta T_3$ [-0.8 to -1.1] (K) (K) & $\mathcal{U}(0, 400)$ & $\rm [C/H]$ & 
$\mathcal{U}(-1.5,1.5)$ \\
$\Delta T_4$ [-1.1 to -1.5] (K) & $\mathcal{U}(100, 850)$ & \logco & $\mathcal{U}(0, 6)$ \\
$\Delta T_5$ [-1.5 to -2.0] (K) & $\mathcal{U}(50, 550)$ & $f_{\rm sed}$ (one for each cloud) & $\mathcal{U}(0, 10)$ \\
$\Delta T_6$ [-2.0 to -3.5] (K) & $\mathcal{U}(200, 750)$ & ${\rm log}(K_{\rm zz}/\rm{cm^2~s^{-1}})$ & $\mathcal{U}(5, 13)$ \\ 
$\Delta T_7$ [-3.5 to -5.0] (K) & $\mathcal{U}(50, 600)$  & $\sigma_{\rm g}$ & $\mathcal{U}(1.05, 3)$ \\ 
log(gray opacity/$\rm{cm}^2~g^{-1}$) & $\mathcal{U}(-6, 6)$  &  ${\rm log}(\tilde{X}_{\rm Al_2O_3})$ & $\mathcal{U}(-2.3, 1)$ \\
Error multiple & $\mathcal{U}(1, 5)$ & ${\rm log}(\tilde{X}_{\rm Fe})$ & $\mathcal{U}(-2.3, 1)$ \\
Comp. flux, $\alpha_c$ (counts) & $\mathcal{U}(0, 100)$ & Speckle flux, $\alpha_s$ (counts) & $\mathcal{U}(0, 100)$  \\
\hline 
HIP 79098 b\\
\hline 
$\Delta T_1$ [0.7 to 0.0] (K) & $\mathcal{U}(300, 800)$ & $T_{\rm anchor}$ [log($P$)=-0.7)]  (K) & $\mathcal{U}(1900, 2800)$  \\
$\Delta T_2$ [0.0 to -0.3] (K) & $\mathcal{U}(0, 400)$ &  Mass ($\Mj$) & $\mathcal{N}(28.0, 13.0)$  \\
$\Delta T_3$ [-0.3 to -0.7] (K) & $\mathcal{U}(50, 500)$ & Radius ($\Rj$) & $\mathcal{N}(2.6, 0.6)$   \\
$\Delta T_4$ [-0.7 to -1.0] (K) & $\mathcal{U}(100, 500)$ & Comp. flux, $\alpha_c$ (counts) & $\mathcal{U}(0, 100)$   \\
$\Delta T_5$ [-1.0 to -1.5] (K) & $\mathcal{U}(200, 650)$ & ${\rm log}(\tilde{X}_{\rm Al_2O_3})$ & $\mathcal{U}(-2.3, 1)$\\
$\Delta T_6$ [-1.5 to -3.0] (K) & $\mathcal{U}(50, 650)$ & ${\rm log}(\tilde{X}_{\rm Fe})$ & $\mathcal{U}(-2.3, 1)$\\ 
$\Delta T_7$ [-3.0 to -4.5] (K) & $\mathcal{U}(100, 450)$  & \logco & $\mathcal{U}(0, 6)$ \\
\hline 
DH Tau b\\
\hline 
$\Delta T_1$ [0.7 to 0.0] (K) & $\mathcal{U}(200, 900)$ & $T_{\rm anchor}$ [log($P$)=-0.7)] (K) & $\mathcal{U}(1800, 3000)$  \\
$\Delta T_2$ [0.0 to -0.3] (K) & $\mathcal{U}(0, 450)$ &  Mass ($\Mj$) & $\mathcal{N}(12.0, 4.0)$  \\
$\Delta T_3$ [-0.3 to -0.7] (K) & $\mathcal{U}(50, 550)$ & Radius ($\Rj$) & $\mathcal{N}(2.6, 0.6)$   \\
$\Delta T_4$ [-0.7 to -1.0] (K) & $\mathcal{U}(0, 450)$ & Comp. flux, $\alpha_c$ (counts) & $\mathcal{U}(0, 100)$   \\
$\Delta T_5$ [-1.0 to -1.5] (K) & $\mathcal{U}(50, 650)$ & ${\rm log}(\tilde{X}_{\rm Al_2O_3})$ & $\mathcal{U}(-2.3, 1)$\\
$\Delta T_6$ [-1.5 to -3.0] (K) & $\mathcal{U}(200, 900)$ & ${\rm log}(\tilde{X}_{\rm Fe})$ & $\mathcal{U}(-2.3, 1)$\\ 
$\Delta T_7$ [-3.0 to -5.0] (K) & $\mathcal{U}(100, 700)$ & \logco & $\mathcal{U}(0, 6)$  \\
\hline 
ROX 12 b\\
\hline 
$\Delta T_1$ [0.7 to 0.0] (K) & $\mathcal{U}(150, 850)$ & $T_{\rm anchor}$ [log($P$)=-0.7)] (K) & $\mathcal{U}(2200, 3000)$  \\
$\Delta T_2$ [0.0 to -0.3] (K) & $\mathcal{U}(0, 450)$ &  Mass ($\Mj$) & $\mathcal{N}(19.0, 5.0)$  \\
$\Delta T_3$ [-0.3 to -0.7] (K) & $\mathcal{U}(50, 550)$ & Radius ($\Rj$) & $\mathcal{N}(2.2, 0.35)$   \\
$\Delta T_4$ [-0.7 to -1.0] (K) & $\mathcal{U}(0, 450)$ & Comp. flux, $\alpha_c$ (counts) & $\mathcal{U}(0, 100)$   \\
$\Delta T_5$ [-1.0 to -1.5] (K) & $\mathcal{U}(50, 650)$ & ${\rm log}(\tilde{X}_{\rm Al_2O_3})$ & $\mathcal{U}(-2.3, 1)$\\
$\Delta T_6$ [-1.5 to -3.0] (K) & $\mathcal{U}(200, 900)$ & ${\rm log}(\tilde{X}_{\rm Fe})$ & $\mathcal{U}(-2.3, 1)$\\ 
$\Delta T_7$ [-3.0 to -5.0] (K) & $\mathcal{U}(100, 700)$ & \logco & $\mathcal{U}(0, 6)$  \\
\hline 
GSC 6214-210 b\\
\hline 
$\Delta T_1$ [1.0 to 0.5] (K) & $\mathcal{U}(200, 650)$ & $T_{\rm anchor}$ [log($P$)=-0.1)] (K) & $\mathcal{U}(1700, 3000)$  \\
$\Delta T_2$ [0.5 to 0.2] (K) & $\mathcal{U}(100, 500)$ &  Mass ($\Mj$) & $\mathcal{N}(21.0, 6.0)$  \\
$\Delta T_3$ [0.2 to -0.1] (K) & $\mathcal{U}(100, 450)$ & Radius ($\Rj$) & $\mathcal{N}(1.55, 0.25)$   \\
$\Delta T_4$ [-0.1 to -0.4] (K) & $\mathcal{U}(0, 550)$ & Comp. flux, $\alpha_c$ (counts) & $\mathcal{U}(0, 100)$   \\
$\Delta T_5$ [-0.4 to -1.0] (K) & $\mathcal{U}(400, 1000)$ & ${\rm log}(\tilde{X}_{\rm Al_2O_3})$ & $\mathcal{U}(-2.3, 1)$\\
$\Delta T_6$ [-1.0 to -2.0] (K) & $\mathcal{U}(100, 750)$ & ${\rm log}(\tilde{X}_{\rm Fe})$ & $\mathcal{U}(-2.3, 1)$\\ 
$\Delta T_7$ [-2.0 to -4.7] (K) & $\mathcal{U}(0, 500)$  &  \\ 
\enddata
\end{deluxetable*}

\begin{deluxetable*}{llll}[t!]
\tabletypesize{\footnotesize}
\tablehead{Parameter & Prior & Parameter & Prior}
\startdata
\hline 
2M 0122 b\\
\hline 
$\Delta T_1$ [1.0 to 0.5] (K) & $\mathcal{U}(350, 800)$ & $T_{\rm anchor}$ [log($P$)=-0.3)]  (K) & $\mathcal{U}(950, 1800)$  \\
$\Delta T_2$ [0.5 to 0.2] (K) & $\mathcal{U}(100, 500)$ &  Mass ($\Mj$) & $\mathcal{N}(25.0, 12.0)$  \\
$\Delta T_3$ [0.2 to -0.3] (K) & $\mathcal{U}(0, 500)$ & Radius ($\Rj$) & $\mathcal{N}(1.2, 0.2)$   \\
$\Delta T_4$ [-0.3 to -0.7] (K) & $\mathcal{U}(0, 300)$ & Comp. flux, $\alpha_c$ (counts) & $\mathcal{U}(0, 100)$   \\
$\Delta T_5$ [-0.7 to -1.0] (K) & $\mathcal{U}(0, 300)$ & ${\rm log}(\tilde{X}_{\rm MgSiO_3})$ & $\mathcal{U}(-2.3, 1)$\\
$\Delta T_6$ [-1.0 to -2.5] (K) & $\mathcal{U}(0, 400)$ & ${\rm log}(\tilde{X}_{\rm Fe})$ & $\mathcal{U}(-2.3, 1)$\\ 
$\Delta T_7$ [-2.5 to -4.5] (K) & $\mathcal{U}(0, 400)$ &  \\
\hline 
ROXs 42 Bb\\
\hline 
$\Delta T_1$ [0.3 to -0.4] (K) & $\mathcal{U}(200, 1200)$ & $T_{\rm anchor}$ [log($P$)=-1.0)]  (K) & $\mathcal{U}(1300, 3000)$  \\
$\Delta T_2$ [-0.4 to -0.7] (K) & $\mathcal{U}(50, 550)$ &  Mass ($\Mj$) & $\mathcal{N}(13.0, 5.0)$  \\
$\Delta T_3$ [-0.7 to -1.0] (K) & $\mathcal{U}(50, 550)$ & Radius ($\Rj$) & $\mathcal{N}(2.1, 0.35)$   \\
$\Delta T_4$ [-1.0 to -1.3] (K) & $\mathcal{U}(50, 550)$ & Comp. flux, $\alpha_c$ (counts) & $\mathcal{U}(0, 100)$   \\
$\Delta T_5$ [-1.3 to -2.0] (K) & $\mathcal{U}(50, 750)$ & ${\rm log}(\tilde{X}_{\rm Al_2O_3})$ & $\mathcal{U}(-2.3, 1)$\\
$\Delta T_6$ [-2.0 to -3.5] (K) & $\mathcal{U}(50, 750)$ & ${\rm log}(\tilde{X}_{\rm Fe})$ & $\mathcal{U}(-2.3, 1)$\\ 
$\Delta T_7$ [-3.5 to -5.0] (K) & $\mathcal{U}(50, 650)$  &  \\ 
\enddata
\tablecomments{For definitions and table notes see Table~\ref{tab:param_prior}. In the table entries for companions after GQ Lup b, we omit a few common parameters that use the same priors. The different cloud species considered for each companion are indicated. Note that we only fit for speckle flux in the GQ Lup b and kap And b data, as explained in \S~\ref{sec:forward_m}.}
\end{deluxetable*}

\onecolumngrid
\FloatBarrier
\section{Best-fit Models from Retrievals}\label{app:model_bestfit}
\restartappendixnumbering
Here, we plot the KPIC spectra and best-fit models for the seven other companions. The corresponding plot for GQ Lup b is shown in Fig.~\ref{fig:kpic_model_gqlup}.

\begin{figure*}[t!]
    \centering
\includegraphics[width=\linewidth]{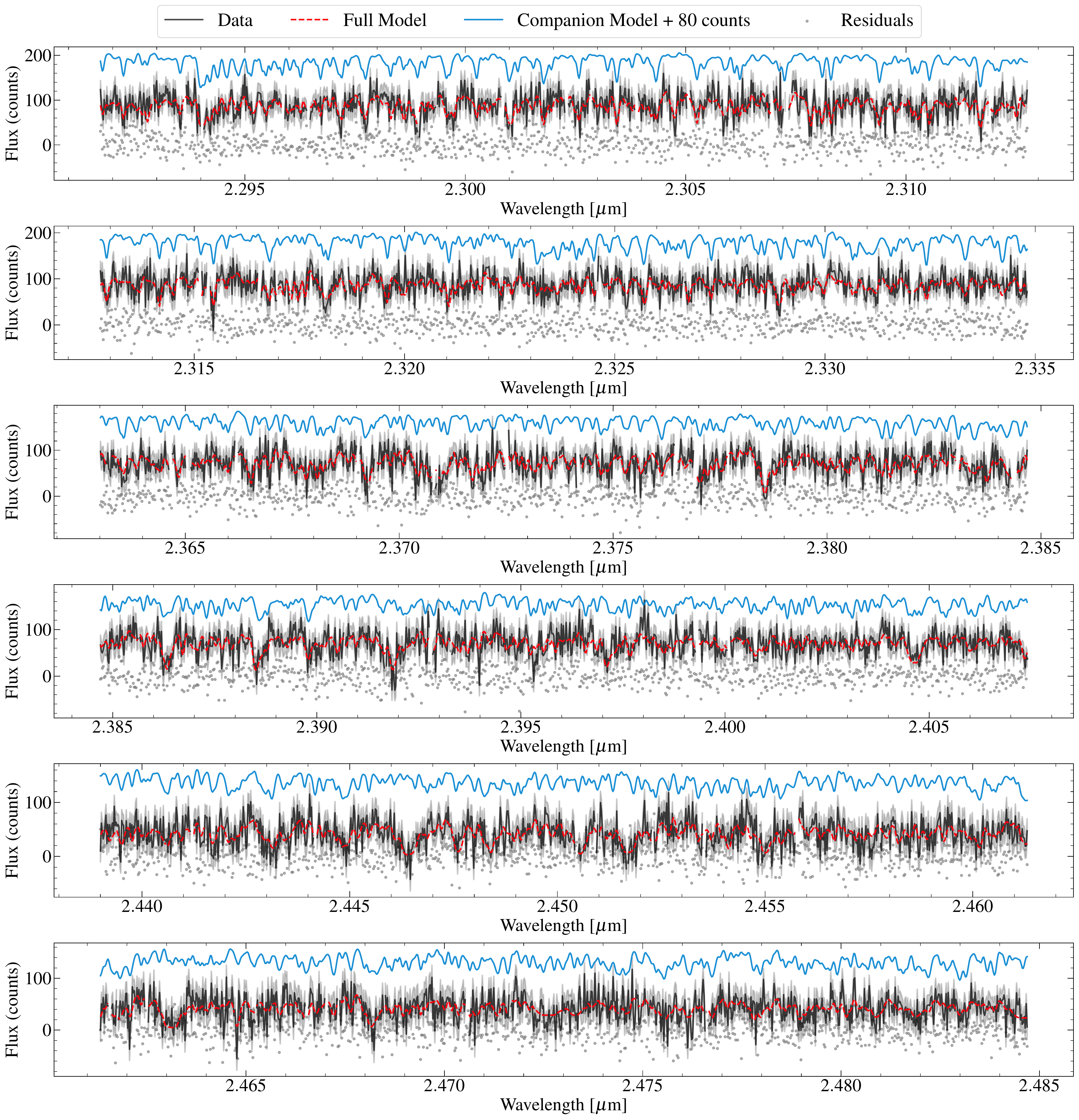}
    \caption{Same as Fig.~\ref{fig:kpic_model_gqlup}, but for HIP~79098~b.}
    \label{fig:kpic_model_hip}
\end{figure*}

\begin{figure*}[t!]
    \centering
\includegraphics[width=\linewidth]{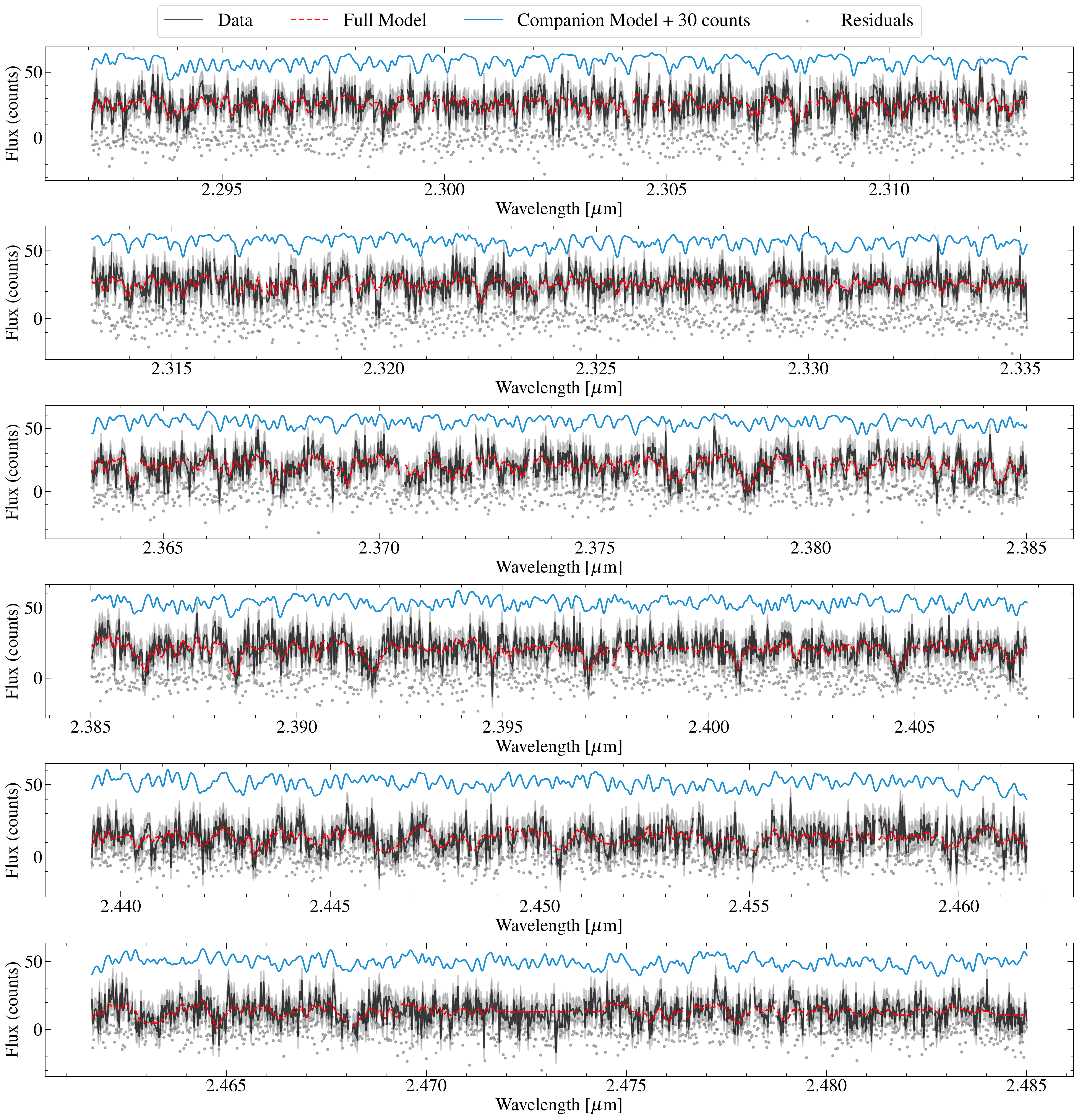}
    \caption{Same as Fig.~\ref{fig:kpic_model_hip}, but for DH~Tau~b.}
\end{figure*}

\begin{figure*}[t!]
    \centering
\includegraphics[width=\linewidth]{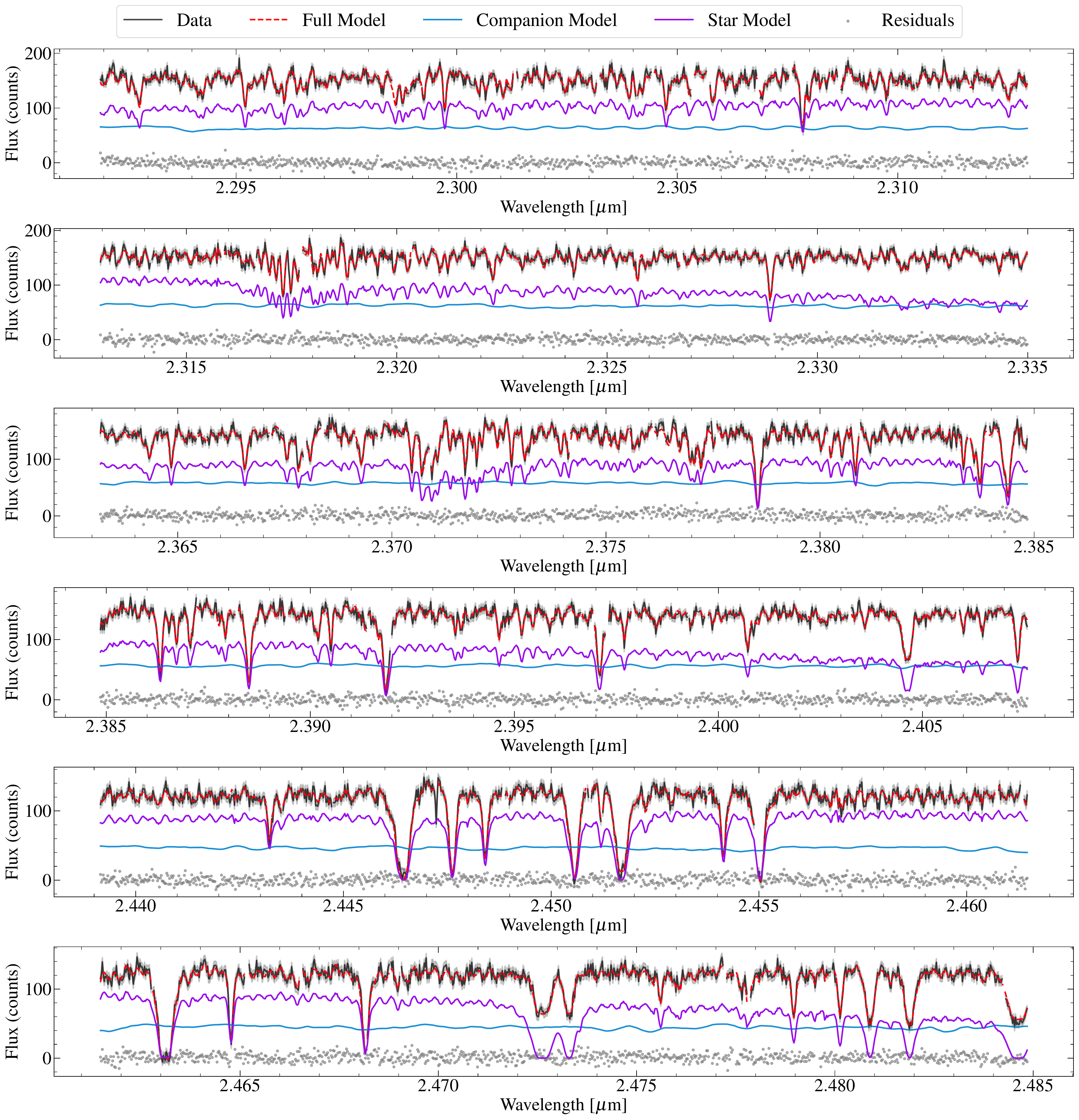}
    \caption{Same as Fig.~\ref{fig:kpic_model_hip}, but for kap~And~b. }
\end{figure*}

\begin{figure*}[t!]
    \centering
\includegraphics[width=\linewidth]{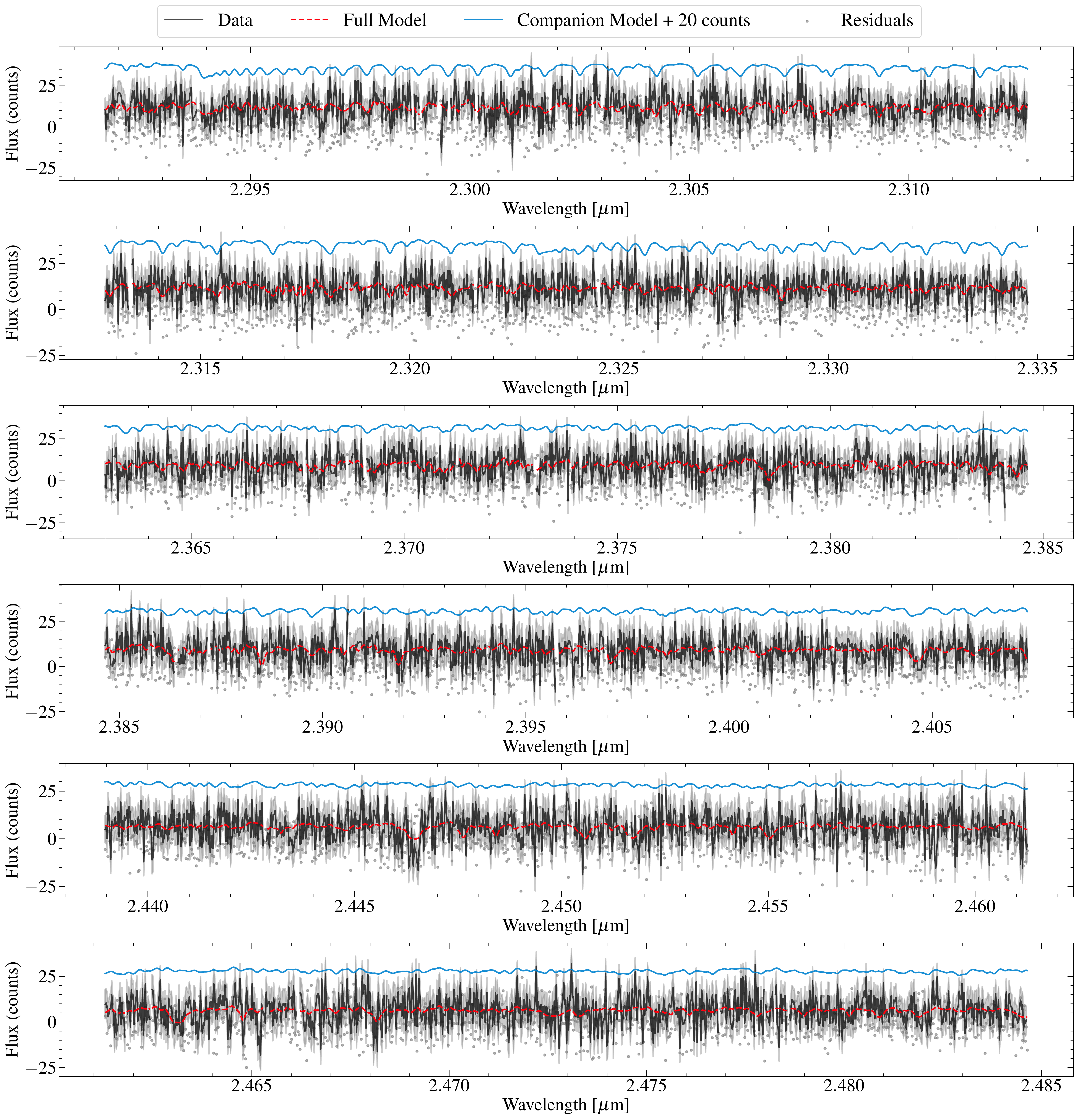}
    \caption{Same as Fig.~\ref{fig:kpic_model_hip}, but for GSC~6214-210~b.}
\end{figure*}

\begin{figure*}[t!]
    \centering
\includegraphics[width=\linewidth]{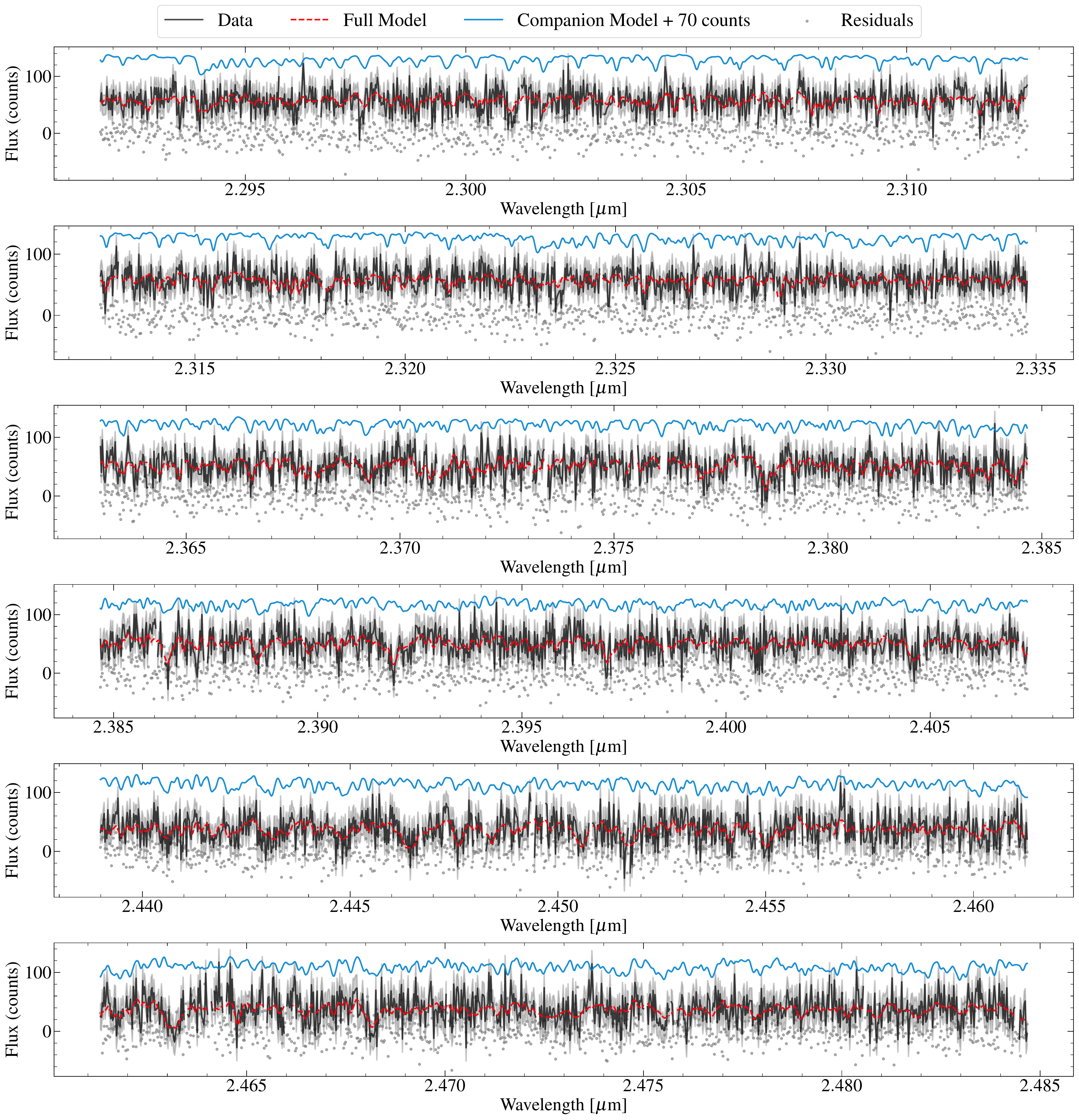}
    \caption{Same as Fig.~\ref{fig:kpic_model_hip}, but for ROXs~12~b.}
\end{figure*}

\begin{figure*}[t!]
    \centering
\includegraphics[width=\linewidth]{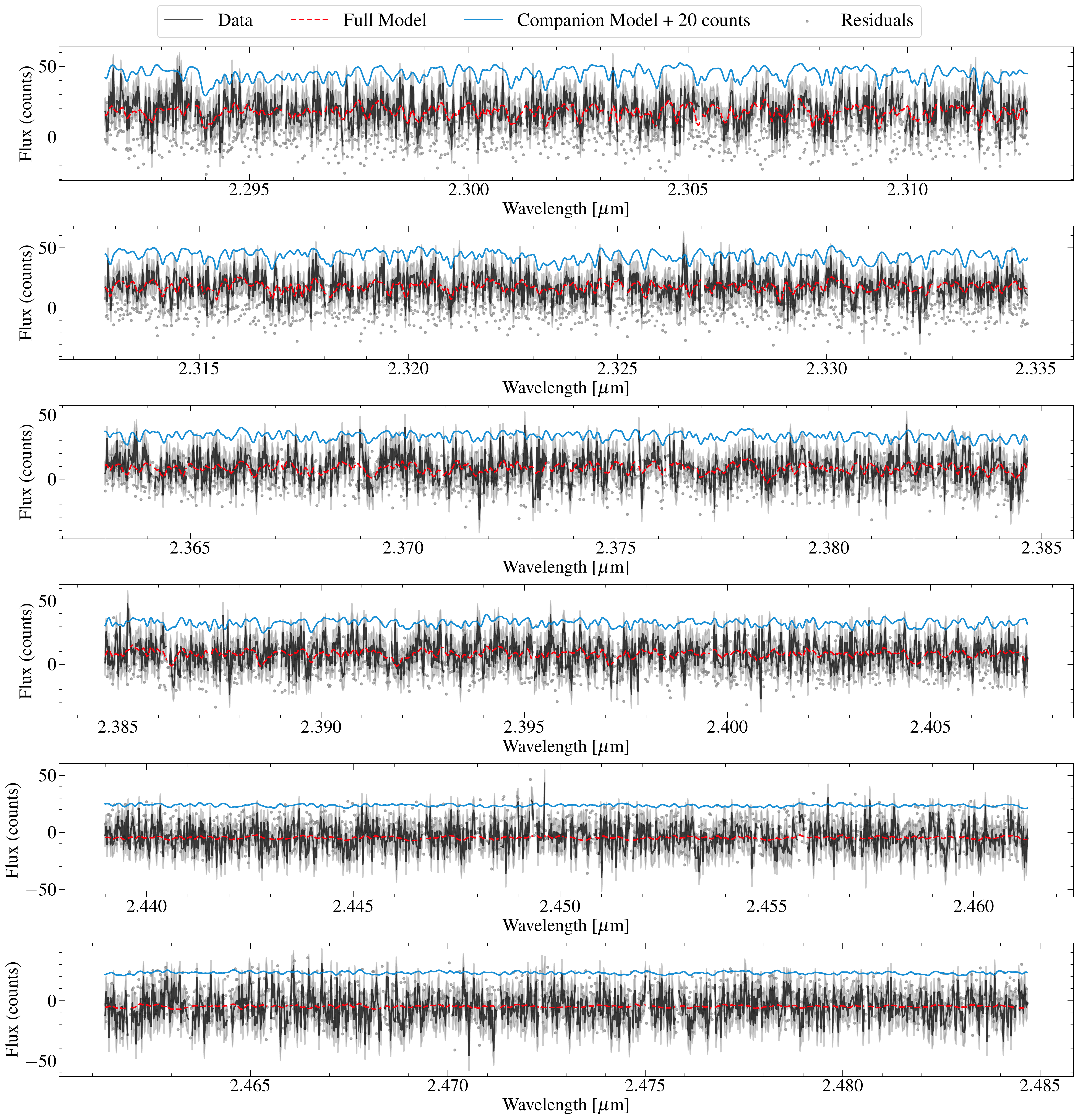}
    \caption{Same as Fig.~\ref{fig:kpic_model_hip}, but for ROXs 42B b.}
\end{figure*}

\begin{figure*}[t!]
    \centering
\includegraphics[width=\linewidth]{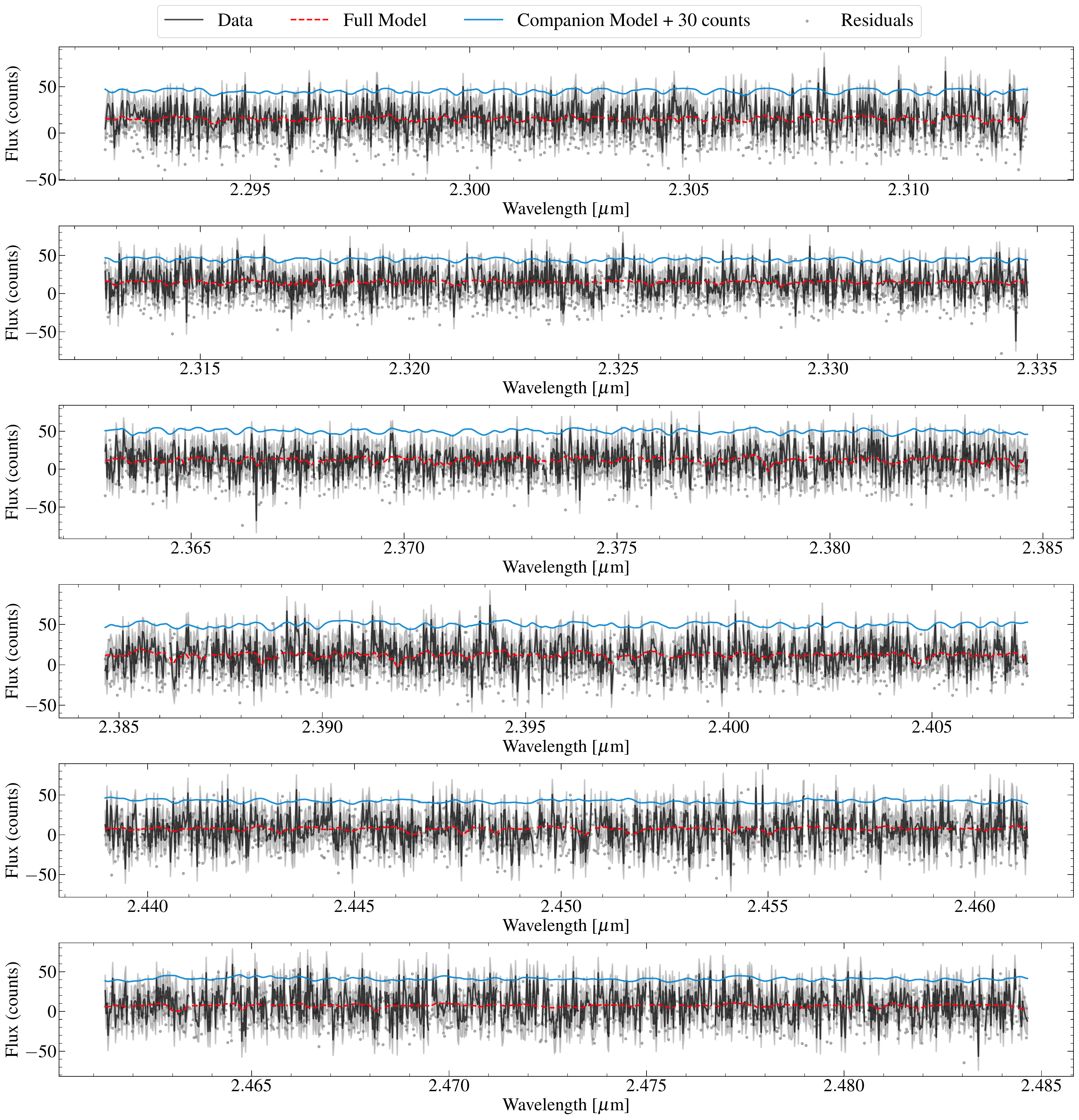}
    \caption{Same as Fig.~\ref{fig:kpic_model_hip}, but for 2M0122~b. Visually, and from the CCF analysis (Fig.~\ref{fig:ccf_detect}), this is our lowest S/N dataset. Retrievals for 2M0122~b should be re-visited with improved S/N data.}
\end{figure*}

\FloatBarrier
\section{Pressure-temperature profiles and emission contribution functions}\label{app:pt}
\restartappendixnumbering
Here, we plot the retrieved temperature profiles and emission contribution functions for the five other companions. The corresponding plots for \kapandb, GSC 6214-210 b, and HIP~79098~b are shown in Fig.~\ref{fig:ptprofile}. We note that a super-adiabatic region is visible in the P-T profile of GQ Lup b around $5\times10^{-1}$ bars. Deviations from self-consistent 1D models such as SPHINX and Sonora are not unexpected due to 3D effects such as rotation-induced horizontal transport \citep{Tan2021b}. However, our data alone is insufficient to show that this super-adiabatic region is real. For purposes of this paper, we check that the retrieved P-T profile of GQ Lup b does not bias its retrieved parameters by running a separate retrieval with a fixed P-T profile matching the $\Teff=2600$~K, $\logg=4.0$ SPHINX profile. The resulting posteriors are consistent to within $1\sigma$ for all parameters, demonstrating that the results are not sensitive to the exact shape of the P-T profile.

\begin{figure*}[b!]
    \centering
    \includegraphics[width=.46\linewidth]{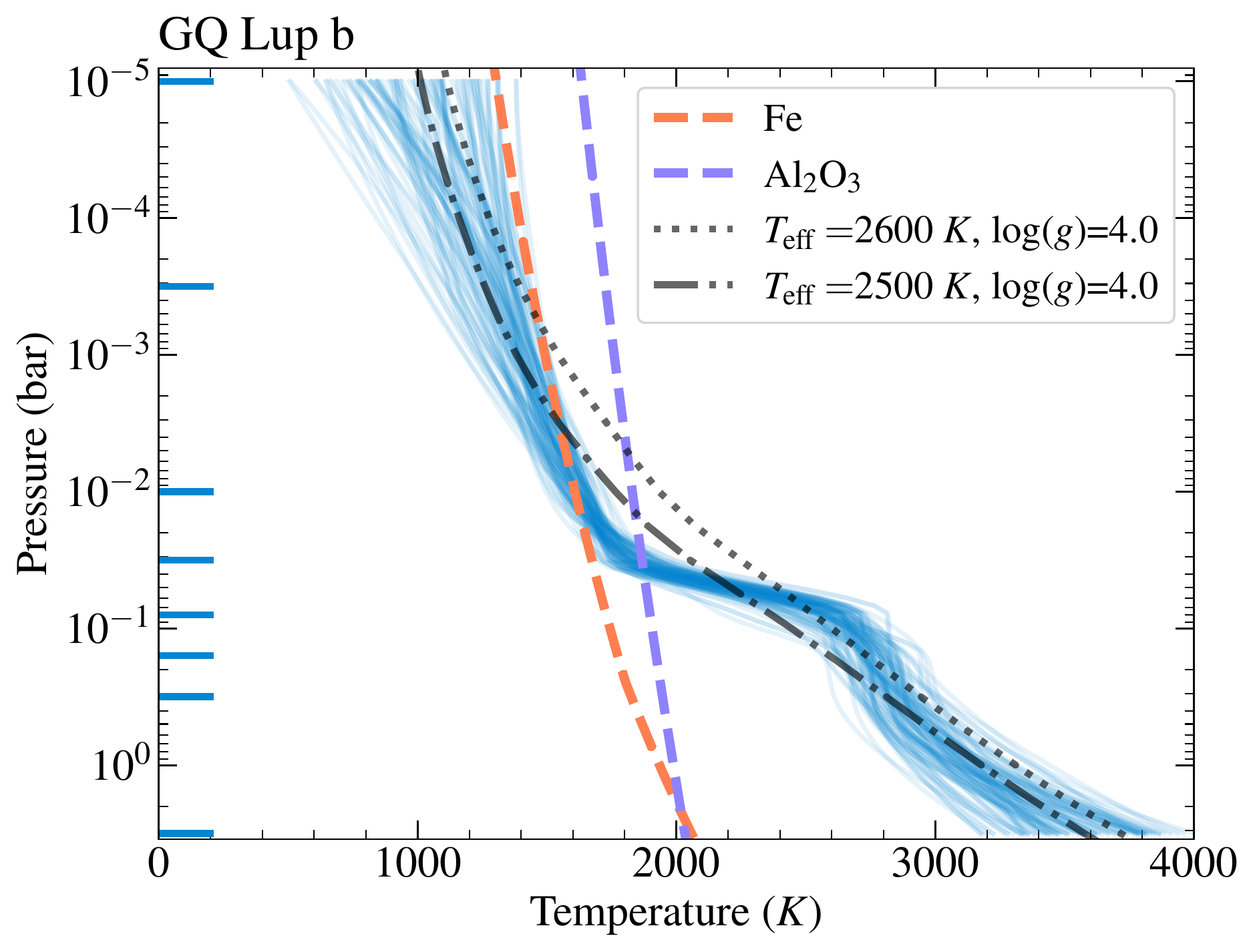}
    \centering
    \includegraphics[width=.44\linewidth]{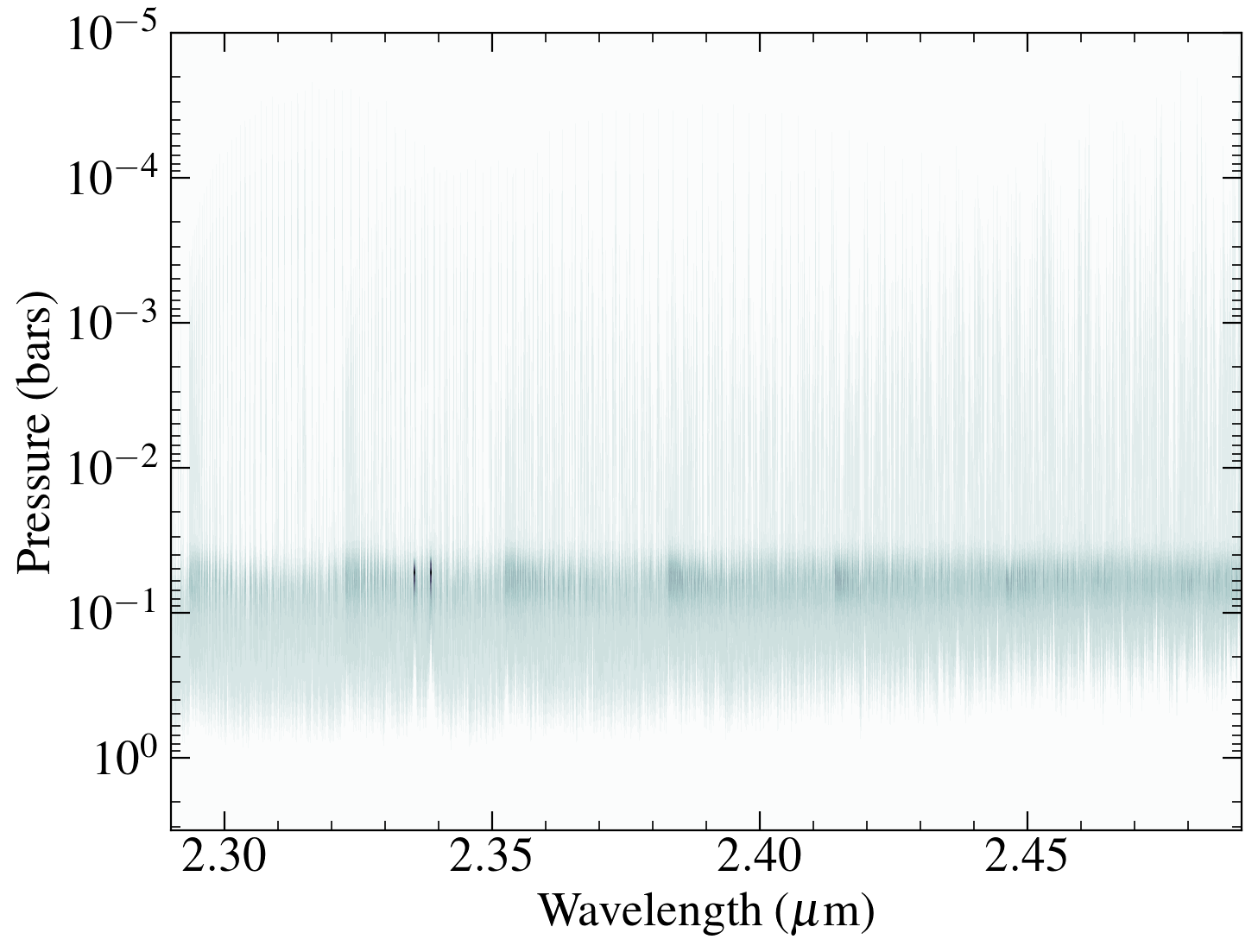}
    \centering
    \includegraphics[width=.46\linewidth]{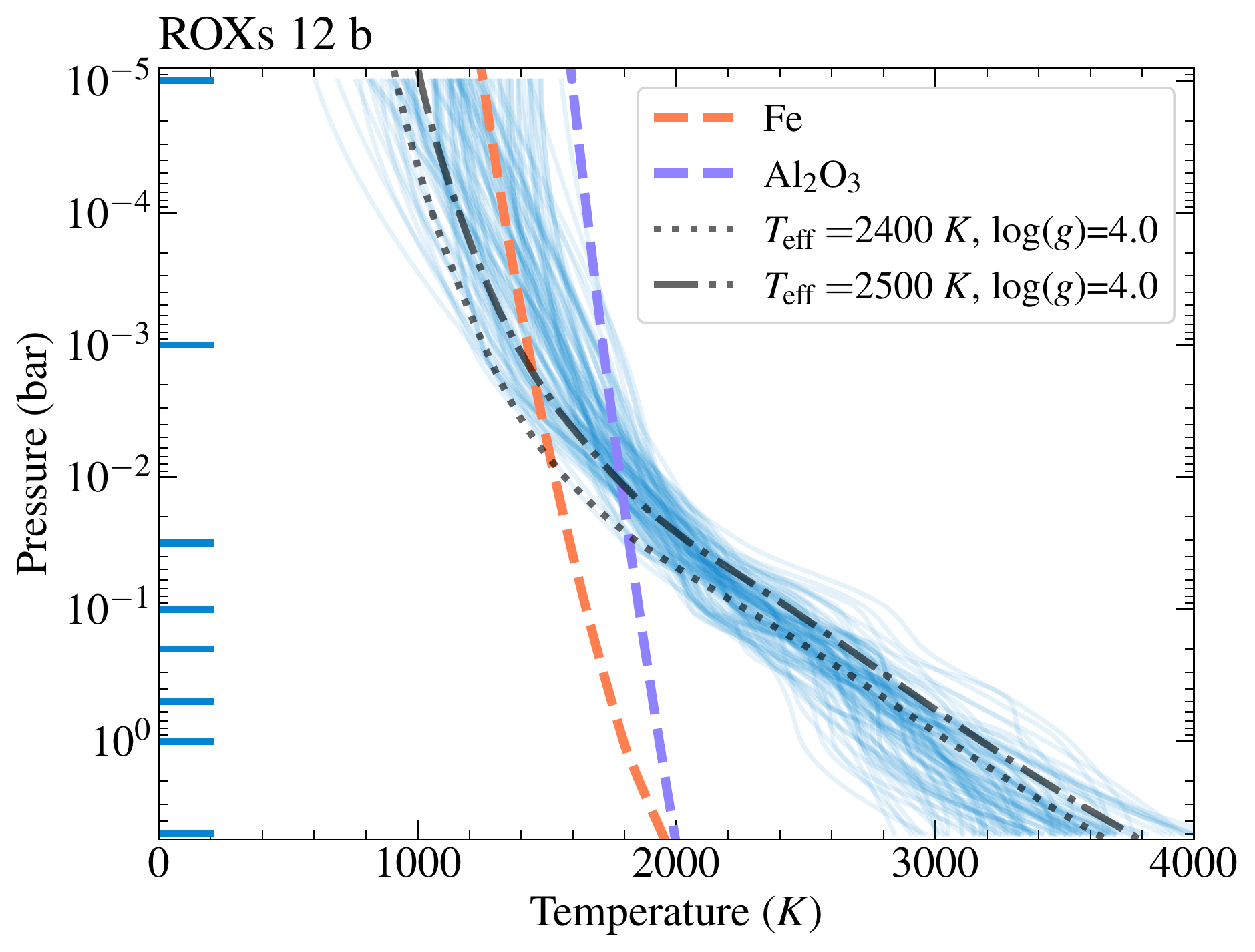}
    \centering
    \includegraphics[width=.44\linewidth]{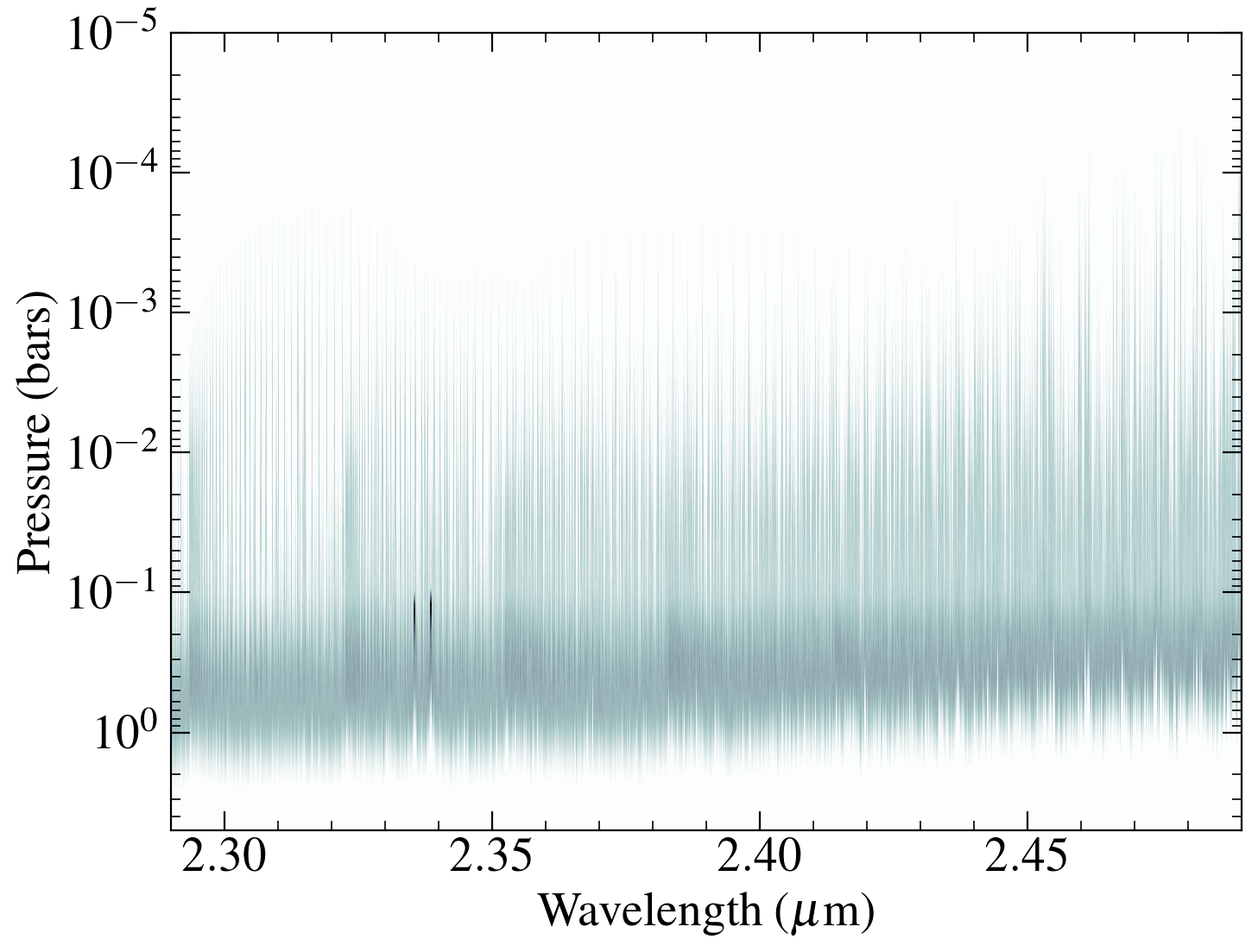}
    \centering
    \includegraphics[width=.46\linewidth]{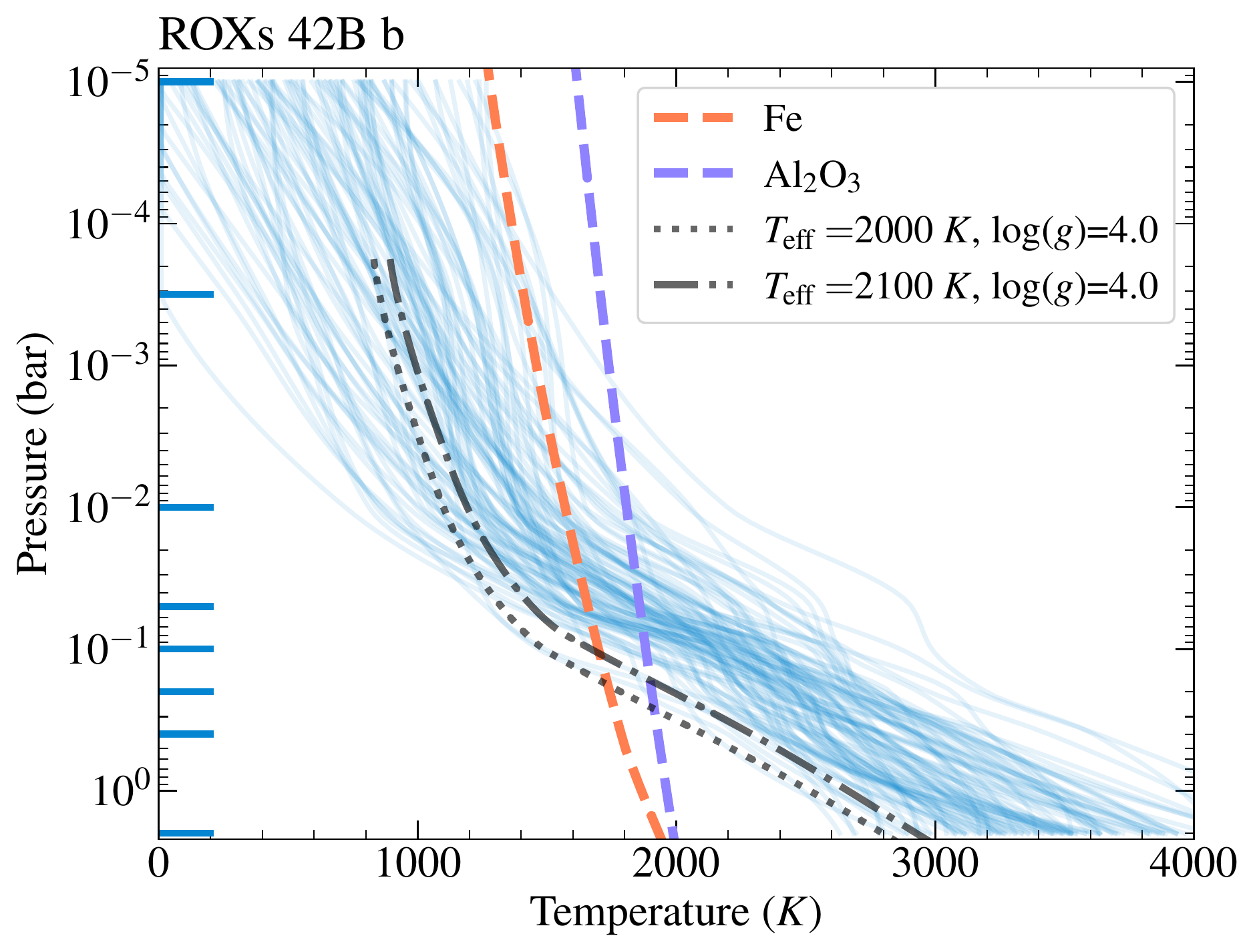}
    \centering
    \includegraphics[width=.44\linewidth]{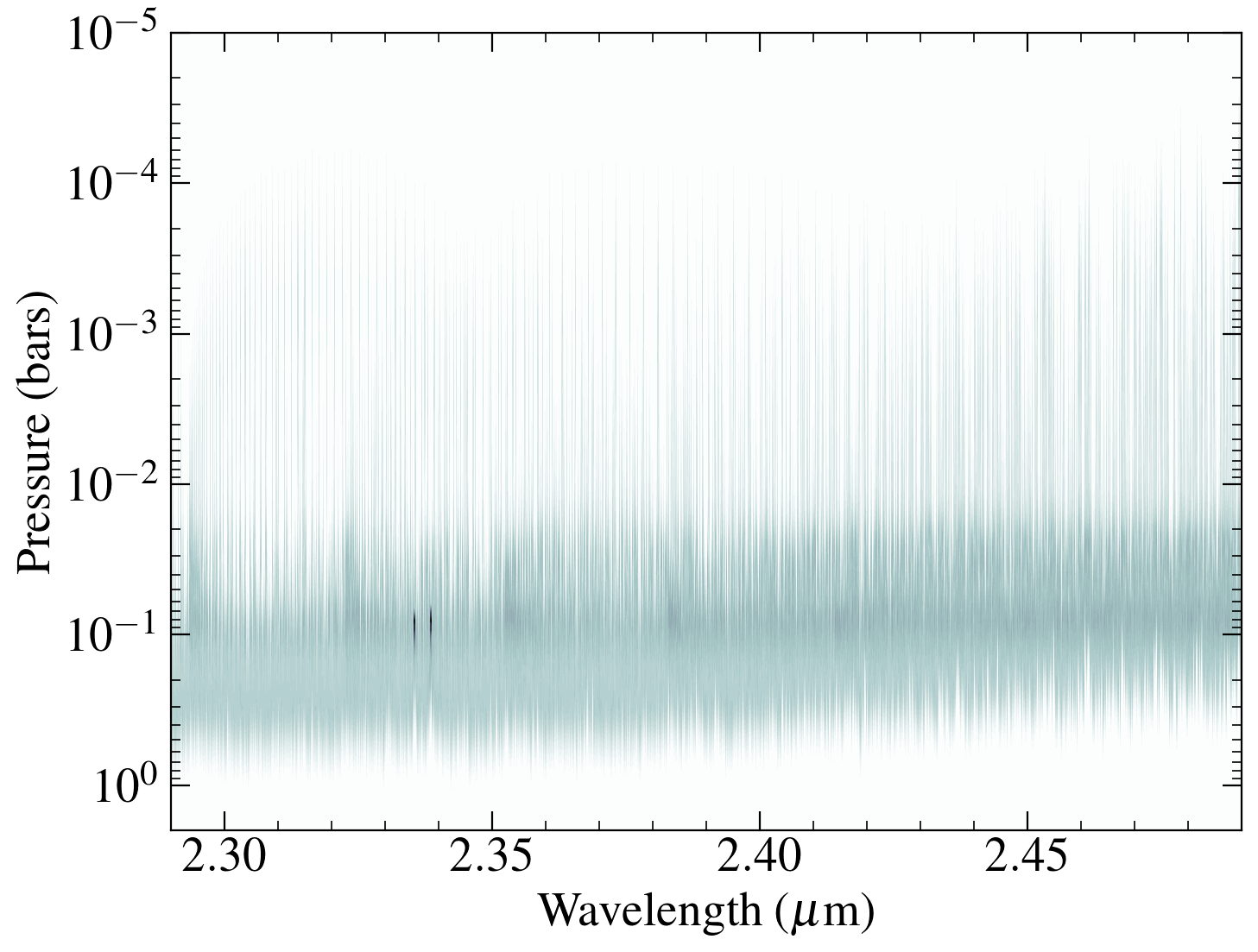}
\end{figure*}

\begin{figure*}
    \centering
    \includegraphics[width=.46\linewidth]{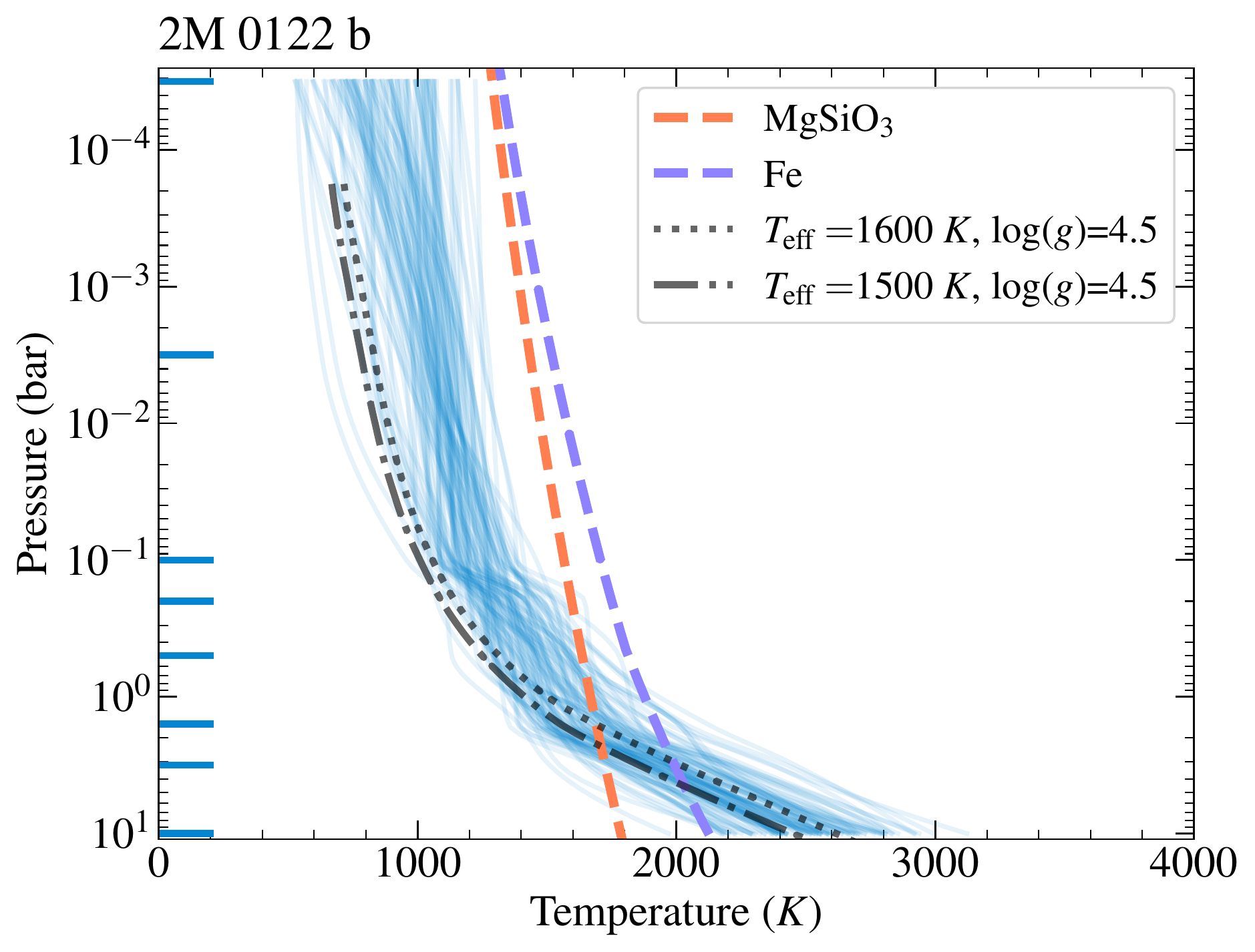}
    \centering
    \includegraphics[width=.44\linewidth]{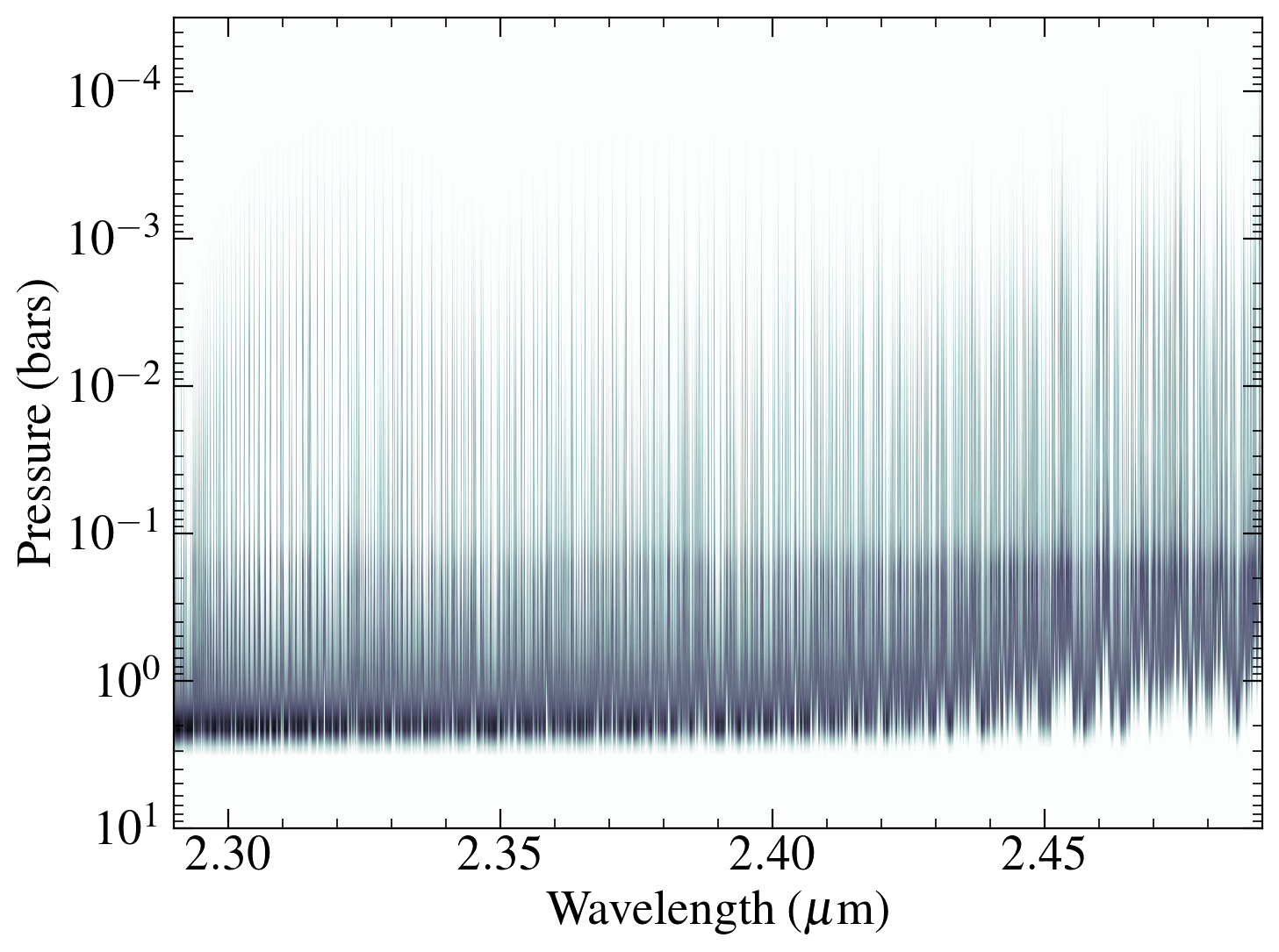}
    \centering
    \includegraphics[width=.46\linewidth]{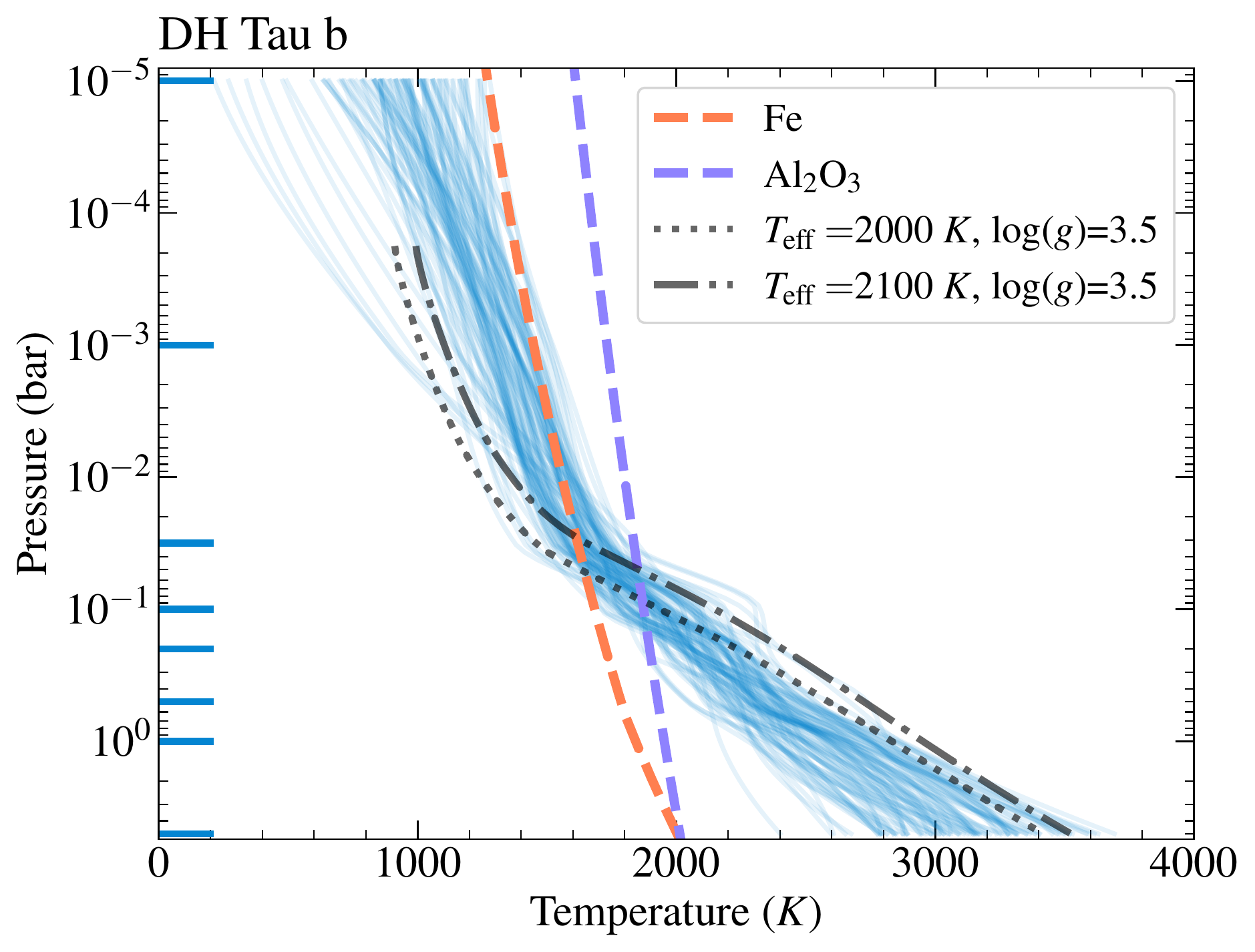}
    \centering
    \includegraphics[width=.44\linewidth]{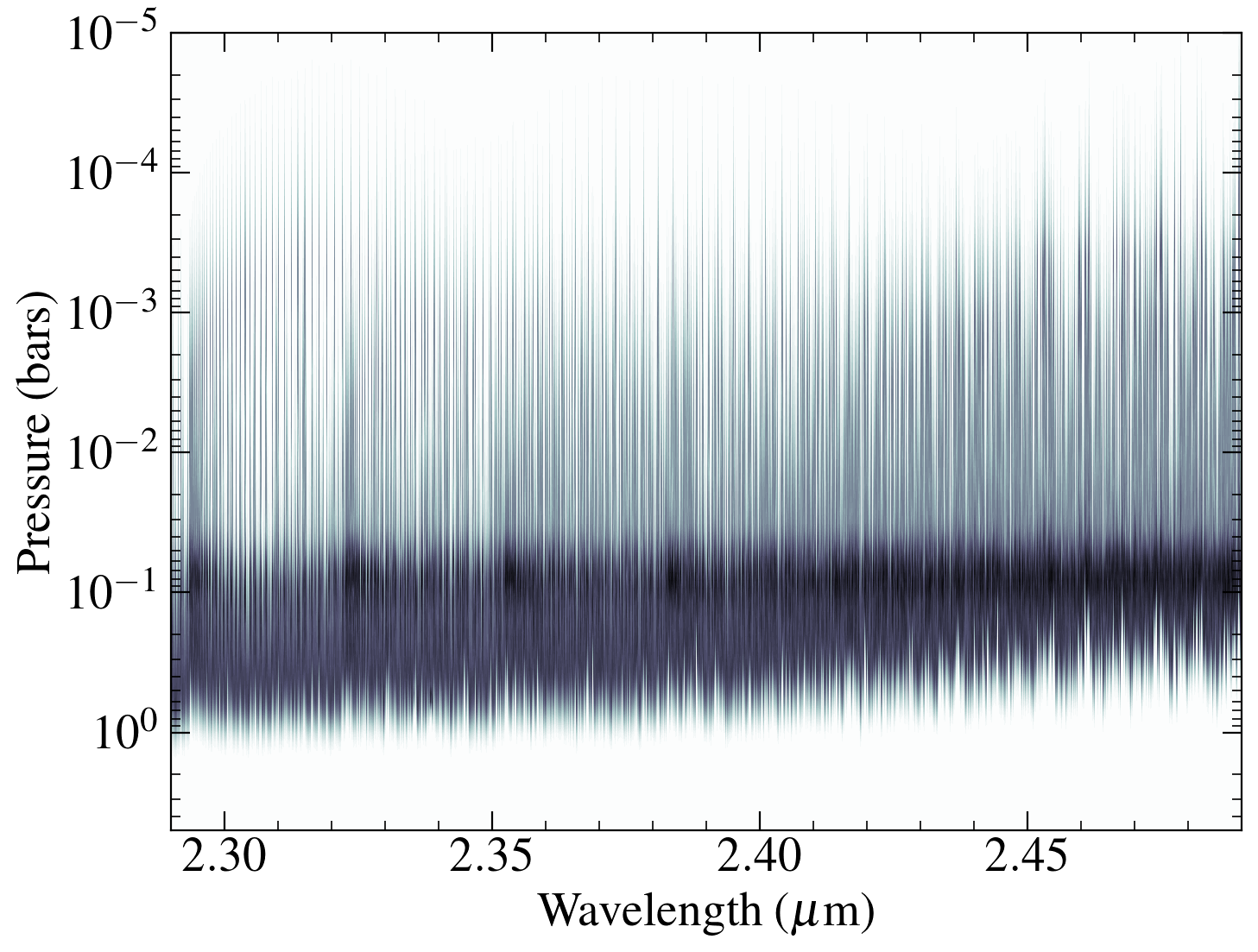}
    \centering
    \caption{Same as Fig.~\ref{fig:ptprofile}, but for the other five companions.}
\end{figure*}

\FloatBarrier
\section{Radial velocities for the substellar companions}\label{app:rvs}
\restartappendixnumbering
Here, we list the measured RVs of the eight companions, which have been corrected for barycentric motion.  

\begin{deluxetable*}{ccccc}[h]
    \tabletypesize{\small}
    \tablecaption{KPIC radial velocity measurements for the eight substellar companions studied in this work. We have applied the barycentric correction to the RVs, so their reference is the solar system barycenter.}\label{tab:rvs}    
    \tablehead{\colhead{Object} & \colhead{UT Date} & \colhead{BJD-2400,000} & \colhead{RV} (\kms)}
    \startdata
    \hline
        kap~And~b & 2022 Nov 12 & 59895.347 & $-17.7\pm0.9$ \\
        GSC~6214-210~b & 2023 June 23 & 60118.407 & $-5.5\pm0.7$ &  \\
        GQ~Lup~b & 2023 June 23 & 60118.296 & $0.0\pm0.1$ &  \\
        HIP~79098~b & 2022 July 18 & 59778.331 & $-6.0\pm0.1$ &  \\
        2M0122~b & 2021 Nov 19 & 59537.338 & $10.8\pm1.5$ &  \\
        ROXs~12~b & 2020 July 3 & 59033.347 & $-2.4\pm0.2$ &  \\
        ROXs 42B b & 2020 July 2 & 59032.347 & $-3.3\pm0.4$ &  \\
        DH~Tau~b & 2022 Oct 12 & 59864.434 & $14.9\pm0.3$ &  \\
    \enddata
\end{deluxetable*}

\clearpage

\bibliography{main.bib}{}
\bibliographystyle{aasjournal}

\end{document}